\documentclass[%
reprint,
amsmath,amssymb,
aps,
]{revtex4-2}

\usepackage{xcolor}
\usepackage{graphicx}
\usepackage{dcolumn}
\usepackage{bm}
\usepackage{siunitx}
\usepackage{multirow}
\usepackage{subfigure}
\usepackage{amsthm,amsmath,amssymb}
\usepackage{mathrsfs}
\usepackage{balance}
\usepackage{verbatim}

\begin{document}

\preprint{APS/123-QED}

\title{Evaluation of radiative depolarization in the future Circular Electron-Positron Collider}

\author{Wenhao Xia$^{1,2}$}
\altaffiliation[Present address: ]{State Key Laboratory of NBC Protection for Civilian, Beijing 102205, China}
\author{Zhe Duan$^1$}%
\email{duanz@ihep.ac.cn }
\author{Desmond P. Barber$^{3,4}$}%
\author{Yiwei Wang$^1$, Bin Wang$^1$}%
\author{Jie Gao$^{1,2}$}%
\email{gaoj@ihep.ac.cn}
\affiliation{$^1$ Key Laboratory of Particle Acceleration Physics and Technology,
Institute of High Energy Physics, Chinese Academy of Sciences,
19B Yuquan Road, Beijing 10049, China \\
$^2$ University of Chinese Academy of Sciences,
19A Yuquan Road, Beijing 10049, China \\
$^3$ Deutsches Elektronen-Synchrotron (DESY) Hamburg, 22607, Germany \\
$^4$ Department of Mathematics and Statistics, University of New Mexico Albuquerque, NM 87131, USA
}%
\date{\today}

\begin{abstract}
Polarized lepton beams are an important aspect in the design of the future 100 \si{km}-scale Circular Electron-Positron Collider (CEPC). Precision beam energy calibration using resonant depolarization, as well as longitudinally polarized colliding beams are being actively investigated. The achievable beam polarization level for various beam energies and application scenarios depends on the radiative depolarization in the collider rings. In this paper
the radiative depolarization effects are evaluated for a CEPC collider ring lattice with detailed machine imperfections and corrections. Simulations with the SLIM and Monte-Carlo approaches using the Bmad/PTC codes are compared with the theory of  the effects of spin diffusion for ultra-high beam energies and the validity of the theories is thereby addressed. The paper concludes with a summary and suggestions for further investigations. 
\end{abstract}

\maketitle


\section{Introduction}
After the discovery of the Higgs particle at the Large Hadron Collider (LHC), a study of the Circular Electron Positron Collider (CEPC) was launched as one of the global design efforts of future electron-positron circular colliders ~\cite{FCCcdr,shilt}, for precision measurements of Higgs boson properties and electroweak interactions, as well as direct searches for new physics with an unprecedented accuracy.
In the Conceptual Design Report (CDR)~\cite{CEPCcdr} released in November 2018, the CEPC was designed as a double-ring collider with a circumference of about 100 \si{km}, 
and to be operated at beam energies of around 45 \si{\GeV} as a Z factory, 80 \si{\GeV} as a W factory and 120 \si{\GeV} as a Higgs factory. After the CDR, the CEPC accelerator entered the phase of the Technical Design Report (TDR) endorsed by the CEPC International Advisory Committee (IAC).  
Recent updates of the CEPC accelerator R\&D are summarized in \cite{cepcsnowmass}. 

Beam polarization is an important design aspect of the CEPC. On one hand, the resonant depolarization technology~(RD)~\cite{vepp} is essential for precision beam energy calibration
at the Z-pole and the WW threshold. RD utilizes a horizontal oscillating magnetic field to 
excite a vertically polarized beam. As the excitation frequency is scanned, the beam gets depolarized when the perturbation is resonant with the spin precession frequency, which is closely related to the beam energy. RD requires a vertical beam polarization level of 5\%$-$10\% at the CEPC, according to experience at LEP~\cite{lep}. On the other hand,
colliding beam experiments with longitudinal beam polarization, which would greatly broaden the potential of the physics program, are also being explored at the CEPC~\cite{duanz2}. For this purpose,
a longitudinal beam polarization of at least 50\% for one or both $e^+$ and $e^-$ beams at the interaction points is desired. These require preparation and maintenance and 
then
manipulation of the polarized $e^+/e^-$ beams with spin rotators.

Spin rotators have been successfully implemented into 
the CEPC CDR lattice for the Z-pole energy in a separate study~\cite{xia_investigation_2022}. Note that since the 
$e^+$ and $e^-$ beams in the CEPC are in separate rings, their spin helicities can be controlled independently.

The precession of the spin expectation value $\vec S$, and of 
 $\hat S = \vec S /|\vec S|$, of a relativistic charged particle, in electric and magnetic fields follows the Thomas-BMT equation~\cite{thomas,BMT},
which, for a circular accelerator or storage ring,  can be expressed in the following form,
\begin{equation}
\frac{d\hat S}{d\theta}=\left [ \vec{\Omega}_0(\theta)+ \vec{\omega}(\vec u;\theta) \right] \times \hat S. 
\label{eq:tbmt}
\end{equation}
with 
\begin{equation}
\vec{\Omega}_0(\theta)= \vec{\Omega}_{00}(\theta)+\Delta \vec{\Omega}(\theta)
\label{eq:Yospinmotionrev}
\end{equation}
where $\vec{\Omega}_{0}(\theta)$  is the spin-precession vector on the closed orbit at the azimuthal angle $\theta$.  $\vec{\Omega}_{00}(\theta)$ and $\Delta \vec{\Omega}(\theta)$ denote the contribution of the fields on the design orbit and the magnet errors/correction fields to the spin precession vector, respectively. $\vec{\omega}(\vec u;\theta)$ is due to the orbital oscillations $\vec u$
relative to the closed orbit. We use
a Frenet-Serret coordinate system, where
 $\vec{e}_x,\vec{e}_y,\vec{e}_z$ are the unit vectors pointing radially outwards, vertically upwards and longitudinally (clockwise), respectively, so that they form a right-handed unit-vector basis.
In this coordinate system, 
the particle's phase-space coordinate can be expressed as
$\vec u=(x, p_x, y, p_y, z, \delta)$. $x$ and $y$ are the transverse coordinates of the particle. The transverse phase-space momenta $P_x$ and $P_y$ are normalized by the reference momentum $P_0$, namely $p_x=P_x/P_0$ and $p_y=P_y/P_0$, respectively. $z=-\beta c \Delta t$ where $\Delta t$ is the time difference between the particle and the reference particle arriving at  the azimuthal angle $\theta$. $\beta=v/c$, where $v$ and $c$ are the velocity of the particle and light, respectively. $\delta=\Delta{E}/{E}_0$ is the relative energy deviation.
The spin polarization of an electron bunch is the ensemble average of the $\hat S$.

In a circular accelerator running at a fixed energy
$\vec{\Omega}_{0}$ is a periodic function of $\theta$, for which there is  a unit-length periodic solution 
of Eq.~(\ref{eq:tbmt}), $\hat{n}_0(\theta)$, satisfying $\hat{n}_0(\theta+2\pi)=\hat{n}_0(\theta)$. Then a spin vector $\hat S$ perpendicular to $\hat{n}_0$ precesses by a rotation angle $2\pi\nu_0$ in one revolution around $\hat{n}_0$, where $\nu_0$ is called the closed-orbit spin tune. In a storage ring designed with a planar geometry and no solenoids (hereafter referred as
the ``planar ring''), 
 $\hat{n}_{0}(\theta)$ is close to the vertical direction, and $\nu_0 \approx a \gamma_0$, where $a=0.00115965219$ for electrons (positrons), and $\gamma_0$ is the relativistic factor
 for the design energy.

These key concepts of the spin motion on the closed orbit can be extended to the more general phase-space coordinates. 
One special unit-length solution of Eq.~(\ref{eq:tbmt}) is the invariant spin field (ISF)~\cite{barber}, $\hat{n}(\vec u;\theta)$, satisfying 
the periodicity condition $\hat{n}(\vec u;\theta+2\pi)=\hat{n}(\vec u;\theta)$ and satisfying 
Eq.~(\ref{eq:tbmt})
along particle trajectories, ${\vec u}(\theta)$, obeying Hamilton's
equations.  
Assuming that the orbital motion is integrable, so that the
orbital motion can be expressed in terms of action-angle variables, the analogue of $\nu_0$ for the combined betatron and synchrotron motion with orbital actions  ${\vec I} \equiv (I_{x}, I_{y}, I_{z})$,
is the amplitude-dependent spin tune $\nu_s(\vec I)$ 
describing the rate of spin precession around $\hat{n}$ \cite{barber}. The projection of the spin vector of a particle on the invariant spin field $J_S = \hat S \cdot \hat{n} $ is an adiabatic invariant of its spin motion~\cite{hoffstaetter_adiabatic_2006}.
On the closed orbit, $\hat{n}(\vec u;\theta)$ and $\nu_s(\vec J)$ reduce to $\hat{n}_{0}(\theta)$ and $\nu_0$, respectively, but for typical
orbital amplitudes with electrons $\nu_s(\vec I) \approx \nu_0$ in any case.

Machine imperfections and orbital oscillations contribute to $\Delta \vec{\Omega}(\theta)$
and $\vec{\omega}(\vec u;\theta)$ respectively, and may perturb the spin motion in a resonant manner when the following condition of spin-orbit coupling resonances (spin resonances in short) is nearly satisfied,
\begin{equation}
\nu_s(\vec I) =k+k_x \nu_x +k_y \nu_y+ k_z \nu_z,~~~ k, k_x, k_y, k_y \in Z
\label{eq:res}
\end{equation}
where $\nu_x, ~\nu_y$ and $\nu_z$ are the horizontal, vertical betatron tunes and the synchrotron tune, respectively.  $\hat{n}(\vec u;;\theta)$ deviates from $\hat{n}_{0}(\theta)$ significantly near these spin resonances.
As will be seen below, in perturbative estimates 
of spin motion, $\nu_s(\vec I)$ is replaced by $\nu_0$ in the above  resonance condition.
Note that the orbital tunes $\nu_x, \nu_y$ and $\nu_z$ in the above definition of the spin resonances, are 
conventionally used in the case of weak couplings, reduced from the more general orbital tunes $\nu_{I}$, $\nu_{II}$ and $\nu_{III}$ obtained from the orbital eigen-analysis~\cite{chao_evaluation_1979}.
Spin resonances with $\nu_0=k, k \in Z$,  
are called integer spin resonances.
In a perfectly aligned planar ring with no solenoids $\hat{n}_{0}(\theta)$ 
is vertical. In a misaligned ring $\hat{n}_{0}(\theta)$ can deviate
strongly from the design direction when $\nu_0$ is near an integer.
Spin resonances with $|k_x|  +|k_y| + |k_z|=1$ and $|k_x|  +|k_y| + |k_z| > 1$ are called  first-order spin resonances and 
higher-order spin resonances, respectively. 

Unlike polarized proton beams which must be polarised at the source, $e^+/e^-$ beams can also become spontaneously polarized in a storage ring by the
emission of synchrotron radiation.
In a perfectly aligned ring with no solenoids with vertical $\hat n_0$ the polarization is vertical but in general the polarization of the beam is along $\hat n_0$. 
The evolution of the beam polarization $P(t)$ in an electron storage ring 
for an initially unpolarised beam is
\begin{equation}
P(t)= P_{\infty}[1-\exp{(-t/\tau_p)}]
\label{eq:pvstp}
\end{equation}
where 
\begin{equation}
P_{\infty} = \frac{-8}{5\sqrt{3}}\frac{\oint d\theta \frac{  \hat{n}_0\cdot\hat{b}}{|\rho|^3}}  {\oint d\theta  \frac{1-\frac{2}{9}(\hat{n}_0 \cdot \hat{s})^2}{|\rho|^3} }
\label{eq:p2}
\end{equation}
where $\hat{s}$ is a unit vector along the particle's direction of motion, and $\hat{b}= \hat{s} \times \dot{\hat{s}}/|\dot{\hat{s}}|$, is the direction of
the guiding magnetic field, which is normally vertical.
$P_{\infty}$ is the equilibrium beam polarization taking into account the orbital imperfections, but disregarding the radiative depolarization effect~\cite{baier_kinetics_1970}.
The time constant of the Sokolov-Ternov effect, $\tau_p$, is
\begin{equation}
\tau_p^{-1} \approx \frac{5 \sqrt{3}}{8}\frac{r_e \gamma_0^5\hbar}{m_e}\frac{1}{2\pi}\oint d\theta   \frac{ 1-\frac{2}{9}(\hat{n}_0\cdot \hat{s})^2}{|\rho|^3}
\label{eq:tao_p}
\end{equation}
In an ideal planar ring,  $P_{\infty}$ is 92.4\%.
In contrast, as mentioned, for an
imperfect storage ring near integer values of $\nu_0$, $\hat{n}_{0}$ may deviate from
the vertical direction, and lead to a reduced level of $P_{\infty}$.

In addition to the Sokolov-Ternov effect, synchrotron radiation also causes radiative depolarization in electron storage rings, 
namely the spin diffusion effect.
It is implicit in  standard  literature~\cite{eq,dkm} that the polarization at each point in phase space $\vec u$ is 
parallel to $\hat n (\vec u; \theta)$. This is the the so-called ISF approximation
which we discuss again in the Conclusions.
A photon emission perturbs the orbital motion and leads to a change in $\hat{n}$ and thus a slightly different $\hat S \cdot \hat{n} $. The stochastic nature of synchrotron radiation, then leads to a diffusion in $\hat S \cdot \hat{n} $ and thus a reduction of beam polarization~\cite{baier_quantum_1966}, 
which has  also been called
``non-resonant spin diffusion''~\cite{dks}.
However, 
in the proximity of spin resonances where spin motion becomes far from adiabatic,
$\hat S \cdot \hat{n} $ is no longer an invariant of motion, and  
it has  been suggested that
the above physics picture
is no longer complete.
Then a second mechanism, 
which could be important at very high energy, 
namely the repetitive fast but ``uncorrelated'' crossing of the spin resonances, due to stochastic photon emissions, 
would be
another source
of radiative depolarization~\cite{K74,DK75},
which 
has also been called
``resonant spin diffusion''~\cite{dks}. In any case, the evolution of beam polarization in an electron storage ring is a compromise between the spontaneous polarization build-up
and the radiative depolarization effects.

Analytical estimates of the equilibrium beam polarization 
in the presence of just non-resonant spin diffusion
are generally made with the Derbenev-Kondratenko (DK) formula ~\cite{eq,dkm}: 
\begin{equation}
P_{eq}= \frac{-\frac{8}{5\sqrt{3}} \times \oint d\theta \left\langle \frac{1}{|\rho|^3} \hat{b} \cdot (\hat{n}-\frac{\partial \hat{ n}}{\partial \delta})\right \rangle}{\oint d\theta \left\langle\frac{1}{|\rho|^3}\left[1-\frac{2}{9}(\hat{n}\cdot \hat{s})^2 +\frac{11}{18}(\frac{\partial \hat{n}}{\partial \delta})^2\right]\right \rangle}
\label{eq:dk}
\end{equation}
where the $\left\langle\right\rangle$ brackets denote an average over the orbital phase space at azimuth $\theta$. 
The spin-orbit coupling function
$\partial \hat{n}/\partial \delta$  quantifies the depolarization. This can be rather large near spin resonances, leading to a reduced level of equilibrium polarization, as well as a smaller self-polarization build-up time $\tau_\text{tot}$ relative to $\tau_p$
\begin{equation}
\tau_{\text{tot}}^{-1}=\tau_{p}^{-1}+\tau_{d}^{-1}
\label{eq:tao_tot}
\end{equation}
where $\tau_{d}$ is the time constant of the spin diffusion effect,
\begin{equation}
\tau_d^{-1}=\frac{5 \sqrt{3}}{8}\frac{r_e \gamma_0^5\hbar}{m_e}\frac{1}{2\pi}\oint d\theta \left\langle \frac{\frac{11}{18}(\frac{\partial \hat{n}}{\partial \delta})^2}{|\rho|^3}\right \rangle     
\label{eq:tao_d}
\end{equation}

The key to evaluation of the DK formula lies in the evaluation of $\partial \hat{n}/\partial \delta$.  
As already indicated, in the derivation of the DK formula, 
only the first mechanism 
(non-resonant spin diffusion)
of radiative depolarization was included. 
Then as suggested in~\cite{DK75,dks},
the DK formula should be augmented to take into account the effects of resonant spin diffusion too.

In a mainly-planar storage ring, the equilibrium polarization can also be approximated by
\begin{equation}
P_{eq}\approx \frac{P_{\infty}}{1+\frac{\tau_{p}}{\tau_{d}}}
\label{eq:p1}
\end{equation}
This equation disregards the tiny effect of kinetic polarization associated with the term $\hat{b} \cdot \frac{\partial \hat{ n}}{\partial \delta}$ in Eq.~\ref{eq:dk}, but captures the essentials of the trade-off between the Sokolov-Ternov effect and the radiative depolarization effect, regardless of the underlying mechanism of radiative depolarization.
The strength of the depolarization effects is quantified by the ratio $\tau_p/\tau_d$.

So far, the evaluation of the equilibrium polarization in real storage rings has involved two classes of computer algorithms, as reviewed in detail in
Ref~\cite{barber2,barber3}.
One class of algorithms numerically evaluates $\hat{n}$ and $\frac{\partial \hat{ n}}{\partial \delta}$ and then applies Eq.~(\ref{eq:dk}) to calculate the equilibrium polarization.
For example,
SLIM~\cite{slim} treats linearized orbital and spin motion, and thus reflects the influence of first-order spin resonances
while taking into account the effects of tilted $\hat n_0$ near 
integer values of $\nu_0$.  It can be classed as a first-order perturbative
algorithm. Several other algorithms
employ full three-dimensional spin motion ~\cite{mane_electron-spin_1987,sodom,sodomII,spinlie} and thus expose the higher-order spin resonances resulting from that.  Among these are the higher-order perturbative algorithm of Ref.~\cite{mane_electron-spin_1987}
and the non-perturbative approaches of the  SODOM algorithms \cite{sodom,sodomII}.

The other class of algorithms evaluate the depolarization time $\tau_d$ through Monte-Carlo simulations, like SITROS~\cite{sitros}, 
SLICKTRACK~\cite{barber_polarisation_2005} and Bmad~\cite{Bmad},
 and then calculate the equilibrium polarization with Eqs.~(\ref{eq:p2}), (\ref{eq:tao_p}) and (\ref{eq:p1}). Unlike the previous class of algorithms, the Monte-Carlo method,
 with its use of full three-dimensional spin motion,
 exposes both first-order and higher-order spin resonances automatically, while not relying
on the calculation of $\hat{n}$ and $\frac{\partial \hat{ n}}{\partial \delta}$
and  Eq.~(\ref{eq:dk}). Nor is it dependent on other theories of spin diffusion.
Therefore, the Monte-Carlo method can be used to check theoretical models of radiative depolarization. The Monte-Carlo method has the additional advantage that it can handle non-linear orbital motion such as that 
caused by beam-beam forces.
 
Resonant and non-resonant spin  diffusion for ultra-high
energy electron storage rings  have been discussed in detail by 
Derbenev, Kondratenko and Skrinsky ~\cite{dks}.
These theories were later compared with the experiments with vertically polarized beams in LEP~\cite{assmann}. 
In the past decade, global interest has arisen in building future $e^+ e^-$ circular colliders.
Mane worked out a scaling of equilibrium beam polarization 
at ultra-high beam energies and studied possibilities for using Siberian snakes to mitigate the depolarization while retaining self-polarization at the same time~\cite{mane2}. Nikitin evaluated the radiative depolarization and the equilibrium transverse
polarization in these super electron-positron circular colliders theoretically, and discussed options of achieving longitudinally polarized colliding beams~\cite{nikitin2019,nikitin2020}. There are also attempts to re-evaluate the radiative depolarization and 
equilibrium beam polarization using an approach based on the Bloch equation~\cite{heinemann_bloch_2019,beznosov_wave_2020,Heinemann_2020pzl}.

There have already been some simulations for the equilibrium polarization for the future circular $e^+ e^-$ colliders at ultra-high energies.
For example, Gianfelice-Wendt launched Monte-Carlo simulations with the SITROS package\cite{sitros} of the achievable self-polarization at
CERN's Future Circular $e^+ e^-$ collider (FCC-ee)~\cite{wendt} at 45 and 80 \si{\GeV} beam energies, 
aiming at energy calibration using RD. 
Quadrupole misalignments and beam-position-monitor errors were introduced in a simplified ring lattice without the interaction regions. It was shown that a reasonable polarization level can be achieved in the collider ring at both energies with a well planned orbit correction scheme, while asymmetric wigglers are 
necessary to boost the self-polarization build-up at 45 \si{\GeV}.
Koop developed a Monte-Carlo simulation code with a simplified 
model including one-turn energy-dependent spin precession and a lumped element to model spin perturbations, in addition to
synchrotron oscillations and synchrotron radiation. This simulation code was also used to study the RD process~\cite{koop_resonant_2018}.

On our side
a  Monte-Carlo simulation of radiative depolarization
based on the Polymorphic Tracking Code (PTC)~\cite{PTC, FPP} 
was developed in Ref.~\cite{duanz}. The equilibrium beam polarization was simulated for a model ring lattice with a circumference of 50 \si{km} and artificial skew quadrupoles to excite betatron spin resonances.
That study supported the suggestion in Ref.~\cite{dks} that the spin dynamics enter an ``uncorrelated regime'' of spin resonance crossing at ultra-high beam energies. However, it was not 
clear if this paradigm shift would exist in a realistic imperfect ring.     

For this paper, we extend our studies to the radiative depolarization effects for the CEPC at the Z pole as well as higher beam energies, and for various operation scenarios. We
use a CEPC CDR lattice that contains two interaction regions, and includes detailed error modeling and corrections.
This lattice does not contain solenoids and has relatively weak orbital correctors with horizontal
magnetic fields, and belongs to the category of the ``planar ring''. We present the theories of radiative
depolarization applicable to this case.
Monte-Carlo simulations based on the PTC code~\cite{PTC, FPP, duanz} were then applied to estimate the radiative depolarization effects, which are compared to those from the theories of radiative depolarization at ultra-high beam energies.

This paper is arranged as follows.
More details of theories of radiative depolarization in ultra-high energy electron storage rings are reviewed in Section~\ref{sec:theory}. The setup of the CEPC lattice is introduced in Section~\ref{sec:lattice}. In Section~\ref{sec:ana}, the theories are applied to the CEPC lattice to evaluate the depolarization effects. Section~\ref{sec:sim},
presents the results of Monte-Carlo simulations for the equilibrium polarization for the standard CEPC lattice as well as for a lattice with
asymmetric wigglers and a double RF system and these are compared with the results from theoretical models. The last section provides a summary and suggestions for future numerical and theoretical work.

\section{Theories of radiative depolarization~\label{sec:theory}}

In the following we focus our analysis on realistic planar electron storage rings.
Severe depolarization may occur when the 
closed-orbit spin tune $\nu_0$ is near an integer $k$, so that 
$\hat n_0$ is tilted from the design direction  and  then as shown below, the two first-order ``parent" synchrotron spin resonances
$\nu_0\pm\nu_z=k$ can therefore be strong, even overlapping with each other. Depolarization also occurs near 
first-order ``parent" betatron spin resonances $\nu_0\pm\nu_r=k, ~r=x,y$.
Moreover, even if $\nu_0$ is not near an integer, its distance to adjacent integer spin resonances
 and first-order spin resonances
is less than 220~\si{MeV}. As the spread of the spin precession frequency
increases with the beam energy, it becomes more difficult to 
attain high polarization. In particular, 
as we show in the perturbation theory sketched below, higher-order synchrotron ``sideband'' spin resonances $\nu_0\pm\nu_r + m\nu_z=k$, where
$m$ is an integer and $r=x,y,z$, become much more prominent, compared to those in lower-energy electron storage rings. The strongest of these resonances are centered on first-order ``parent" spin resonances. 
Note that in this context an integer spin resonance $\nu_0=k$ will be a ``sideband''
spin resonance of the ``parent'' first-order synchrotron spin resonances $\nu_0\pm\nu_z=k$. Note also that this kind of resonance can even occur when the ring is perfectly aligned and $\hat n_0$ is not tilted from the design direction as, for example, when spin rotators make $\hat n_0$ horizontal
in a region with quadrupoles and dispersion.
The higher-order ``parent" resonances also carry 
synchrotron sideband resonances.

Observations of beam polarization in electron storage rings have so far been in broad agreement with the expectations of Eq.~(\ref{eq:dk}), taking into account of 
the depolarization associated with these first-order and higher-order spin resonances.
However, as already mentioned and as  proposed in Ref.~\cite{K74, DK75, dks}, 
the above picture with 
its consequences explored using perturbation theory, belongs to the ``non-resonant spin diffusion'',
while a somehow different picture of  ``resonant spin diffusion'' is required for ultra-high beam energies.
This argument is better appreciated from a ``dynamical'' perspective as described in more detail below.
Under certain circumstances, 
the combined influence of synchrotron oscillation and synchrotron radiation could lead to ``repetitive fast
but uncorrelated crossings'' of the underlying spin resonances. Note that in this case, the influence of the 
resonant spin diffusion is not limited to the close proximity of the underlying spin resonances.

Now, to set the scene, we'll first present the depolarization theory of the ``parent" first-order spin resonances, or ``first-order theory'' in short. We explicitly 
derive the dominating contribution of first-order ``parent" synchrotron spin resonances to radiative depolarization characterized by $\tau_p/\tau_d$, and
establish its connection with the strengths of tilts of $\hat n_0$ near  integer values of $\nu_0$. 
Then we briefly review the theory of (high-order) synchrotron sideband spin resonances, 
referred to as the theory of the ``correlated regime'' in the following context. In these analyses, it will be shown that, at ultra-high beam energies the 
synchrotron sideband spin resonances centered on integers are the most important contributors to depolarization. Following that we'll also refer to other, more detailed, perturbative calculations surrounding  higher-order ``parent" resonances and their sidebands.
Then the theory of ``uncorrelated regime'' 
will also be reviewed. These theories will be compared to simulations in later sections.

To exhibit the depolarization effects of the first-order ``parent''
resonances and their sidebands in a storage ring with practical machine imperfections, we'll now follow 
Yokoya's perturbative approach~\cite{yok,yok2}. 
We denote the right-handed orthonormal set of unit-length solutions to Eq.~(\ref{eq:tbmt}) on the design orbit by $\hat{n}_{00}$, $\hat{m}_{00}$ and  $\hat{l}_{00}$, where $|\Delta \vec{\Omega}|=|\vec{\omega}|=0$, and define  $\hat{k}_{00}=\hat{m}_{00}+i\hat{l}_{00}$. Then
$\hat{k}_{00}(\theta)=e^{i(\Upsilon(\theta)-\Upsilon(\theta')}\hat{k}_{00}(\theta')$,
where $\Upsilon(\theta)$ is the spin precession phase. In a planar ring, $\Upsilon(\theta)\approx\nu_0\Phi(\theta)$,
where $\Phi(\theta)=R\int _0^{\theta}\frac{1}{\rho_x}d\theta$ is the integral of the bending angle between the azimuthal angles 0 and $\theta$,
$R$ is the average radius of the storage ring, $\rho_x$ is the radius of curvature for the local orbit.
In a perfect planar ring, $\hat{n}_{00}$ is 
in the vertical direction while $\hat{k}_{00}$ is in the horizontal plane, respectively.  Considering the perturbed fields on the closed orbit,
$|\Delta \vec{\Omega}|\neq 0, |\vec{\omega}|=0$, the perturbed orthonormal solutions of Eq.~(\ref{eq:tbmt}) are $\hat{n}_0=\hat{n}_{00}+\Delta \hat{n}_{00}$ and $\hat{k}_0=\hat{k}_{00}+\Delta \hat{k}_{00}$. The vector $\hat{k}_0$ is quasi-periodic, 
\begin{equation}
    \hat{k}_0 (\theta+2\pi)=e^{i2\pi \nu_0} \hat{k}_0 (\theta)
\label{eq:peroid_k}    
\end{equation}

When $|\Delta \vec{\Omega}|\neq 0$, $|\vec{\omega}|\neq 0$,
$\hat{n}$ can be expressed by
\begin{equation}
\hat{n}(\vec u; \theta) =\hat{n}_0 \sqrt{1-|\zeta(\vec u; \theta)|^2}+\mathscr{Re}(\hat{k}_0^{\star}\zeta(\vec u; \theta))
\end{equation}
where $\zeta(\vec u(\theta); \theta)$
satisfies
\begin{equation}
\frac{d\zeta}{d\theta}=-i\vec{\omega} \cdot \hat{k}_0 \sqrt{1-|\zeta|^2}+i\vec{\omega}\cdot \hat{n}_0 \zeta
\label{eq:zeta}
\end{equation}
To arrive at our chosen set of resonances we  choose $|\zeta|\ll1$. Then the solution for $\zeta(\vec u; \theta)$ to Eq.~(\ref{eq:zeta}) at 
$(\vec u; \theta)$ with $\zeta(\vec u; \theta)  = \zeta(\vec u; \theta + 2 \pi)$ is

\begin{eqnarray}
    \zeta(\vec u;\theta) &=&  -i e^{-i\chi(\vec u;\theta)}\substack { {\rm Lim} \\ \epsilon \rightarrow +0}    \biggl[
     \int_{-\infty}^{\theta}e^{\epsilon\theta'}e^{i\chi(\vec u(\theta'); \theta')} \nonumber \\
&& \vec{\omega}(\vec u(\theta'); \theta')\cdot \hat{k}_0(\theta')  d\theta'  \biggr]
\label{eq:zeta1}
\end{eqnarray}
where 
\begin{equation}
 \chi(\vec u;\theta)=
 \substack { {\rm Lim} \\ \epsilon \rightarrow +0} \left[-\int_{-\infty}^{\theta}
 {e^{\epsilon \theta'}} \vec{\omega}(\vec u(\theta');\theta')\cdot \hat{n}_0(\theta') d\theta' \right]
\end{equation}
The integral from $-\infty$ in these expressions can be appreciated 
as follows as an instance of the ``anti-damping'' procedure using the adiabatic invariance of $J_S = \hat S \cdot \hat{n}$~\cite{hoffstaetter_high-energy_2006,hoff,vogt_bounds_2000, mane_electron-spin_1987}.
The particle is placed infinitisimally close to the closed orbit in the 
infinite ``past'' with $\hat S \cdot \hat{n}_0 = 1$. The particle
and its spin are then tracked forward up to $\theta$ while the orbital amplitudes and the perturbation to the spin
increase very slowly and exponentially. Then at $\theta$, $\hat n$ is given by the final $\hat S$~\cite{mane_electron-spin_1987}.
 

As will be shown below, the first-order ``parent'' spin resonances are driven by the perturbation term $\vec{\omega} \cdot \hat{k}_0$,
while the higher-order synchrotron ``sideband'' spin resonances are due to the modulation of the rate of spin precession around $\hat{n}_0$
driven by $\vec{\omega}\cdot \hat{n}_0$. The spin-orbit coupling function can be obtained as
\begin{equation}
  \frac{\partial \hat{n}}{\partial\delta}  \approx \mathscr{Re} \left(\hat{k}_0^{\star} \frac{\partial \zeta}{\partial \delta}\right)
\end{equation}

\subsection{First-order ``parent'' spin resonances}
Considering the spin resonances up to the first-order, the solution for $\zeta$ is
\begin{equation}
\zeta(\vec u;\theta) \approx  \substack { {\rm Lim} \\ \epsilon \rightarrow +0} \left[-i \int_{-\infty}^{\theta} 
 {e^{\epsilon \theta'}}\vec{\omega}\cdot \hat{k}_0 d\theta' \right]
\label{eq:zeta0}
\end{equation}
where for this,
$\vec{\omega}$ is  linearized with respect 
to the betatron  coordinates $r_{\beta}, ~r=x,y$,
as well as the synchrotron coordinate $\delta$, and decomposed into three oscillation modes~\cite{yok2}:
\begin{equation}
    \vec{\omega}= \vec{\omega}_z \delta+\vec{\omega}_x x_{\beta}+\vec{\omega}_y y_{\beta}
\end{equation}
with
\begin{eqnarray}
\vec{\omega}_z&=&-R[(1+a\gamma_0)(G_x\eta_x+Q_y\eta_y) -\frac{1}{\rho_x}]\vec{e}_y\nonumber \\
              &&+R[(1+a\gamma_0)(G_y\eta_y+Q_x\eta_x) -\frac{1}{\rho_y}]\vec{e}_x \nonumber \\
\vec{\omega}_x&=&-R(1+a\gamma_0)(G_x+Q_y)\vec{e}_y \nonumber \\
\vec{\omega}_y&=&R(1+a\gamma_0)(G_y+Q_x)\vec{e}_x.
\label{eq:omega}
\end{eqnarray}
The $\rho_y$ is the vertical radius of curvature for the local orbit, which could
be due to vertical quadrupole misalignment errors, dipole roll errors, or orbital correctors. Lastly, $\eta_x,\eta_y$ are the dispersion functions and
the focusing gradients of quadrupoles or dipoles 
are expressed as
$G_x=\frac{e}{P_0}\frac{\partial B_y}{\partial x}+\frac{1}{\rho_x^2}$,
$G_y=\frac{e}{P_0}\frac{\partial B_x}{\partial y}+\frac{1}{\rho_y^2}$,
while the gradients of skew quadrupoles and inclined dipoles are expressed as
$Q_x=\frac{e}{P_0}\frac{\partial B_x}{\partial x}-\frac{1}{\rho_x\rho_y}$,
$Q_y=\frac{e}{P_0}\frac{\partial B_y}{\partial y}+\frac{1}{\rho_x\rho_y}$

Then Eq.~(\ref{eq:zeta0}) can be analyzed separately for each oscillation mode with  the integral of each mode amounting to a resonant term times a one-turn integral.  
The contributions of betatron oscillation modes $\vec{\omega}_r\cdot r_{\beta}, ~r=x,y$ to $\zeta$ are
\begin{eqnarray}
    \zeta_r(I_r,\psi_r; \theta)&=& \substack { {\rm Lim} \\ \epsilon \rightarrow +0} \left[-i\int_{-\infty}^{\theta}{e^{\epsilon \theta'}}\vec{\omega}_r r_{\beta}\cdot \hat{k}_0 d\theta' \right] \nonumber \\
    &=&\sum_{\pm}\frac{i\sqrt{I_r/2}}{e^{-2\pi{i}(\nu_0\pm\nu_r)}-1}F_{{\pm}r}(\psi_r; \theta) 
\label{eq:zetaj}
\end{eqnarray}
with
\begin{equation}
        F_{\pm{r}}(\psi_r; \theta)=\int^{\theta}_{\theta-2\pi}\vec{\omega}_r\cdot\hat{k}_0 \sqrt{\beta_r}e^{\pm i(\psi_r+\widetilde{\Psi}_r)}d\theta'
        \label{eq:intj}
\end{equation}
after using
the quasi-periodicity of $\hat{k}_0(\theta)$, and where the angle $\psi_r$ is evaluated at $\theta$ while $r_{\beta}$ is expressed in terms of action-angle variables $I_r$ and $\psi_r$  as
\begin{equation}
    r_{\beta}=\sqrt{2I_r \beta_r}\cos(\psi_r+\widetilde{\Psi}_r)
\end{equation}
and where $\beta_r$ is the betatron function and $\widetilde{\Psi}_r$ is 
\begin{equation}
    \widetilde{\Psi}_r=R\int^{\theta}_0 \frac{d\theta'}{\beta_r}-\nu_r\theta
\end{equation}
at the azimuth $\theta$.
Similarly, and in the usually
good approximation that synchrotron motion is simply harmonic with a $\theta$-independent $\beta_z$, the contribution from the synchrotron mode to $\zeta$ is
\begin{eqnarray}
    \zeta_z(I_z, \psi_z; \theta)&=& \substack{ {\rm Lim} \\ \epsilon \rightarrow +0}  \left[-i\int_{-\infty}^{\theta}{e^{\epsilon \theta'}}\vec{\omega}_z \delta \cdot \hat{k}_0 d\theta' \right] \nonumber \\
    &=&\sum_{\pm}\frac{i\sqrt{I_z/2}}{e^{-2\pi{i}(\nu_0\pm\nu_z)}-1}F_{\pm{\delta}}(\psi_z;\theta) 
\label{eq:zetaz}
\end{eqnarray}
with 
\begin{equation}
     F_{\pm{\delta}}(\psi_z; \theta)=\int^{\theta}_{\theta-2\pi}\vec{\omega}_z\cdot\hat{k}_0 \sqrt{\beta_z} e^{\pm i\psi_z}d\theta'
     \label{eq:intz}
\end{equation}
where $I_z$ and $\psi_z$ are the action-angle variables of the synchrotron oscillation, $\psi_z=\psi_{z0}+\nu_z\theta$ and
\begin{equation}
    \delta=\sqrt{2I_z \beta_z}\cos(\psi_z)
    \label{eq:sync_action_angle}
\end{equation}

Eq.~(\ref{eq:zetaj}) and Eq.~(\ref{eq:zetaz}) reveal first-order ``parent" betatron spin resonances and first-order ``parent" synchrotron spin resonances, respectively. The one-turn integrals $F_{\pm{r}}$ and $F_{\pm{\delta}}$
reflect the strengths of these spin resonances. Comparison of the above formulae for first-order 
resonances  with those under the topic ``Reformulation in terms of beta functions and dispersion" in Ref.~\cite{barber2}, shows that we  are, in effect, using the ``betatron-dispersion'' version of the SLIM formalism.

 In ultra-high energy electron storage rings, the betatron oscillations complete many periods in one revolution, namely $\nu_x,\nu_y\gg 1$. Hence the one-turn integrals 
$F_{\pm{r}}$  tend to be small. In contrast, $\nu_z \ll 1$.
Then the synchrotron phase changes relatively little over one turn 
so that the integrals $F_{\pm{\delta}}$ need not be so small. Therefore, first-order ``parent" betatron  spin 
resonances can be much weaker than the first-order ``parent" synchrotron spin resonances.

So we next focus on the pair of
first-order ``parent" synchrotron spin resonances centered on an 
integer value, $k$, of $\nu_0$. 
The spin-orbit coupling function can be approximated by
\begin{equation}
\frac{\partial \widehat{n}}{\partial\delta}\approx\frac{1}{2}\mathscr{Re}[ \widehat{k}_0^{\star}\cdot (D_{+z}+D_{-z})]
\label{eq:spinorbit2}
\end{equation}
with
\begin{equation}
    D_{\pm z}(\theta)=\frac{{i}e^{\mp i\nu_z \theta}}{e^{-2\pi{i}(\nu_0\pm\nu_z)}-1}\int^{\theta}_{\theta-2\pi}\vec{\omega}_z\cdot\hat{k}_0 e^{\pm i\nu_z\theta'}d\theta'. 
\label{eq:Dpm}
\end{equation}
There are two different driving terms in the presence of machine imperfections.
The first is from the projection of the vertical component of $\vec{\omega}_z$ onto $\hat{k}_0$. 
Near an integer value of $\nu_0$,
$\hat{k}_0$ deviates from the horizontal plane, and $\vec{e}_y\cdot\hat{k}_0$ is actually
related to the strength, ${\tilde \omega}_k$,  of the integer resonance in the tilt
of $\hat n_0$.
In fact, as shown in detail in the Appendix~(Section~\ref{eywk}),
\begin{eqnarray}
   \vec{e}_y\cdot\hat{k}_0&=&i\sum_{k=-\infty}^{\infty} \frac{{\tilde \omega}_k e^{i(\nu_0-k)\theta'}}{\nu_0-k} \nonumber \\
   {\tilde \omega}_k &=& \frac{1}{2\pi }\int^{2\pi}_{0}\Delta \Omega_x e^{i\nu_0(\Phi(\theta')-\theta')+ik\theta'}d \theta' \nonumber \\
   \Delta\Omega_x&=&-R(1+a\gamma_0)(y_{\text{COD}}\cdot G_y \nonumber \\
   &+&x_{\text{COD}}\cdot Q_x+\frac{1}{\rho_y})
   \label{eq:intresdef}
\end{eqnarray}
where $x_{\text{COD}}$
and $y_{\text{COD}}$ are the horizontal and vertical closed-orbit distortions, respectively,
and the feed-down effect of normal quadrupoles generally dominates $\Delta\Omega_x$, and where the denominator  ${\nu_0-k}$ accounts for the resonant behaviour. In proton and electron rings, in the context of acceleration, the ${\tilde \omega}_k$ are also the strengths of the so-called ``imperfection resonances".
With neglect of the terms $1/\rho_x$ and $Q_y\eta_y$, the vertical component of $\vec{\omega}_z$ can be expanded into a Fourier series,
\begin{equation}
    -(1+a\gamma_0)RG_x\eta_x=\sum_{j=-\infty}^{\infty} \xi_j e^{-ij\theta}
    \label{eq:xidef}
\end{equation}
Then by inserting  Eq.~(\ref{eq:xidef}), Eq.~(\ref{eq:intresdef}) and Eq.~(\ref{eq:Dpm}) into Eq.~(\ref{eq:spinorbit2}), we obtain
\begin{equation}                                                                                                            
\frac{\partial \widehat{n}}{\partial\delta}(\theta)\approx\mathscr{Re}[ \widehat{k}_0^{\star}\cdot\sum_{\substack{k=-\infty \\ j=-\infty}}^{+\infty}
\frac{-i{\tilde \omega}_k\xi_j(\nu_0-k-j)e^{i(\nu_0-k-j)\theta}}{(\nu_0-k)((\nu_0-k-j)^2-\nu_z^2)}]
\label{eq:D+-D}
\end{equation}
and 
\begin{equation}
|\frac{\partial \widehat{n}}{\partial\delta}(\theta)|^2\approx|\sum_{\substack{k=-\infty \\ j=-\infty}}^{+\infty}
\frac{{\tilde \omega}_k\xi_j(\nu_0-k-j)e^{i(\nu_0-k-j)\theta}}{(\nu_0-k)((\nu_0-k-j)^2-\nu_z^2)}|^2
\label{eq:spinorbit3}
\end{equation}
Note that 
\begin{equation}
    \frac{\tau_p}{\tau_d} \approx\frac{11}{18}\oint d\theta  \frac{(\frac{\partial \hat{n}}{\partial \delta})^2}{|\rho|^3}/\oint d\theta \frac{1}{|\rho|^3}
\end{equation}
When averaged around the ring, the cross-terms vanish in Eq.~(\ref{eq:spinorbit3}),  
so that  the depolarization effect is described as
\begin{equation}
     \frac{\tau_p}{\tau_d} \approx\frac{11}{18}\sum_{\substack{k=-\infty \\ j=-\infty}}^{+\infty}
  \frac{|{\tilde \omega}_{k-j}|^2|\xi_j|^2(\nu_0-k)^2}{(\nu_0-k+j)^2[(\nu_0-k)^2-\nu_z^2]^2}
  \label{eq:1stvert}
\end{equation}


Generally, evaluation of Eq.~(\ref{eq:1stvert}) requires retaining a 
sufficient number of $k$ and $j$ terms, after a check of convergence.
However, in ultra-high energy electron storage rings with 
typical layouts such as that of the CEPC, and as shown in the Appendix~(Section~\ref{xi_0}),
where we study the structure of $\xi_j$ in a simplified model lattice
following the approach in Ref.~\cite{leeSpinDynamicsSnakes1997},
 in Eq.~(\ref{eq:1stvert})
the Fourier harmonic $\xi_0 \approx -(1+a\gamma_0)$, is the major contributor in the Fourier expansion 
for the contribution of the horizontal dispersion to depolarization.
Let's denote the integer part of $\nu_0$ by $n$, so that only a few Fourier harmonics of ${\tilde \omega}_k$ with $|k-n|\leq{l}$ have a strong influence, 
where $l$ is 
a small positive integer to be determined as a result of a check of convergence.
Then the depolarization effect can be approximated by
\begin{equation}
    \frac{\tau_p}{\tau_d} \approx \frac{11}{18}\sum_{k=n-l}^{n+l}\frac{ \nu_0^2 |{\tilde \omega}_k|^2}{\left[(\nu_0-k)^2-\nu_z^2\right]^2}
\label{eq:Peq1storder}
\end{equation}

Normally,  $\nu_z\ll1$ in high-energy electron storage rings. Then when $\nu_0$ is chosen to be far from the first-order ``parent" synchrotron spin resonances,
Eq.~(\ref{eq:Peq1storder})  can be approximated by
\begin{equation}
 \frac{\tau_p}{\tau_d} \approx  \frac{11}{18}\sum_{k=n-l}^{n+l} \frac{\nu_0^{2} \left|{\tilde \omega}_k\right|^{2}}{(\nu_0-k)^{4}}
\label{eq:p3}
\end{equation}
Eq.~(\ref{eq:p3}) agrees with the result~\footnote{Note, however, that the Eq. 3.3 in Ref.~\cite{dks}
suffers from an error
whereby the exponent in the denominator should be 4 instead of 2.
} in Ref.~\cite{dks,kondratenko_radiative_1982}.
Eq.~(\ref{eq:p3}) does not apply when $\nu_0$ is near first-order ``parent" synchrotron spin resonances while Eq.~(\ref{eq:Peq1storder}) is
still applicable.

The second driving term  arises from nonzero vertical dispersion, mainly due to vertical closed orbit 
distortions in quadrupoles, and also due to roll errors of dipoles, as well as vertical orbital correctors,
contributes to the projection of the horizontal component of $\vec{\omega}_z$ onto $\hat{k}_0$~\cite{montaguePolarizedBeamsHigh1984}.
With neglect of the terms $1/\rho_y$ and $Q_x \eta_x$, the projection of the horizontal component of $\vec{\omega}_z$ on $\hat{k}_0$ can
be expanded into a Fourier series,
\begin{equation}
   (1+a\gamma_0)RG_y\eta_y\vec{e}_x\cdot\hat{k}_0=\sum_{k=-\infty}^{\infty} {\tilde\lambda}_k e^{i(\nu_0-k)\theta}
\end{equation}
where
\begin{equation}
   {\tilde\lambda}_k = \frac{1}{2\pi }\int^{2\pi}_{0} (1+a\gamma_0)RG_y\eta_y e^{i\nu_0(\Phi(\theta')-\theta')+ik\theta'}d \theta'
   \label{eq:lambdak}
\end{equation}
Then the depolarization effect is described by~\cite{montaguePolarizedBeamsHigh1984}
\begin{equation}
    \frac{\tau_p}{\tau_d} \approx \frac{11}{18}\sum_{k=n-l}^{n+l}\frac{ (\nu_0-k)^2 |{\tilde\lambda}_k|^2  }{\left[(\nu_0-k)^2-\nu_z^2\right]^2}
    \label{eq:p31}
\end{equation}

Now in this regime of weak coupling we combine  these two different contributions to $\tau_p/\tau_d$ to evaluate the depolarization effect due to the ``parent" first-order synchrotron resonances as
\begin{equation}
    \frac{\tau_p}{\tau_d} \approx \frac{11}{18}\sum_{k=n-l}^{n+l}\frac{ \nu_0^{2} \left|{\tilde \omega}_k\right|^{2} + (\nu_0-k)^2 |{\tilde\lambda}_k|^2  }{\left[(\nu_0-k)^2-\nu_z^2\right]^2}
    \label{eq:firstorder}
\end{equation}

In practice only a few harmonics in the above formulas dominate and
they could be partially compensated, and then the tilt of $\hat n_0$ reduced, with specially arranged correction coils forming closed-orbit vertical bumps~\cite{rossmanithCompensationDepolarizingEffects1985} chosen to generate 
``anti-harmonics". These so-called harmonic 
closed-orbit spin matching techniques~\cite{barberHighSpinPolarization1994,assmannLeptonBeamPolarization1995a} have  been shown to be essential for improving the equilibrium beam polarization level.

In this paper, we focus on the comparison between the polarization theory and simulations, applied to the imperfect CEPC lattices without taking account of the harmonic closed-orbit spin matching.
Considering only the first-order ``parent" spin resonances, the equilibrium beam polarization can be analytically evaluated using Eqs.(\ref{eq:p2}), (\ref{eq:p1}) and (\ref{eq:firstorder}). 

\subsection{Higher-order synchrotron sideband spin resonances}
As shown previously, the first-order ``parent'' spin resonances arise from the perturbation term $\vec{\omega}\cdot\hat{k}_0$. In addition, the spin precession rate is dependent on the coordinates of orbital motion, and orbital oscillations induce modulations of the spin precession rate, which lead to the 
higher-order ``sideband'' resonances of these first-order
``parent'' spin resonances. 
This can be seen from Eq.~(\ref{eq:zeta1}), where $\chi(\theta)$, as an integral of $\vec{\omega}\cdot\hat{n}_0$,
can also be expressed as a resonant term times a one-turn integral, which represents the average modulation over one turn~\cite{yok,yok2}.
Then betatron tunes are so large that their one-turn integrals are small and the corresponding sideband spin resonances are insignificant. In contrast,
the sychrotron tune is very small and the average modulation cannot be ignored.  

In particular, around each integer $k$ there can be an infinite number of higher-order synchrotron sideband spin resonances in the form of $\nu_0\pm{m}\nu_z=k$. 
The overall depolarization effects of this family of spin resonances, can be incorporated by multiplying the $\tau_p/\tau_d$ contribution of 
the first-order ``parent" synchrotron spin resonances around integer $k$ by a depolarization
enhancement factor $\mathcal{F}_k$~\cite{yok} 
\begin{equation}
\mathcal{F}_k =  ((\nu_0-k)^{2}-\nu_z^2)^{2} \sum_{m} \frac{e^{-\sigma^2}I_m(\sigma^2)
}{\left[\left(\nu_0-k-m \nu_{z}\right)^{2}-\nu_{z}^{2}\right]^{2}} 
\label{eq:highersideband}
\end{equation}
where $I_m$ is the modified Bessel function 
with the modulation index $\sigma$ defined as
\begin{equation}
 \sigma=\frac{\sigma_0}{\nu_z}=
\frac{\nu_0 \sigma_{\delta}}{\nu_z}
\end{equation}
and where $\sigma_0= \nu_0 \sigma_{\delta}$
measures the spread of the instantaneous spin precession frequencies in a planar ring for 
the rms relative energy spread, $\sigma_{\delta}$, resulting from synchrotron radiation.
Note that in the derivation of Eq.(\ref{eq:highersideband}), it was also assumed
that synchrotron motion is approximately harmonic so that Eq.(\ref{eq:sync_action_angle}) holds.

The modulation index $\sigma$ reflects the spread of spin
precession frequencies in a beam, relative to the spacing of adjacent synchrotron sideband spin resonances. For a fixed $\nu_z$, $\sigma$ scales with ${\gamma_0}^2$,
and higher-order synchrotron sideband spin resonances become more prominent at higher beam energies. 
To achieve a higher equilibrium beam polarization level, it is essential to mitigate the influence of these important synchrotron sideband spin resonances. It is customary to choose $[a\gamma_0]\approx0.5$ where here
and later $[x]$ denotes the fractional part of $x\in R$.

Considering both the first-order and higher-order synchrotron spin resonances,
the depolarization effect can be evaluated as
\begin{eqnarray}
    \frac{\tau_p}{\tau_d} &\approx& \frac{11}{18}\sum_{k=n-l}^{n+l}\sum_{m=-\infty}^{\infty} \left( \frac{  \nu_0^{2}\left|{\tilde \omega}_k\right|^{2}e^{-\sigma^2}I_m(\sigma^2) }{\left[\left(\nu_0-k-m \nu_{z}\right)^{2}-\nu_{z}^{2}\right]^{2}} \right. \nonumber \\
    &+&\left. \frac{(\nu_0-k)^2 |{\tilde\lambda}_k|^2 e^{-\sigma^2}I_m(\sigma^2)}{\left[\left(\nu_0-k-m \nu_{z}\right)^{2}-\nu_{z}^{2}\right]^{2}} \right)
    \label{eq:highertorder}
\end{eqnarray}

For further treatments of higher-order resonances using perturbation theory see Ref.~\cite{mane_electron-spin_1987,mane_synchrotron_1990,mane_polarization_1992,mane_polarization_1992_2}.

\subsection{The correlated and uncorrelated regimes of resonance crossing}


Ref.~\cite{dks} distinguishes between two regimes, namely the so-called
   non-resonant and resonant spin diffusion. In a modern treatment we would focus on the 
   invariant spin field $\hat n(\vec u;\theta)$ and the  
   amplitude-dependent spin tune $\nu_s(\vec J)$  for the combined
   betatron and synchrotron motion.
The amplitude-dependent spin tune is constant during such 
   motion for fixed orbital amplitudes but changes if, for example, the amplitude of
   the synchrotron motion changes due to photon emission.
   However, in Ref.~\cite{dks} a hybrid approach is adopted whereby the invariant spin field
   $\hat n$ is explicitly time-independent and the synchrotron motion is added by hand, so that the 
    instantaneous spin-precession rate $\nu$ is dependent on the instantaneous energy deviation of a particle, i.e., $\nu\approx\nu_0(1+\delta)$. Since $\nu_z\ll1$ in most electron storage rings, $\nu$ looks like a slowly varying $\nu_0$.
    In this ``dynamical'' picture, some underlying spin resonances could be crossed as a result of synchrotron oscillations, or synchrotron radiation, or the combined effect.
   As far as we know Ref.~\cite{dks} is the most recent work on these topics. So to facilitate 
   comparison we adopt that approach and argumentation too. 

   As already mentioned in subsection B, stochastic photon emissions cause a random walk in
   particle energies
   and thus lead to a spread in particle energies and a spread in spin precession frequencies characterized by $\sigma_0=\nu_0\sigma_{\delta}$. 
 The parameter $\sigma_0$ also characterizes the amplitude of variation of the spin precession rate of an arbitrary particle in the beam, as a result of synchrotron radiation and synchrotron oscillation. 
   At low energy, typically $\sigma_0 \ll 1$, in particular when $\sigma_0 \ll \nu_z$, the design energy can be set so that
   $\nu_s(\vec J)$ of beam particles sit between important spin-orbit resonances. Then the perturbation theory of subsection B above can be used
   to get first estimates of $\tau_p/\tau_d$ and the sidebands will be narrow according to subsection B.
   This is the regime called non-resonant spin diffusion in Ref.~\cite{dks}.

   At higher beam energies $\sigma_0$ becomes larger so that some beam particles inevitably
   cross some
   important spin-orbit resonances in the process of synchrotron radiation and synchrotron oscillation.
   Then there can be so-called resonant spin diffusion. In particular, it was argued 
   in Ref.~\cite{dks} that when the diffusion of the spin-precession phase advance introduced by stochastic photon emissions
   is negligible in one synchrotron period, the crossings of an underlying spin resonance 
   during precession-rate oscillations driven by synchrotron oscillations, are correlated.
   Then, according to  Ref.~\cite{dks}, the perturbation theory of subsection B  can still be used. This poses a constraint on the ``correlation index'' $\kappa$, which characterizes the spread in the spin-precession phase advance in a synchrotron period,

\begin{equation}
\kappa=\frac {\nu_0^2 \lambda_p}{  \nu_z^3 } \ll 1
\label{eq:limit}
\end{equation}
where $\lambda_p$ is the normalized polarizing rate,
\begin{equation}
\lambda_p= \tau_p^{-1} \frac{R}{c} 
\label{eq:lambda}
\end{equation}
\noindent
Note that following Ref.~\cite{dks} we have 
\begin{equation}
 \sigma^2=\frac{\sigma_0^2}{\nu_z^2} = 
\frac{\nu_0^2 \sigma_{\delta}^2}{\nu_z^2} = \frac{11}{36}
\frac{\nu_0^2 \lambda_p}{\nu_z^2\Lambda_z}
\nonumber
\end{equation}
where $\Lambda_z$ is the damping decrement for longitudinal motion,
\begin{equation}
\Lambda_z= \tau_z^{-1} \frac{R}{c} 
\label{eq:lambda_z}
\end{equation}
where $\tau_z$ is the longitudinal damping time.
Then since $\Lambda_z \ll \nu_z$  the above constraint on $\kappa$ is 
more relaxed than the earlier constraint on ${\sigma_0}/{\nu_z}$.

The perturbative depolarization theory of higher-order synchrotron sideband spin resonances will be hereafter referred to as the theory of ``correlated regime'' of spin resonance
crossing (or in short the ``correlated regime'').

In contrast, if the radiation is extremely violent, or the synchrotron tune is very small, the condition in Eq.~(\ref{eq:limit}) could be violated. 
Then if $\sigma_0 \gg \nu_z$ also holds, successive crossings of the underlying spin resonance in synchrotron oscillations become completely uncorrelated. 
In this case, it was suggested in Ref.~\cite{dks} that, the normalized depolarization rate $\lambda_d$ due to the photon emissions can be evaluated by~\cite{K74}
\begin{equation}
    \lambda_d=\pi\sum_k \left\langle|{\tilde \epsilon}_k|^2 \delta(\nu-k)\right\rangle
\end{equation}
where $\delta(\nu-k)$ is the delta function, 
$\nu$ is the instantaneous spin-precession rate,
${\tilde \epsilon}_k$ denotes the strength of spin resonances on general synchro-betatron trajectories and $\left\langle \right\rangle$ is the average over the beam particles.
In ultra-high energy planar electron storage rings, the ${\tilde \omega}_k$,  of the integer resonances in the tilt of $\hat n_0$,
were regarded~\cite{dks} as the major contributions to ${\tilde \epsilon}_k$. Then 
assuming $\nu\approx\nu_0(1+\delta)$
and considering a Gaussian distribution for $\delta$, the depolarization effect can be evaluated as~\cite{dks}
\begin{equation}
\frac{\tau_p}{\tau_d} \approx \frac{ \sqrt{\pi/2}}{\lambda_p}\sum_{k=n-l}^{n+l}\frac{ \left |{\tilde \omega}_k \right|^2 }{\sigma_{0}} \exp{[-\frac{(\nu_0-k)^2}{2\sigma^2_{0}}]}
\label{eq:uncorrelated}
\end{equation}
This spin diffusion theory will be hereafter referred to as the ``uncorrelated regime''. 

It was also suggested in Ref.~\cite{dks} that, when $\sigma_{0}\ll1$, the spin diffusion due to the uncorrelated resonance crossing is comparable 
to that due to the ``trembling'' of the $\hat{n}$ associated with photon emissions, and that the total depolarization effect
can then be evaluated by
adding Eq.~(\ref{eq:uncorrelated}) to Eq.~(\ref{eq:p3}). Note that Eq.~(\ref{eq:p3}) doesn't include the depolarization effect from the synchrotron
sideband spin resonances.
In addition, when $\sigma_{0}\gg1$, 
it was also predicted in Ref.~\cite{dks} that, there would be no resonant dependence of spin diffusion on energy, and beam polarization would grow with energy.   

The theories of the ``correlated regime'' and ``uncorrelated regime'' lead to quite distinct depolarization effects, while the condition of
their application is quite vaguely defined. In particular,
there is no definite boundary of $\kappa$ at which the depolarization effect enters 
the ``uncorrelated regime''. Moreover, it is not clear either what theory of radiative depolarization
is applicable between the ``correlated regime'' and the ``uncorrelated regime''.
Note that the theory of the ``correlated regime'' has been verified in LEP experiments~\cite{assmannLeptonBeamPolarization1995a}, but the theory of the ``uncorrelated regime'' seems to be beyond the parameter space of LEP~\cite{assmannLeptonBeamPolarization1995a}.

\section{CEPC Lattice Setup ~\label{sec:lattice}}

We now continue by describing simulations of  
the equilibrium beam polarization for the CEPC CDR lattice~\cite{CEPCcdr}. For these, the major contributions of misalignment errors and relative field errors of magnets, listed in Table~\ref{tab:1},
were introduced into the lattice. These errors follow a Gaussian distribution truncated at $\pm3\sigma$. Then, a detailed closed orbit and optics correction scheme was developed~\cite{wangbi} to recover the lattice performance, using the SAD~\cite{sad} and Accelerator Toolbox (AT)~\cite{at} codes. BPM errors were not yet included in the correction scheme.
\begin{table}[!htb]
\caption{CEPC magnets' error settings in the simulations.}
{\begin{tabular}{@{}l|ccc|c@{}} \hline \hline
\multirow{2}*{Component}&\multicolumn{3}{c|}{Misalignment error}&\multirow{2}*{Field error} \\ \cline{2-4}
~&$\Delta x (\si{\um})$&$\Delta y (\si{\um})$&$\Delta \theta_z (\si{\micro\radian})$&~\\ \hline
Dipole&-&-&-&0.01\% \\
Arc quadrupole&100&100&100&0.02\% \\
IR quadrupole&50&50&50&-\\
Sextupole&100&100&100&- \\
\hline \hline
\end{tabular} \label{tab:1}}
\end{table}

About 1500 correctors in both the horizontal and vertical planes were used for the closed-orbit distortion (COD) correction. The horizontal correctors and vertical correctors were placed next to focusing quadrupoles and defousing quadrupoles, respectively. About 1500 beam position monitors were placed next to quadrupoles, four per betatron wave period. Both the root-mean squared (rms) values of the horizontal and vertical closed orbits are smaller than 50 \si{\um} after the COD correction. Skew quadrupole coils on sextupoles and some independent skew quadrupoles were used to minimize the vertical dispersion and the betatron coupling. Since the detector solenoids in the interaction regions are not yet included in the lattce, the vertical emittance after correction is $2.5\times 10^{-3}$ \si{pm}$\cdot$\si{rad} at the Z pole after the coupling correction. It is much smaller than the target value, $1.6$ \si{pm}$\cdot$\si{rad}, in the CDR. 

\begin{figure}[htbp]
\centering

\subfigure[Straight sections of the CEPC]{
\begin{minipage}{6cm}
\centering
\includegraphics[width=5.5cm]{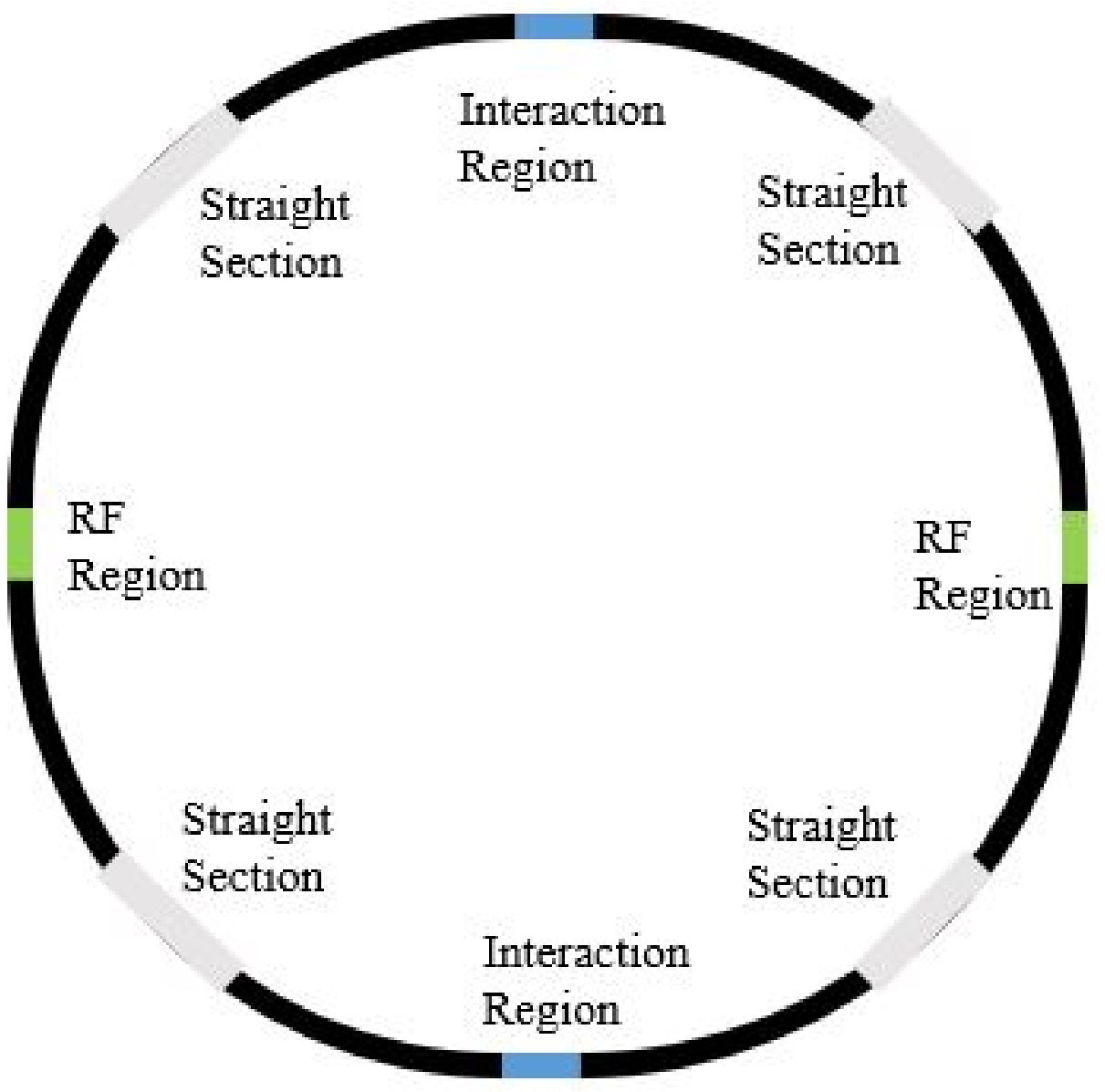} \\
\end{minipage}
}

\subfigure[Locations of SQ1 and SQ2 in one interaction region]{
\begin{minipage}{6cm}
\centering
\includegraphics[width=5.5cm]{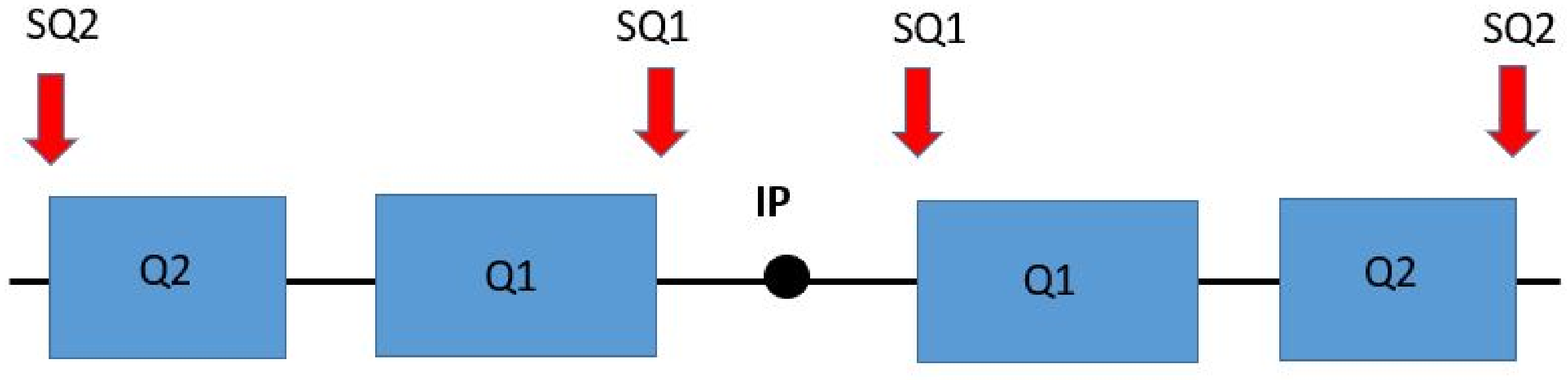} \\
\end{minipage}
}

\caption{The layout of the CEPC straight sections and locations of coupling adjustment components. In the top figure, the 8 straight sections of the CEPC include 2 interaction regions (blue), 2 RF regions (blue), and 4 straight sections (Grey) for other purposes. The quadrupoles in these 4 straight sections are artificially rotated to generate the transverse coupling. In the bottom figure, zero-length skew quadrupoles SQ1 and SQ2 are inserted next to the final
focus quadrupoles Q1 and Q2 around each of the two interaction points, to account for the transverse coupling introduced by the detector solenoids.}
\label{fig:Q1Q2}
\end{figure}

To properly account for the "parent" betatron spin resonances in the equilibrium polarization simulations, the vertical emittance of the imperfect lattice was adjusted to the target value.
According to a separate analysis~\cite{CEPCcdr}, the coupling contribution due to the detector solenoid fringe field is around 28\%  of the target vertical emittance assuming that the detector solenoid field is 2 \si{\tesla} during Z operation. Zero-length skew quadrupoles SQ1 and SQ2 were added next to the final focusing doublet Q1 and Q2, respectively, to account for the solenoid fringe field contribution, as illustrated in Figure~\ref{fig:Q1Q2}(b). The integral field gradients of SQ1 and SQ2 are \SI{6.4e-5}{\per\metre} and \SI{3.7e-6}{\per\metre} respectively.
To generate the remaining 72\% targeted vertical emittance, all quadrupoles in the four straight section regions, as shown in  Figure~\ref{fig:Q1Q2}(a), were rotated by the same angle, 0.2 degrees in the lattice. 

\begin{figure*}[!htb]
\centering
\subfigure[Horizontal closed orbit .]{
\includegraphics[width=6.5cm]{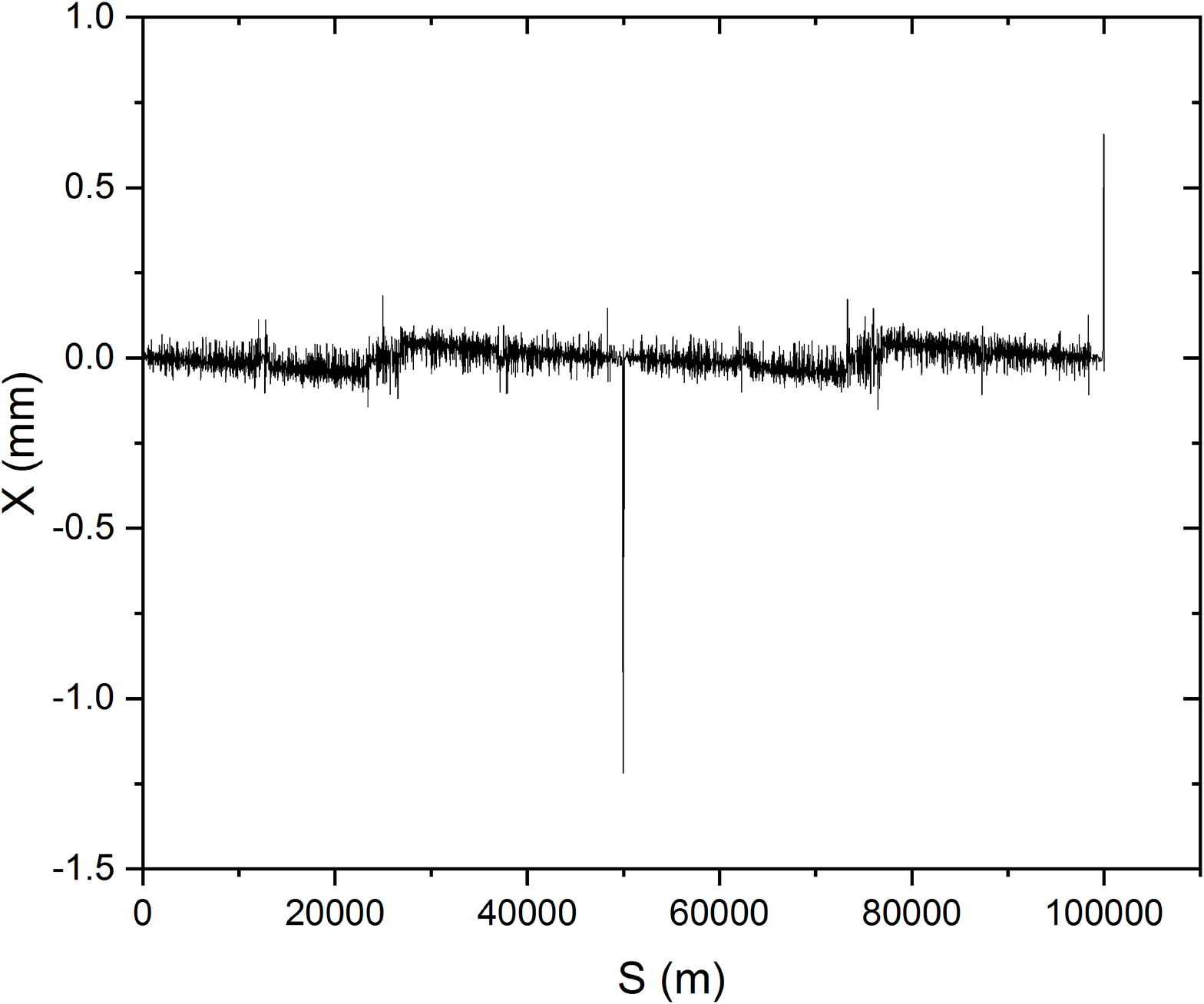}
}
\quad
\subfigure[Vertical closed orbit.]{
\includegraphics[width=6.5cm]{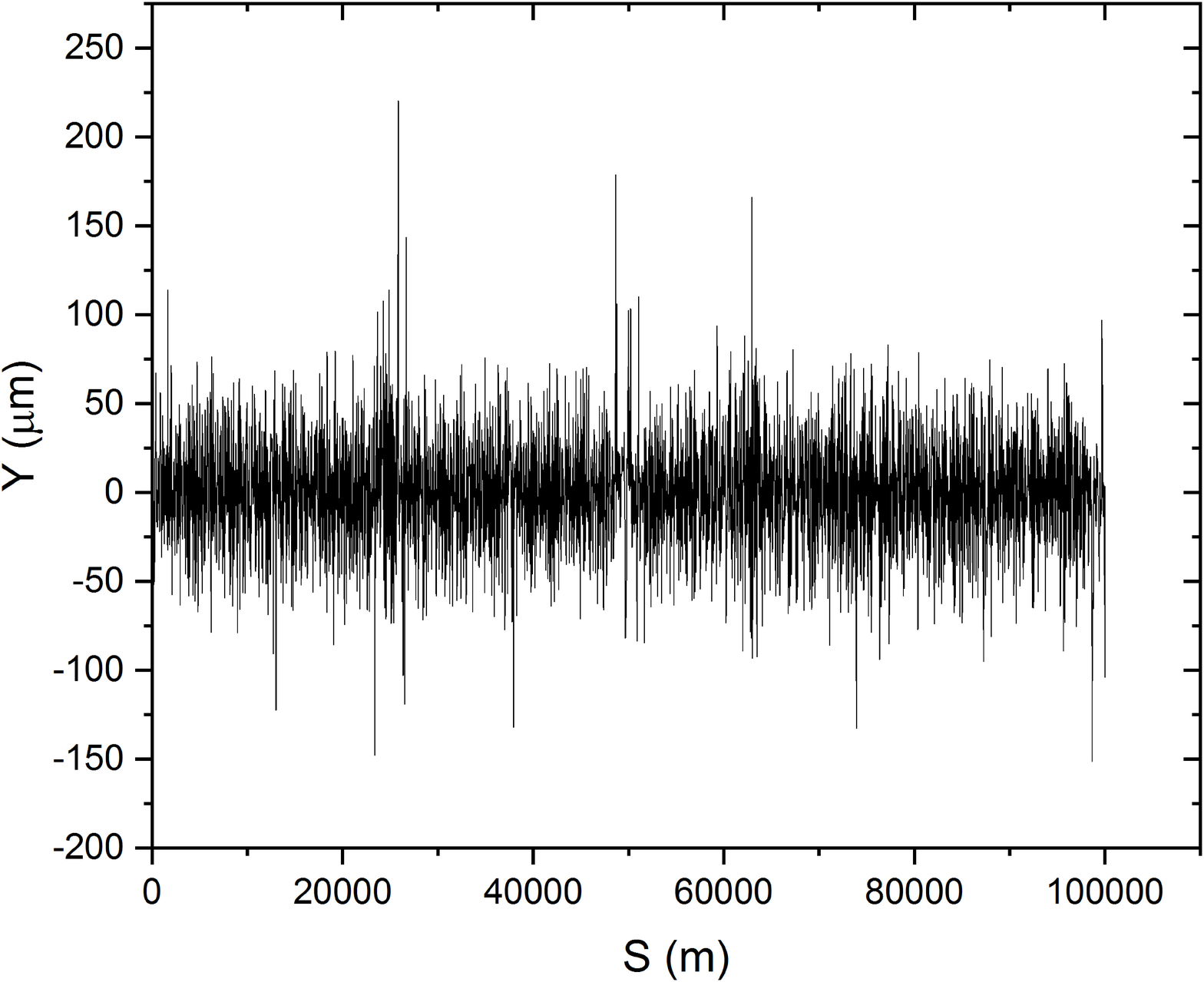}
}
\quad
\subfigure[Horizontal beta beating .]{
\includegraphics[width=6.5cm]{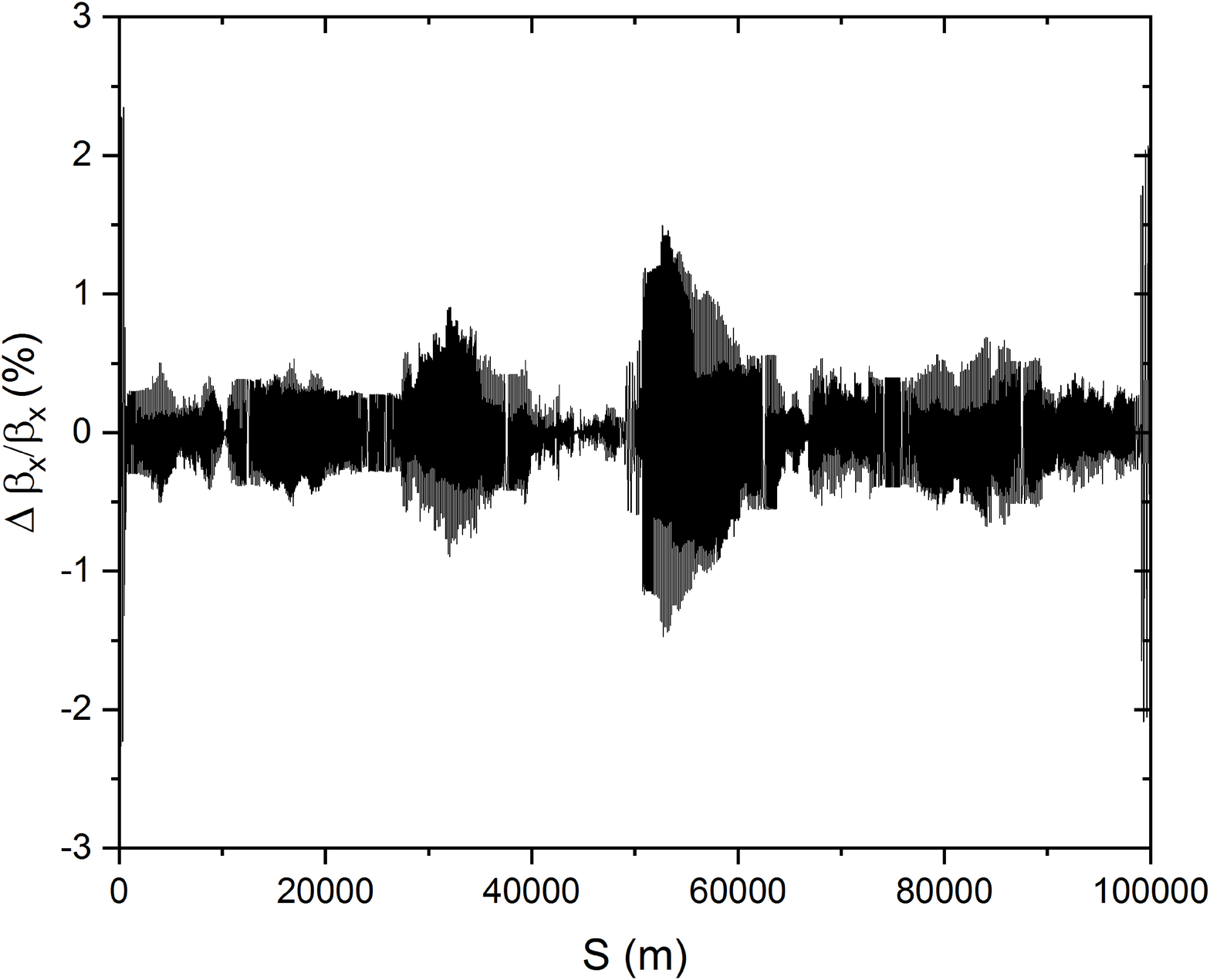}
}
\quad
\subfigure[Vertical beta beating.]{
\includegraphics[width=6.5cm]{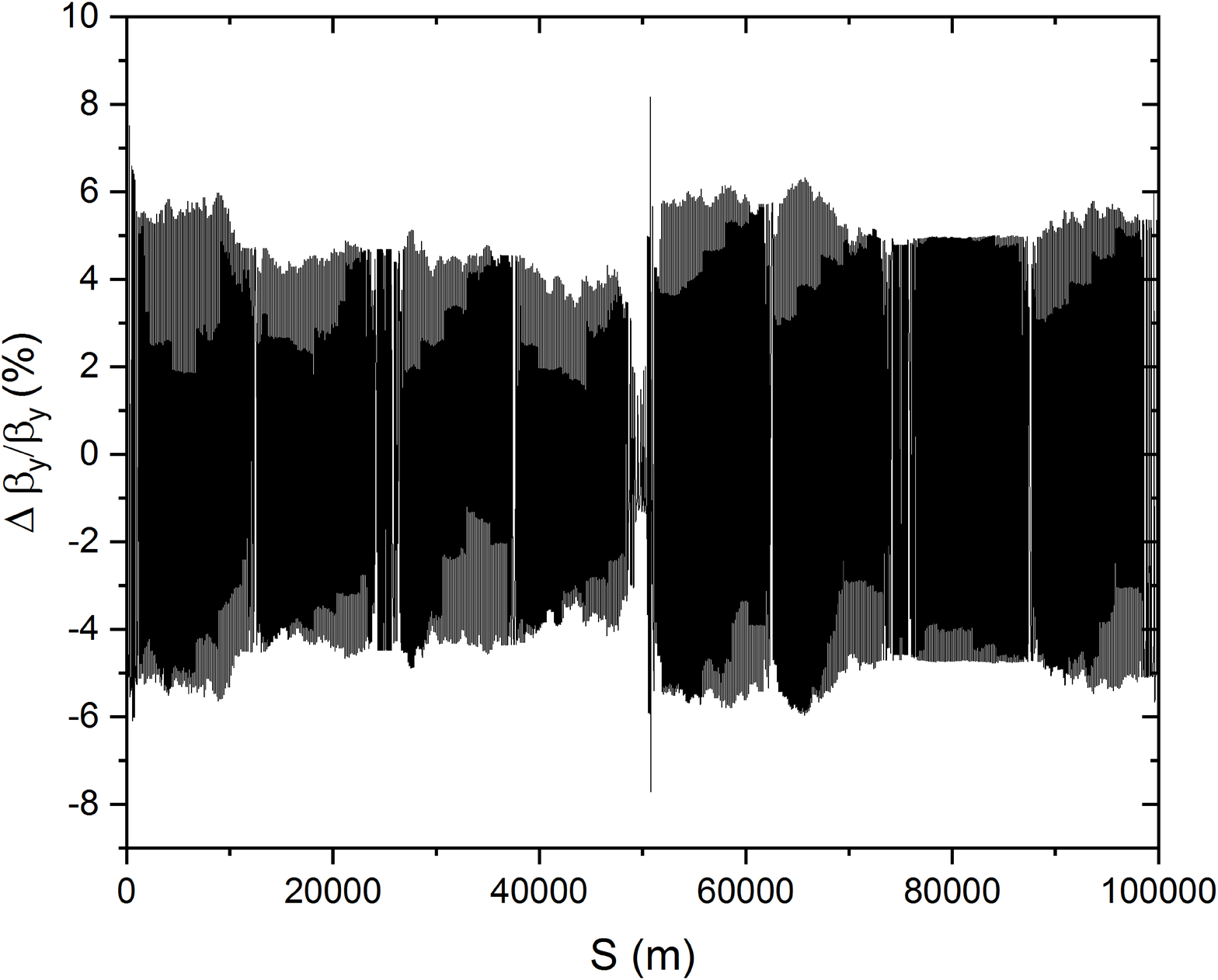}
}
\caption{The closed orbit (a, b) and the beta beating (c, d) of the CEPC lattice computed by SAD at 45.6 \si{\GeV} as a function of the longitudinal position $S$ from one of the two interaction points. 
The rms values of the horizontal closed orbit and vertical closed orbit are $37$~\si{\um} and $28~\si{\um}$, respectively. The rms values of the horizontal beta beating and vertical beta beating are $3.6\times 10^{-3} $ and $3.4 \times 10^{-2} $, respectively. Note that in order to reduce the critical energy of synchrotron radiation from the
last bending magnet upstream of each interaction point, these bending magnets are 93.4~\si{m} 
long. The local horizontal closed-orbit distortions are above 0.5~\si{mm} in (a), but can be reduced by 
implementing more orbital correctors in these regions.}
\label{fig:sad_cod_twiss}
\end{figure*}

Figure~\ref{fig:sad_cod_twiss} shows the closed orbit and the beta beating along the CEPC lattice in the horizontal and vertical directions obtained by SAD. The rms values of the horizontal closed orbit and vertical closed orbit are 37 \si{\um} and 28 \si{\um}. The rms values of the horizontal and vertical beta beatings are $0.36\%$ and $3.4\%$, respectively. We are aware that BPM errors, in particular the relative offset 
errors between BPMs and adjacent quadrupole magnets, once introduced, would worsen
the achievable level of closed-orbit distortions after correction, and also affect
the performance after optics correction. These effects would in turn affect the
equilibrium beam polarization level. These BPM errors and other error sources will be included properly
in future studies, and set the basis for a more practical prediction of the achievable equilibrium beam polarization level.
In this study, the focus is to reveal the
mechanisms of radiative depolarization, rather than to predict the achievable equilibrium beam polarization level.

Most of the CEPC accelerator design and simulations were based on SAD~\cite{sad}. However, the capability to simulate beam polarization is still under development in SAD.
Then, in this paper, 
Bmad~\cite{Bmad} and PTC~\cite{PTC} were used in the simulation of equilibrium beam polarization. PTC is embedded inside Bmad as a library, and much  effort has been invested 
by the authors of both codes, to make the interface almost seamless. Bmad has a built-in module to simulate the equilibrium
beam polarization with contributions from just the first-order spin resonances. This module calls PTC internally to construct the linearized orbital and spin transport matrices, and then
evaluate $\hat{n}_0$ and $\frac{\partial \hat{n}}{\partial \delta}$, using the SLIM formalism~\cite{slim}. To go further, Monte-Carlo simulations of the depolarization effects were studied using PTC~\cite{duanz}, so that 
the influence of high-order spin resonances were exposed. Both approaches were adopted in the study of the equilibrium beam polarization of the CEPC.

The SAD version of the CEPC lattice was converted to the Bmad format using a Python script inside the Bmad package~\cite{Bmad}, and then a lattice file in PTC format was generated within Bmad. Previous studies~\cite{zhoudm} indicated a good match of the calculated bare lattice parameters between SAD and Bmad. The imperfect CEPC lattice was converted to the Bmad format using this Python script with minor modifications. The zero-length correctors modeled by the BEND type in SAD, had to be converted to the KICKER type in Bmad. In addition, the different conventions of the RF phase and the quadrupole roll were taken into account.

\begin{table*}[!htb]
\caption{The differences of the closed orbit (CO) and the relative difference of the beta function ($\Delta \beta/\beta$) of the CEPC at 45.6 \si{\GeV}  calculated by SAD, Bmad and PTC. The minus sign indicates the difference between the two codes. ``rms" is the root mean square of the difference around the ring. ``max" means the maximum absolute value of the difference around the ring.}
{\begin{tabular}{@{}lccccc@{}} \hline \hline
     &&$\textrm{CO}_{\textrm{Bmad-SAD}} (\si{\metre})$&$\textrm{CO}_{\textrm{PTC-SAD}} (\si{\metre})$&$(\Delta \beta/\beta)_{\textrm{Bmad-SAD}}$ &$(\Delta \beta/\beta)_{\textrm{PTC-SAD}} $ \\ \hline
\multirow{2}*{Horizontal direction}&rms&$1.2\times 10^{-8}$&$9.5\times 10^{-8}$&$6.9\times 10^{-8}$&$1.6\times 10^{-5}$ \\
                                   &max&$6.8\times 10^{-8}$&$3.3\times 10^{-7}$&$9.9\times 10^{-8}$&$4.2\times 10^{-4}$\\ \hline
\multirow{2}*{Vertical direction}&rms&$1.4\times 10^{-10}$&$1.1\times 10^{-9}$&$9.0\times 10^{-7}$&$9.5\times 10^{-5}$\\
                                 &max&$1.3\times 10^{-9}$&$7.8\times 10^{-9}$&$1.4\times 10^{-6}$&$6.5\times 10^{-4}$\\
\hline \hline
\end{tabular} \label{tab:sad_bmad_ptc}}
\end{table*}

Then, we compared the closed orbit and beta functions of the CEPC lattice using SAD, Bmad and PTC, respectively. The differences are summarized in Table~\ref{tab:sad_bmad_ptc}. The relative difference of beta function $\Delta \beta /\beta$ is defined as:
\begin{equation}
(\Delta \beta /\beta)_{\mathrm{code-SAD}}=\frac{\beta^1_{\mathrm{code}}-\beta^1_{\mathrm{SAD}}}{\beta^0_{\mathrm{SAD}}}
\label{eq:beta}
\end{equation} 
where the superscripts ``0" and ``1" denote the CEPC bare lattice and the CEPC imperfect lattice after the error corrections, respectively. And the subscripts ``code" represents either Bmad or PTC. For the closed orbit and the beta beating, both the rms and the maximum values of the differences in Table~\ref{tab:sad_bmad_ptc} are much smaller than the rms values of the lattice obtained by SAD in Figure~\ref{fig:sad_cod_twiss}. We also compared the emittances, betatron and synchrotron tunes using these three codes, as shown in Table~\ref{tab:sadBmad}.

\begin{table}[!htb]
\caption{Emittances and fractional tunes of the CEPC at 45.6 \si{\GeV}  calculated by SAD, Bmad and PTC.}
{\begin{tabular}{@{}lccc@{}} \hline \hline
     &SAD&Bmad&PTC \\ \hline
Horizontal emittance~(\si{\nm}$\cdot$\si{rad})&0.1731&0.1738&0.1733 \\
Vertical emittance~(\si{\pm}$\cdot$\si{rad})& 1.615&1.623&1.612\\
Longitudinal emittance~(\si{\um}$\cdot$\si{rad})&0.9017&0.8956&0.9028 \\
Fractional horizontal tune&0.108&0.108&0.108 \\
Fractional vertical tune&0.217&0.217&0.216\\
Fractional synchrotron tune&0.028&0.028&0.028 \\
\hline \hline
\end{tabular} \label{tab:sadBmad}}
\end{table}

Although the three codes have different models, the results of simulations of the closed orbits, beta functions, emittances, tunes and other parameters match very well. The relative difference
between these three codes, are much smaller than the relative difference between the imperfect lattice and the bare lattice. Therefore, the simulations using Bmad and PTC can truthfully reflect the influence of the machine imperfections and the correction scheme on the CEPC lattice.

\section{Application of the theories} \label{sec:ana}
In this section, we apply the theories outlined in Section~\ref{sec:theory} to evaluate the depolarization effects for the CEPC lattice.
We first calculate
the Fourier harmonics ${\tilde\omega}_k$ and ${\tilde\lambda}_k$
near the working beam energy, and apply the theory of the radiative depolarization taking into account spin resonances up to the first order. Then the calculated equilibrium beam polarization is compared with the simulations via the SLIM formalism. Finally, we discuss how the
the theories of the correlated and uncorrelated regimes are applied to evaluate the depolarization effects for the CEPC lattice.

As shown in Section~\ref{sec:theory}, the amplitudes of Fourier harmonics ${\tilde \omega}_k$ and ${\tilde\lambda}_k$, are the major lattice-dependent input to the theories of depolarization effects at ultra-high beam energies. Their evaluation only requires the closed-orbit and some optical parameters which are accessible from 
the SAD calculations.
According to the definition of the strength of the integer spin resonance in Eq.~(\ref{eq:intresdef}), ${\tilde \omega}_k$ can be approximately evaluated using
\begin{eqnarray}
{\tilde \omega}_k &\approx&\frac{1}{2\pi  }\int^{2\pi R}_{0} (1+k) y_{0}''(s) e^ {ik\Phi(s)}d s \nonumber\\
        &\approx& \frac{1+k}{2\pi }\sum_{h=1}^{M} [ p_{y,0}(s_{h,2})-p_{y,0}(s_{h,1})]  e^ {ik\Phi(s_{h,1})}
\label{eq:wk_integer}
\end{eqnarray}
where $y_0$ is the vertical displacement on the closed orbit and $y_{0}''(s)=d^2 y_0(s)/d s^2$ with  $ds=R d\theta'$.
 We approximate $y_{0}'=dy_0/ds$ by $p_{y,0}$, the vertical canonical momentum on the closed orbit,
 and replace the integral by a sum over
 the $M$ magnet elements in the lattice. $s_{h,1}$ and $s_{h,2}$ are the longitudinal
 positions of the entrance and the exit of the $h$-th magnet, respectively. Similarly, ${\tilde\lambda}_k$ in Eq.~(\ref{eq:lambdak}) can be approximately calculated using
\begin{eqnarray}
{\tilde\lambda}_k &=& \frac{1+k}{2\pi }\int^{2\pi R}_{0}G_y(s)\eta_y (s) e^{ik\Phi(s)}d s \nonumber\\
&\approx& \frac{1+k}{2\pi}\sum_{h=1}^{M} G_y (s_{h,1})\Delta s_h\eta_y (s_{h,1}) e^ {ik\Phi(s_{h,1})} 
\label{eq:lambdakapprox}
\end{eqnarray}
where $\Delta s_h$ is the length of the $h$-th magnet.

\begin{figure}[htbp]
\centering
\subfigure[$|{\tilde \omega}_k|^2$]{
\begin{minipage}{1.0\columnwidth}
\centering
\includegraphics[width=1.0\columnwidth]{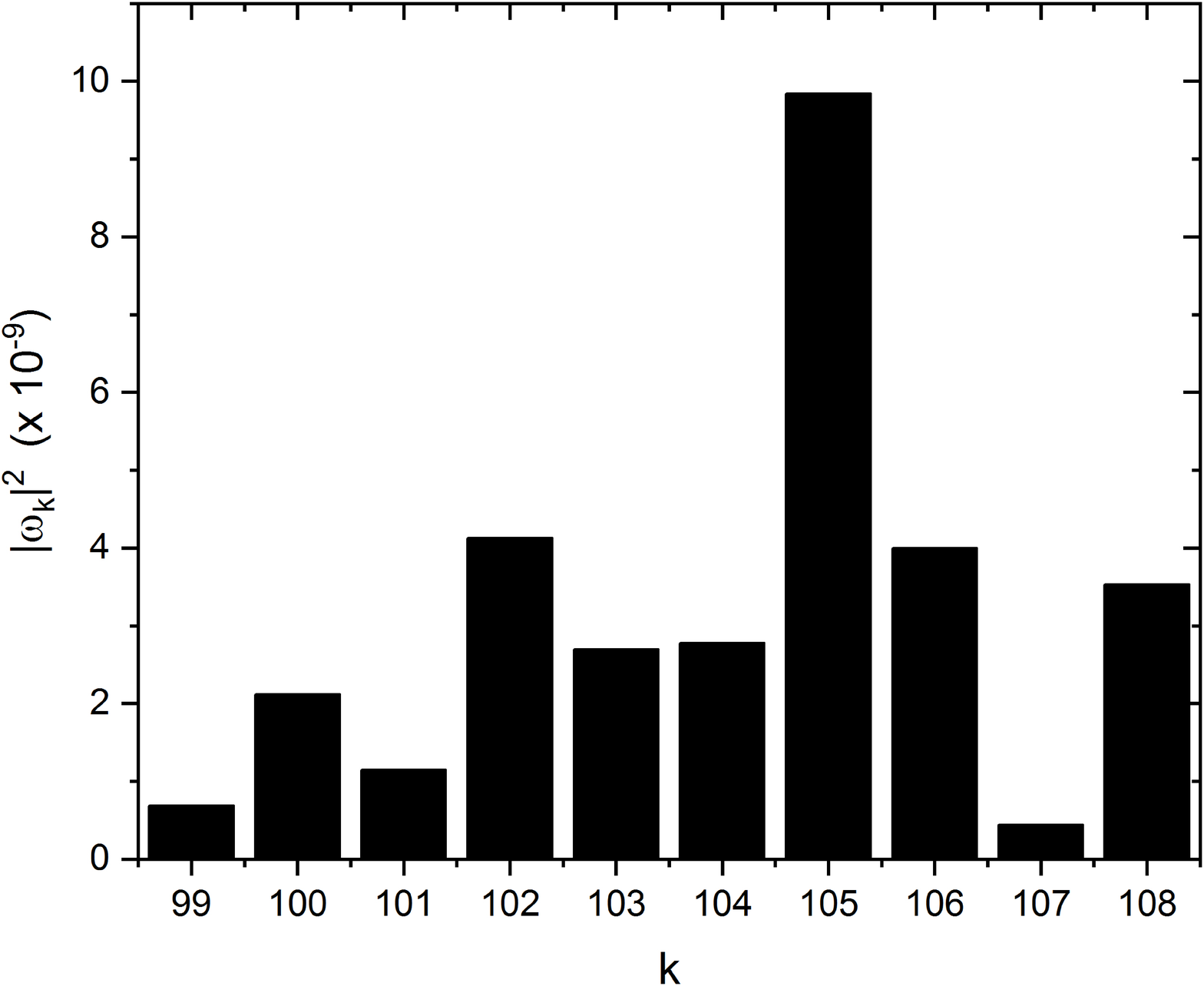} \\
\end{minipage}
}
\subfigure[$|{\tilde\lambda}_k|^2$]{
\begin{minipage}{1.0\columnwidth}
\centering
\includegraphics[width=1.0\columnwidth]{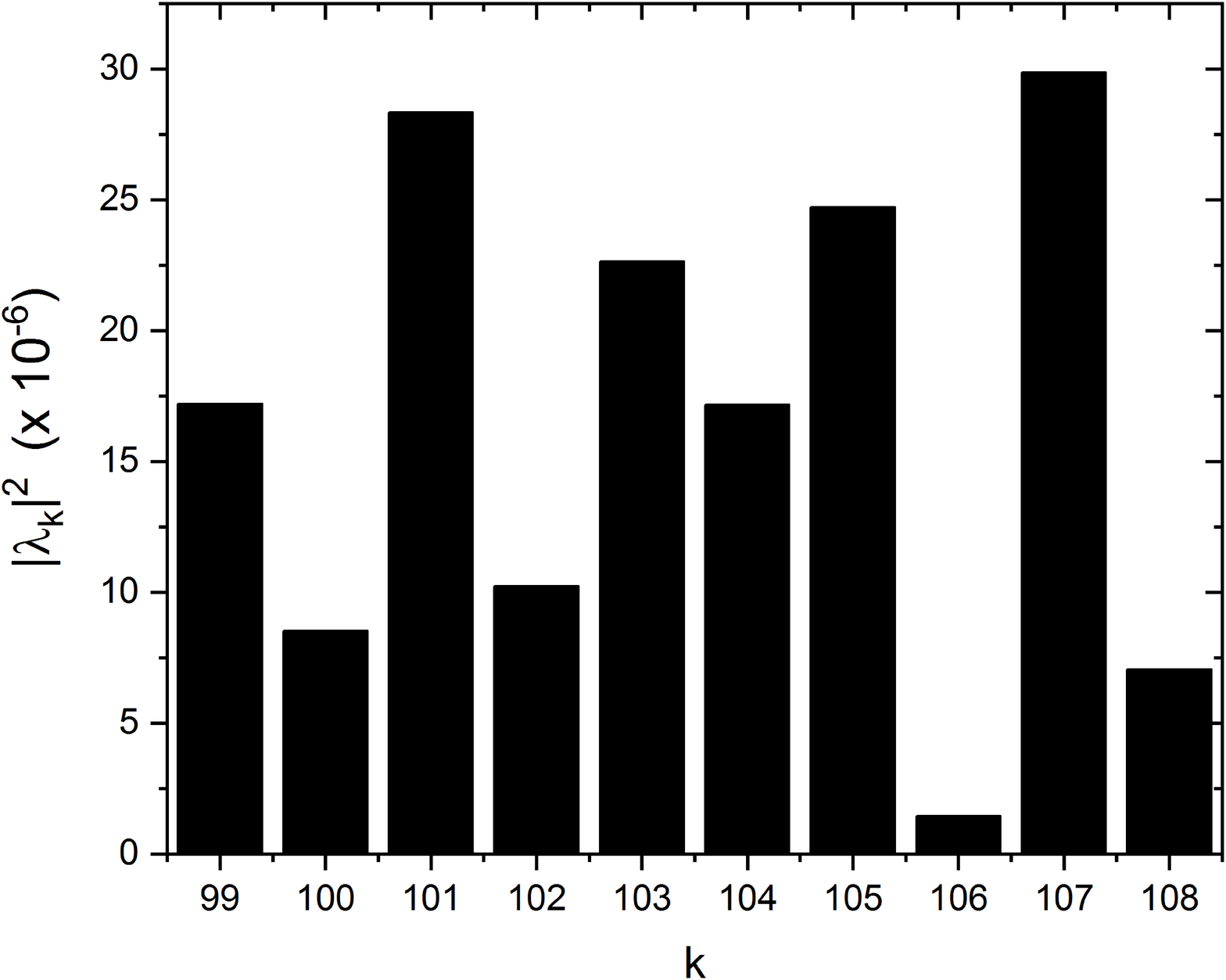} \\
\end{minipage}
}
\caption{Harmonics of $|{\tilde \omega}_k|^2$ (a) and $|{\tilde\lambda}_k|^2$ (b) near $a\gamma_0=103.5$ for the CEPC lattice.}
\label{fig:wk}
\end{figure}

Fig.~\ref{fig:wk} shows the  $\left|{\tilde \omega}_k\right|^2$ and $\left|{\tilde \lambda}_k\right|^2$ calculated for the CEPC lattice, for various integers $k$ near $a\gamma_0=103.5$~(corresponding to the working beam energy at 45.6 \si{\GeV}).
The amplitude variations are mostly within one order of magnitude.
Moreover, evaluation with Eq.~(\ref{eq:firstorder}) of the rate of depolarization with just the 
two harmonics nearest to the integer part of $a\gamma_0$ 
shows that retaining just those harmonics suffices in most cases in the 
working energy range of the CEPC.
\begin{figure}[htbp]
\centering
\subfigure[Z]{
\begin{minipage}[t]{0.9\columnwidth}
\centering
\includegraphics[width=7cm]{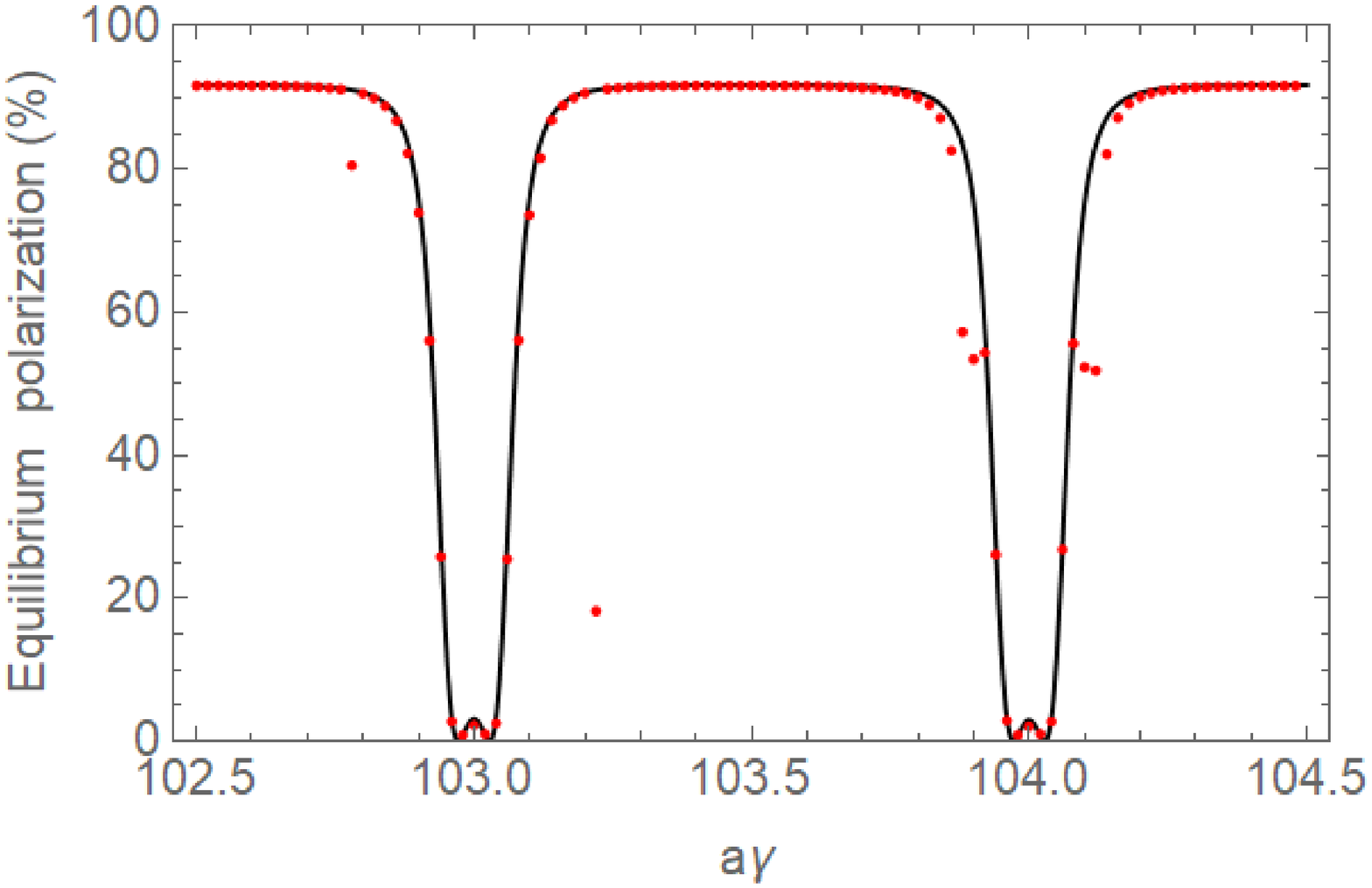}
\end{minipage}%
}%

\subfigure[WW]{
\begin{minipage}[t]{0.9\columnwidth}
\centering
\includegraphics[width=7cm]{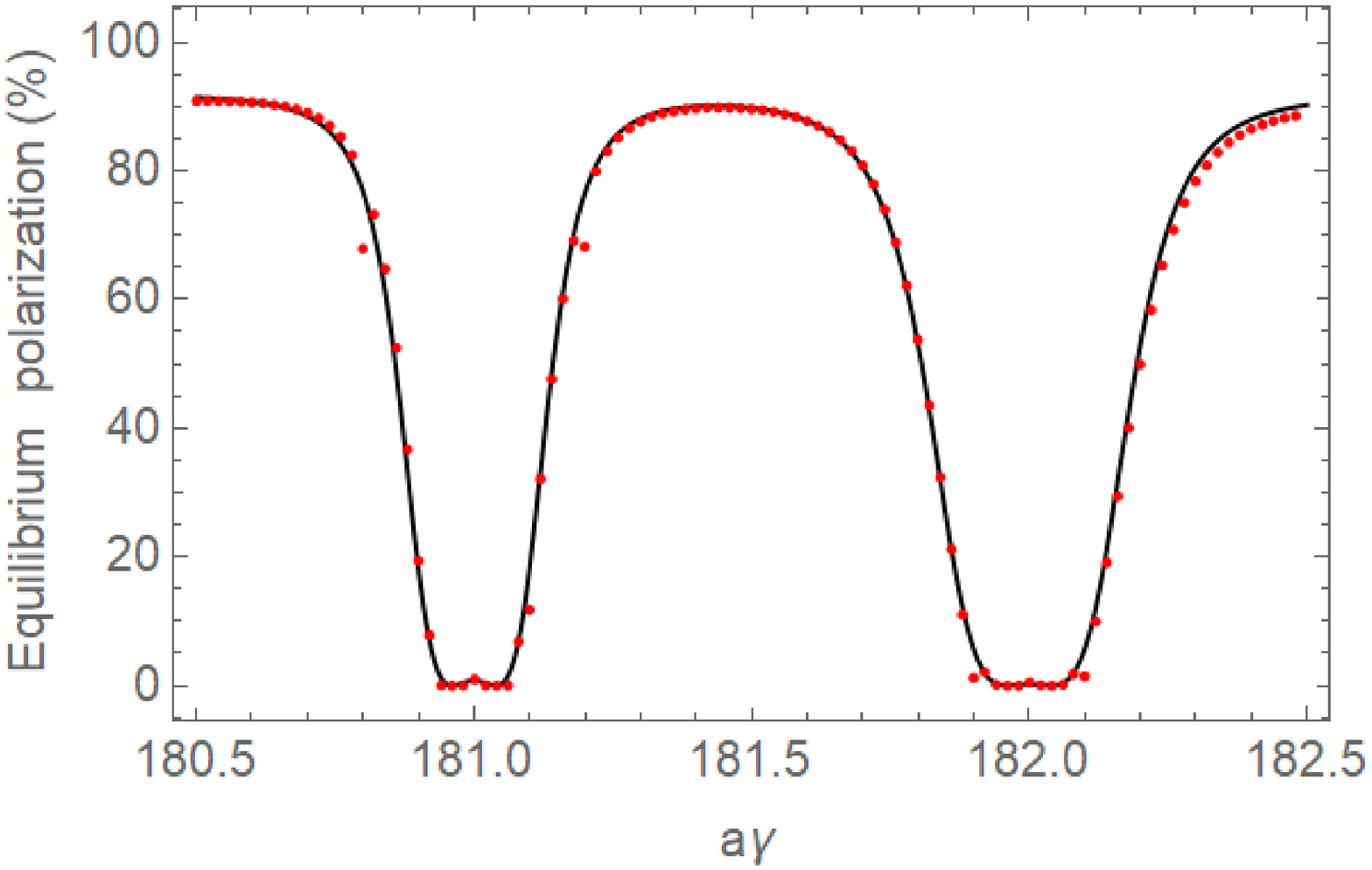}
\end{minipage}%
}%

\subfigure[Higgs]{
\begin{minipage}[t]{0.9\columnwidth}
\centering
\includegraphics[width=7cm]{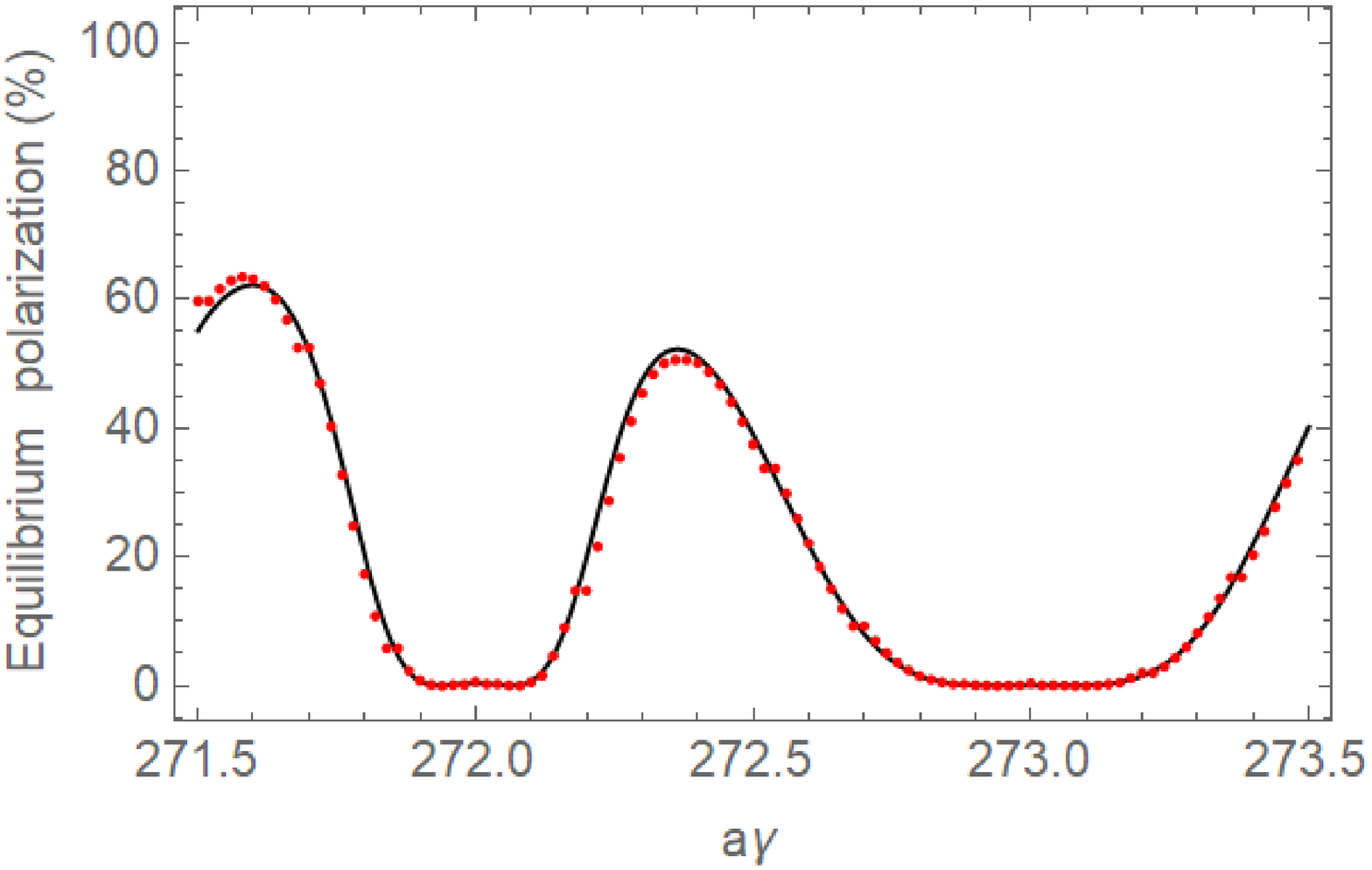}
\end{minipage}
}
\qquad
\caption{The equilibrium beam polarization as a function of $a\gamma_0$ near three working energies of the CEPC.
The red points show the simulation results of 
SLIM, the step size $\Delta (a\gamma_0)=0.02$. 
The black curve shows the analytical calculation results using the first-order
theory. There are dips near first-order ``parent" betatron spin resonances $\nu_0\pm0.11=k$ (horizontal) and $\nu_0\pm0.22=k$ (vertical), but these dips are much narrower compared to those around first-order ``parent" synchrotron spin resonances.}
\label{fig:fitting_plot}
\end{figure}

Then, we employed the SLIM formalism implemented in Bmad to calculate the equilibrium beam polarization as a function of $a\gamma_0$ for the CEPC lattice,
centered around three working beam energies $a\gamma_0=103.5$ (45.6~GeV, Z-pole), $a\gamma_0=181.5$ (80~GeV, WW threshold), and $a\gamma_0=272.5$ (120~GeV, Higgs), as shown in Fig.~\ref{fig:fitting_plot}(a), (b), (c). The step size of $a\gamma_0$ is 0.02 in the simulations. 
For  comparison, we also used Eqs.(\ref{eq:p2}), (\ref{eq:p1}) and (\ref{eq:firstorder}) to
evaluate the equilibrium beam polarization analytically, taking into account spin resonances up to the first order. Only the strengths of the two nearest integer spin resonances were retained in the evaluation of Eq.~(\ref{eq:firstorder}). The (numerical) SLIM simulation results match well with the analytical calculations.
As the integer spin resonances in the tilt of ${\hat n}_0$ become much stronger at the working energy for the Higgs particle, the equilibrium beam polarization for the cases with
$[a\gamma_0]\approx 0.5$ is
also much reduced. 
In addition, there are dips 
in the equilibrium beam polarization near the first-order ``parent" betatron spin resonances, for example when $a\gamma_0$ is near $104\pm 0.108$ at the working energy for Z.  But apparently their influence is quite
localized, and much weaker compared to the first-order ``parent" synchrotron spin resonances. This justifies the analysis of the relative significance of first-order spin resonances in Section~\ref{sec:theory}.

We also evaluated the depolarization effects using the theories of the correlated and uncorrelated regimes of spin resonance crossings.
We used Eq.~(\ref{eq:highertorder}) to estimate the effect of 
the correlated resonance crossing,
and Eq.~(\ref{eq:uncorrelated}) to estimate the effect of 
the uncorrelated resonance crossing, respectively,
Then we evaluated Eq.~(\ref{eq:p2}) and Eq.~(\ref{eq:p1}) to estimate the equilibrium beam polarization of the CEPC lattice.
In both cases, we only retained the two Fourier harmonics nearest to the integer 
part of $a\gamma_0$ and for Eq.~(\ref{eq:highertorder})
we retained $m$ in the range of -100 to 100 
as a result of a check of convergence.

\section{Results of simulations\label{sec:sim}}

In this section we describe the evaluation the depolarization effects of the CEPC lattice using Monte-Carlo Carlo simulations with the PTC code~\cite{duanz},
and compare the results with the predictions of the theories of radiative depolarization, for various lattice settings.
In the Monte-Carlo simulations, a bunch of 54 particles was launched on the closed orbit
with the same vertical spin, and tracked for 10 damping times in the presence of stochastic photon emissions, but disregarding the Sokolov-Ternov effect.
The decay of the beam polarization was recorded, and the data of the calculated beam polarization in the last 8 damping times were
fitted to obtain an estimate of the time constant of the spin diffusion $\tau_d$. 
Then the equilibrium beam polarization was estimated through Eq.~(\ref{eq:p2}) and Eq.~(\ref{eq:p1}).

\subsection{Dependence on beam energy}
\begin{figure}[htbp]
\centering
\subfigure[Z]{
\begin{minipage}{7cm}
\centering
\includegraphics[width=7cm]{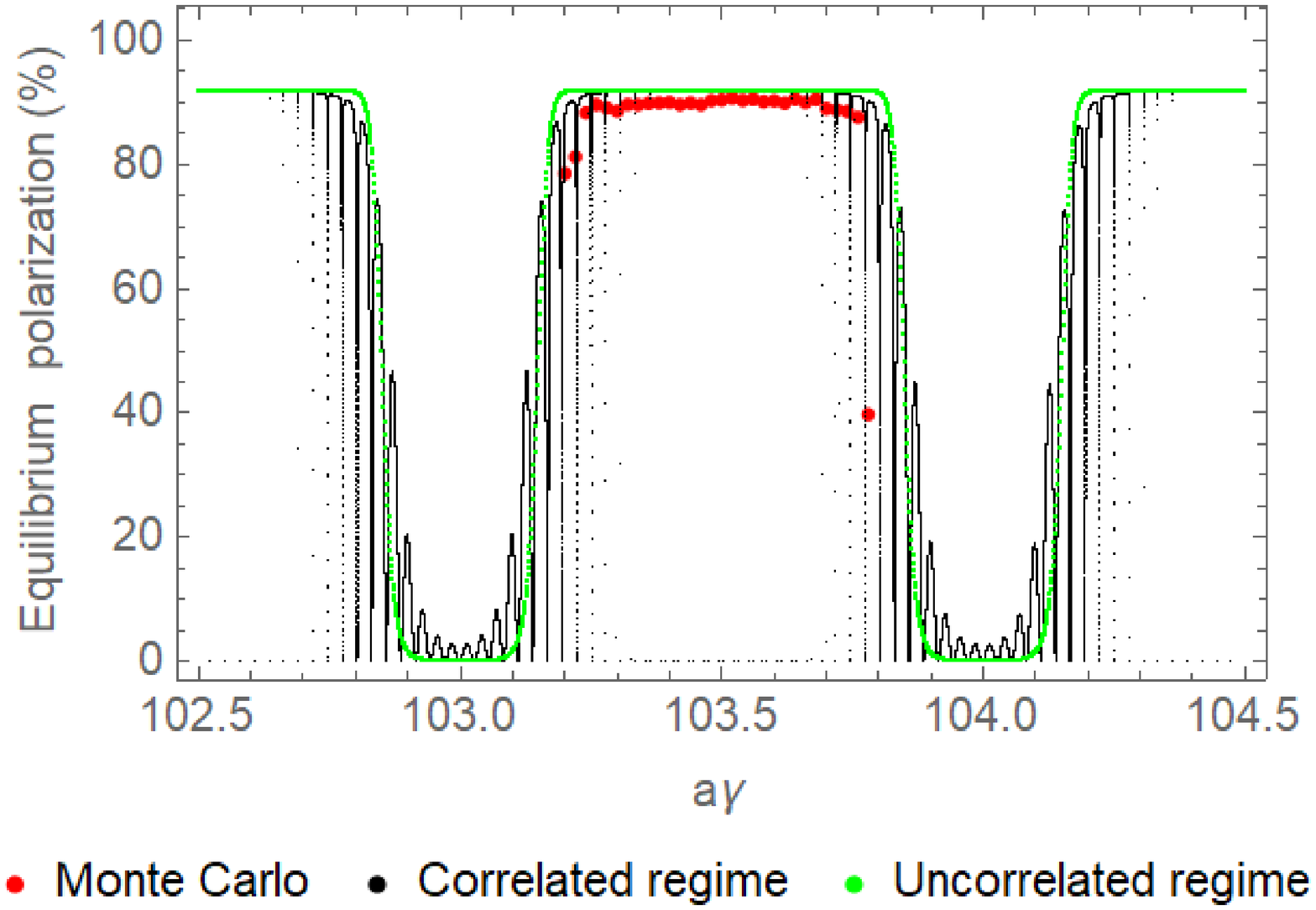} \\
\end{minipage}
}

\subfigure[WW]{
\begin{minipage}{7cm}
\centering
\includegraphics[width=7cm]{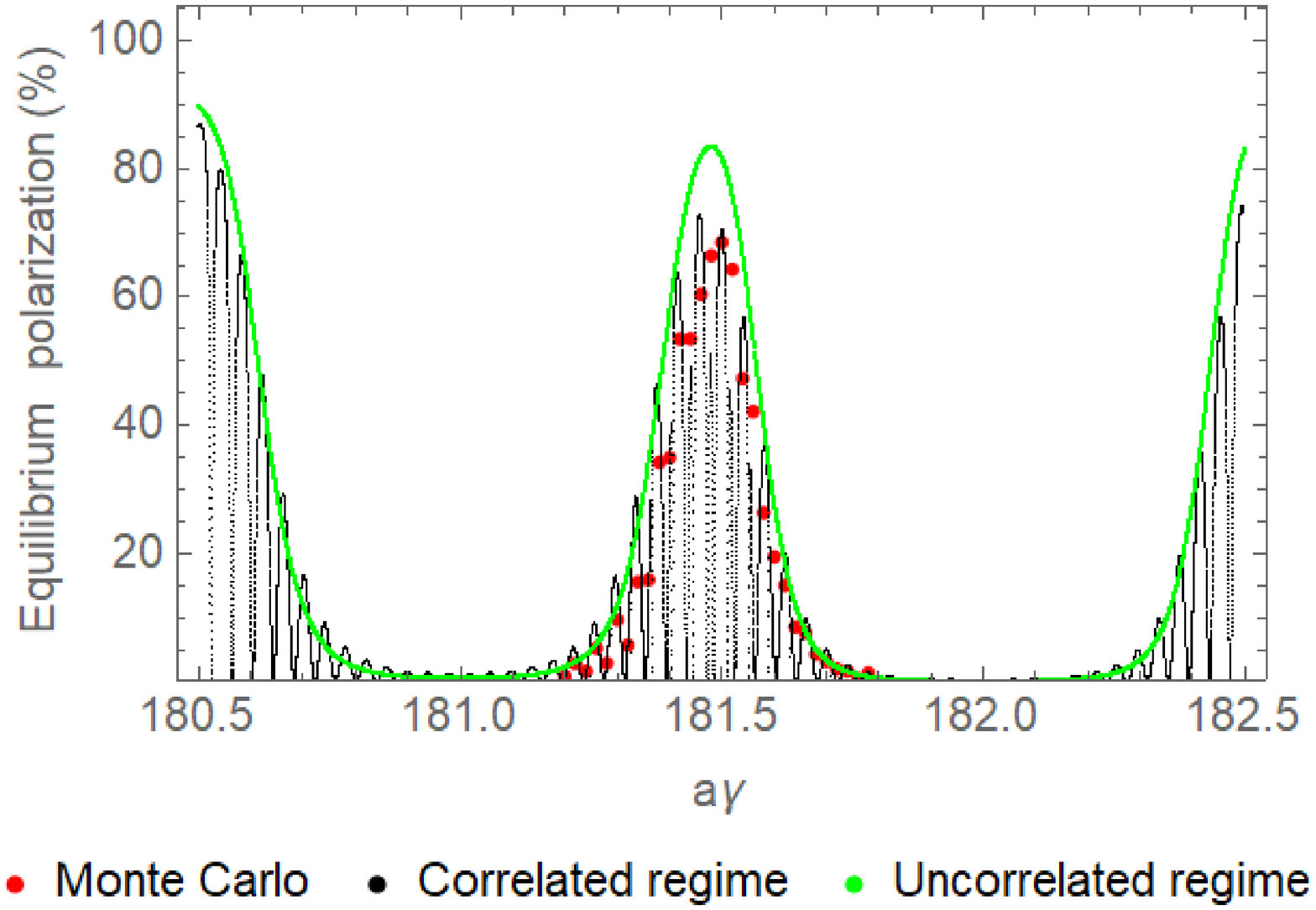} \\
\end{minipage}
}

\subfigure[Higgs]{
\begin{minipage}{7cm}
\centering
\includegraphics[width=7cm]{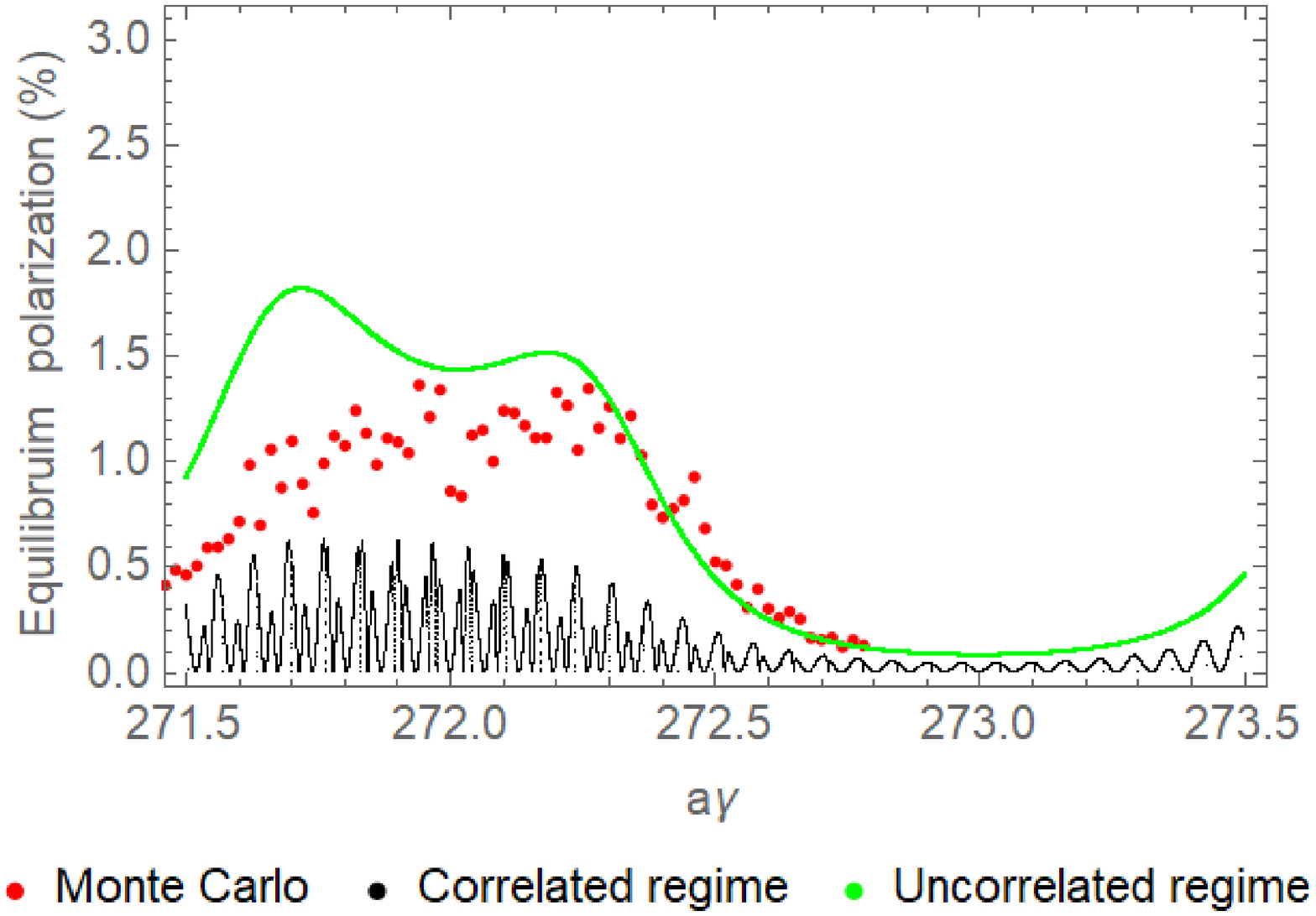} \\
\end{minipage}
}
\caption{ The equilibrium beam polarization as a function of $a\gamma_0$ near three working energies of the CEPC.
The black dots and blue dots show the equilibrium polarization calculated using the theories of correlated and uncorrelated regimes, respectively, the red points represent the Monte-Carlo simulation results with a step size of $a\gamma_0$ of 0.02 in the central region.
Note that in (c) a different y-axis range is used since
the equilibrium polarization is very low at the Higgs energy. }
\label{fig:fitting_plot2}
\end{figure}

We first studied the variation of the equilibrium beam polarization as a function of $a\gamma_0$ near the three operation energies, for Z, WW and Higgs production respectively, as shown in 
Fig.~\ref{fig:fitting_plot2}(a), (b) and (c). For the Z and WW energies, the valleys near integers become much wider relative to those in Fig.~\ref{fig:fitting_plot}(a) and (b), as predicted by both theories, while the Monte-Carlo simulation results agree better with the
theory of the correlated regime, as fine structure of higher-order synchrotron sideband spin resonances is clearly seen. For the Higgs energy,
however, although the equilibrium beam polarization level is very low, the results of the Monte-Carlo simulation are quite close to the predictions of the theory
of the uncorrelated regime, but are generally higher than the prediction of the theory of the correlated regime,
and there are no clear signs of higher-order synchrotron sideband spin resonances. 

\begin{figure}[htbp]
\centering
\subfigure[Polarization time constants]{
\begin{minipage}{0.9\columnwidth}
\centering
\includegraphics[width=0.9\columnwidth]{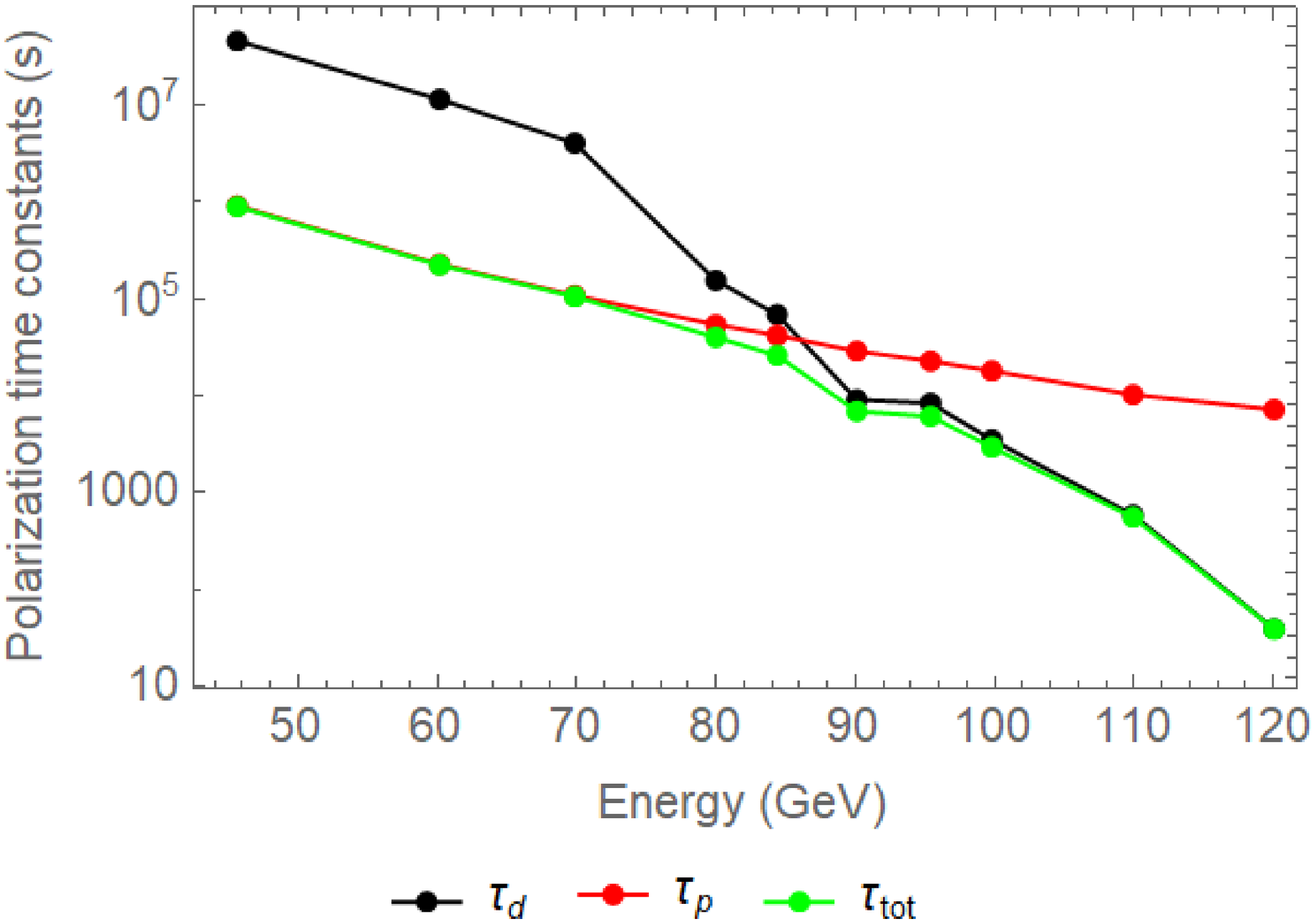} \\
\end{minipage}
}
\subfigure[Equilibrium beam polarization]{
\begin{minipage}{0.9\columnwidth}
\centering
\includegraphics[width=1.0\columnwidth]{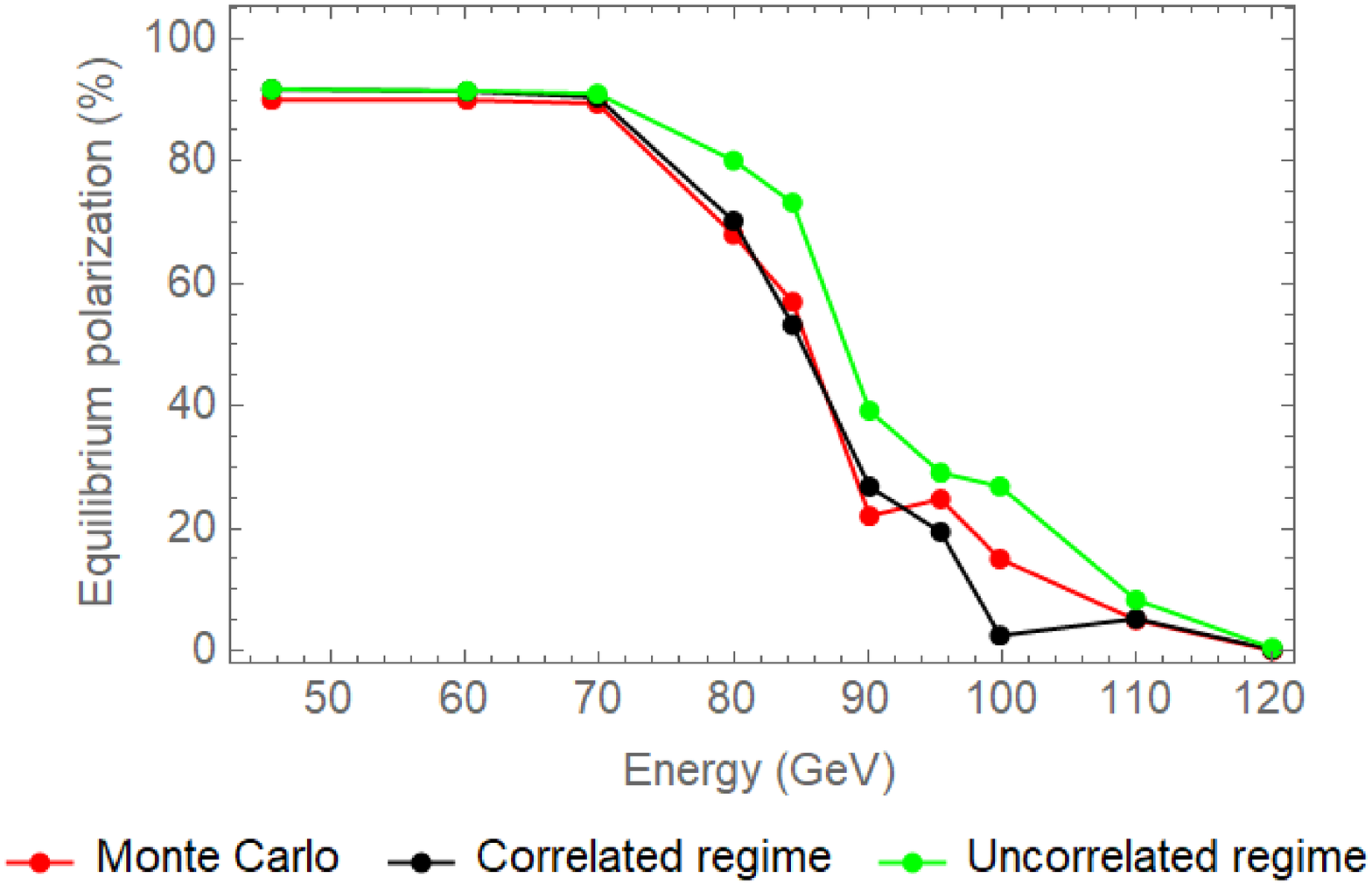} \\
\end{minipage}
}
\caption{Polarization time constants (a) and equilibrium beam polarization (b) as a function of beam energies for the CEPC lattice.
In (b), estimates of the equilibrium beam polarization using the theories of ``correlated regime'' and ``uncorrelated regime'' are also
plotted for comparison with those derived from the Monte-Carlo simulations.
}
\label{fig:pvsE}
\end{figure}

\begin{table*}[!htb]
\caption{The CEPC lattice parameters. (* indicates the planned operation energies in the CEPC CDR.)}
{\begin{tabular}{@{}lcllcccccc@{}} \hline \hline
 Beam energy (\si{\GeV}) &$a\gamma_0$& $|{\tilde \omega}_k|^2$ ($\times 10^{-9}$)& $|{\tilde \lambda}_k|^2$ ($\times 10^{-5}$)   & $\sigma_{\delta} (\times 10^{-4})$ & $\nu_z$&$\tau_p$ (\si{\hour}) & $\kappa$&$\sigma$ \\ \hline
\multirow{2}*{45.6*} & \multirow{2}*{103.5} & $|{\tilde \omega}_{103}|^2=2.7 $ & $|{\tilde \lambda}_{103}|^2=2.2645 $ & \multirow{2}*{3.77} & \multirow{2}*{0.028} & \multirow{2}*{252.72} & \multirow{2}*{0.03}& \multirow{2}*{1.39} \\
&& $|{\tilde \omega}_{104}|^2=2.8 $& $|{\tilde \lambda}_{104}|^2=1.7166 $ &&&& \\
\hline
\multirow{2}*{60.1} & \multirow{2}*{136.5} & $|{\tilde \omega}_{136}|^2=3.7$& $|{\tilde \lambda}_{136}|^2= 0.8178$ & \multirow{2}*{4.96} & \multirow{2}*{0.028} & \multirow{2}*{63.34} & \multirow{2}*{0.20}& \multirow{2}*{2.42} \\
&& $|{\tilde \omega}_{137}|^2=14.7 $& $|{\tilde \lambda}_{137}|^2=5.5717 $ &&&& \\
\hline
\multirow{2}*{69.8} & \multirow{2}*{158.5} & $|{\tilde \omega}_{158}|^2=6.6 $& $|{\tilde \lambda}_{158}|^2=0.9574 $ & \multirow{2}*{5.77} & \multirow{2}*{0.0324} & \multirow{2}*{30.00} & \multirow{2}*{0.36}& \multirow{2}*{2.82} \\
&& $|{\tilde \omega}_{159}|^2=26.4 $& $|{\tilde \lambda}_{159}|^2=4.4585 $ &&&& \\
\hline
\multirow{2}*{80.0*} & \multirow{2}*{181.5} & $|{\tilde \omega}_{181}|^2=14.4 $& $|{\tilde \lambda}_{181}|^2=4.9118 $ & \multirow{2}*{6.61} & \multirow{2}*{0.0395} & \multirow{2}*{15.24} & \multirow{2}*{0.52} & \multirow{2}*{3.04}\\
&& $|{\tilde \omega}_{182}|^2=53.3 $& $|{\tilde \lambda}_{182}|^2=15.6433 $ &&&& \\
\hline
\multirow{2}*{84.4} & \multirow{2}*{191.5} & $|{\tilde \omega}_{191}|^2=16.3 $& $|{\tilde \lambda}_{191}|^2=15.5332 $ & \multirow{2}*{6.97} & \multirow{2}*{0.0425} & \multirow{2}*{11.65} & \multirow{2}*{0.61}& \multirow{2}*{3.14} \\
&& $|{\tilde \omega}_{192}|^2=19.8 $& $|{\tilde \lambda}_{192}|^2= 1.0088$ &&&& \\
\hline
\multirow{2}*{90.1} & \multirow{2}*{204.5} & $|{\tilde \omega}_{204}|^2=19.1 $& $|{\tilde \lambda}_{204}|^2=3.7786 $ & \multirow{2}*{7.43} & \multirow{2}*{0.0467} & \multirow{2}*{8.39} & \multirow{2}*{0.72}& \multirow{2}*{3.25} \\
&& $|{\tilde \omega}_{205}|^2=43.8$& $|{\tilde \lambda}_{205}|^2= 0.6403$ &&&& \\
\hline
\multirow{2}*{95.4} & \multirow{2}*{216.5} & $|{\tilde \omega}_{216}|^2=15.0$& $|{\tilde \lambda}_{216}|^2=6.1547 $ & \multirow{2}*{7.88} & \multirow{2}*{0.0515} & \multirow{2}*{6.31} & \multirow{2}*{0.80}& \multirow{2}*{3.31}  \\
&& $|{\tilde \omega}_{217}|^2=34.8 $& $|{\tilde \lambda}_{217}|^2=1.2292 $ &&&& \\
\hline
\multirow{2}*{99.8} & \multirow{2}*{226.5} & $|{\tilde \omega}_{226}|^2=10.5$& $|{\tilde \lambda}_{226}|^2=35.2604 $& \multirow{2}*{8.24} & \multirow{2}*{0.0550} & \multirow{2}*{5.03} & \multirow{2}*{0.90}& \multirow{2}*{3.39} \\
&& $|{\tilde \omega}_{227}|^2=27.4 $& $|{\tilde \lambda}_{227}|^2=4.9787 $ &&&& \\
\hline
\multirow{2}*{109.9} & \multirow{2}*{249.5} & $|{\tilde \omega}_{249}|^2=56.9 $& $|{\tilde \lambda}_{249}|^2=26.9851 $ & \multirow{2}*{9.08} & \multirow{2}*{0.0585} & \multirow{2}*{3.10} & \multirow{2}*{1.48}& \multirow{2}*{3.87} \\
&& $|{\tilde \omega}_{250}|^2=41.3$& $|{\tilde \lambda}_{250}|^2=46.9963 $ &&&& \\
\hline
\multirow{3}*{120.1*} & \multirow{3}*{272.5} & $|{\tilde \omega}_{271}|^2=770.4 $& $|{\tilde \lambda}_{271}|^2=41.6290 $& \multirow{3}*{9.90} & \multirow{3}*{0.0650} & \multirow{3}*{2.03} & \multirow{3}*{1.95}& \multirow{3}*{4.15} \\
&& $|{\tilde \omega}_{272}|^2=95.8$& $|{\tilde \lambda}_{272}|^2=12.0825$ &&&& \\
&& $|{\tilde \omega}_{273}|^2=1684.2$& $|{\tilde \lambda}_{273}|^2=361.7036 $ &&&& \\
\hline \hline
\end{tabular} \label{tab:p_diff_ener}}
\end{table*}

We then investigated the dependence of the depolarization effects on the beam energy, in the range of 45.6~\si{GeV} to 120~\si{GeV}. The
energy points were selected so that $[a\gamma_0]=0.5$. The
total RF voltage was adjusted to compensate for the increase of synchrotron radiation energy loss at higher beam energies and  the synchrotron tunes
were set to the design values according to the CEPC CDR specifications at 45.6, 80 and 120 \si{GeV}. For other beam energy points, the synchrotron tunes 
were set in a smoothly increasing manner. The related lattice parameters are summarized in Table~\ref{tab:p_diff_ener}.

Fig.~\ref{fig:pvsE} (a) shows the time constant of the spin diffusion $\tau_d$ obtained by Monte-Carlo simulations, 
the time constant of the Sokolov-Ternov effect $\tau_p$ calculated using Bmad, as well as the self-polarization build-up time $\tau_\text{tot}$,
at the various beam energies. $\tau_d$ decreases much faster than $\tau_p$  with increasing beam energy, as predicted by both Eqs.(\ref{eq:highertorder}) and (\ref{eq:uncorrelated}),
and becomes the dominant contributor to $\tau_{\text{tot}}$ at beam energies beyond 85~\si{GeV}, This leads to a much reduced level of equilibrium beam polarization at higher beam energies, as shown in
Fig.~\ref{fig:pvsE} (b). In addition to the results derived from the 
Monte-Carlo simulations, calculations  using the theories of the
correlated  and uncorrelated regimes are also presented in the figure for comparison. Up to 90~\si{GeV}, the simulated equilibrium polarization matches quite well with
the theory of the correlated regime, and is smaller than the theory of the uncorrelated regime. However, as the beam energies get even higher, 
there are nontrivial discrepancies between the results of simulations and the theories.
This might suggest the paradigm shift from the correlated regime towards the uncorrelated regime. As shown previously, the results from 
the Monte-Carlo simulation are close to the theory of the uncorrelated regime at around 120~\si{GeV}.

\subsection{Influence of asymmetric wigglers at the Z pole}

The time constant of the Sokolov-Ternov effect is as large as 253~\si{\hour} at Z pole. Using the self-polarization mechanism to generate sufficient
beam polarization for RD requires a much reduced polarization build-up time. This can be achieved by implementing asymmetric wiggler 
magnets~\cite{dks}. This possibility was investigated in detail for LEP~\cite{lepwiggler}, and also studied for the FCC-ee~\cite{wendt}. 
An asymmetric wiggler consists of three bending magnets. The central bending magnet has a length $L_+$ and a 
magnetic field $B_+$ in the same direction as the guiding magnetic field in the arcs. 
Its bending angle is denoted as $\theta_+$,
while the bending magnets on both sides have equal length $L_-$, equal magnetic field $B_-$ in the opposite direction to $B_+$, and equal bending angles $\theta_-$.
It is required that $\theta_+{+}2\theta_-=0$ to ensure no change in the layout and optics beyond the wiggler insertion. When $N_w$ units of such asymmetric wigglers
are included into the straight sections of the lattice, in 
addition to a decrease of $\tau_p$, there is also an decrease of $P_{\infty}$, as well as
increases in the energy loss per turn $U_0$, and the rms relative energy spread $\sigma_{\delta}$. 
A larger $\sigma_{\delta}$ leads to a larger modulation index $\sigma$, so that the higher-order synchrotron sideband spin resonances are much enhanced. This could
in turn increase $\tau_p/\tau_d$ and further reduce the equilibrium beam polarization level.

To this end, the influence of asymmetric wigglers on radiative depolarization was evaluated quantitatively. 
Ten identical units of asymmetric wigglers were inserted into the straight
sections of the CEPC lattice~\cite{xiawh}, with $L_-=2$ \si{\metre}, $L_+=1$ \si{\metre}, i.e., $B_+/B_-=4$. 
In this paper, various  settings of
$B_+$ and thus $\theta_+$ were used to study the influence on the depolarization effects.
The synchrotron tune was kept fixed at 0.028 by adjusting the total RF voltages. 

\begin{table}[hbt!]
\caption{Beam parameters for  various  wiggler settings}
{\begin{tabular}{@{}lcccccc@{}} \hline \hline
wigglers &$\theta_+$ ({rad})&$U_0$ (\si{\MeV})&$\sigma_{\delta} (\times 10^{-4})$&$\tau_p$ (\si{\hour})&$\kappa$&$\sigma$ \\ \hline
w/o  &  -           & 36.1      & 3.77  & 252.8&0.03 &1.39 \\
Case 1   & 0.0033      & 43.9      & 9.53  & 32.3&0.22 &3.52\\
Case 2   & 0.0056     & 60.0      & 17.26 & 7.2 &1.00 &6.38\\
Case 3   & 0.0080     & 84.8      & 24.55 & 2.5 &2.85 &9.07\\
\hline \hline
\end{tabular} \label{tab:wiggler}}
\end{table}

Table~\ref{tab:wiggler} shows the beam parameters for the lattice without wigglers, and three different wiggler settings
with increasing $\theta_+$. For example, in Case 1,
the time constant of the Sokolov-Ternov effect $\tau_p$ is reduced to about 
32~\si{h}, and  then about 2~\si{h} are needed to obtain 5\% beam polarization,
more or less sufficient for energy calibrations. Case 2 and Case 3 use even stronger wiggler magnets,
and the increase in $U_0$ and $\sigma_{\delta}$ is more substantial. Though not relevant for practical use, these
cases are employed here out of academic interest. 
In these three cases, the locations of higher-order synchrotron sideband spin resonances are the same while the theory of the
correlated regime predicts lower levels of equilibrium polarization
and wider valleys near these resonances for increasing wiggler strengths characterized by $\theta_+$.
In contrast, as predicted by the theory of the uncorrelated regime,
there are no higher-order synchrotron sideband spin resonances,
and the equilibrium polarization level is higher for $a\gamma_0$ near 103.5 compared to the prediction of the theory of the
correlated regime.

\begin{figure}[htbp]
\centering
\subfigure[Case 1: $\theta_+=0.0033$~rad]{
\begin{minipage}{8cm}
\centering
\includegraphics[width=7.5cm]{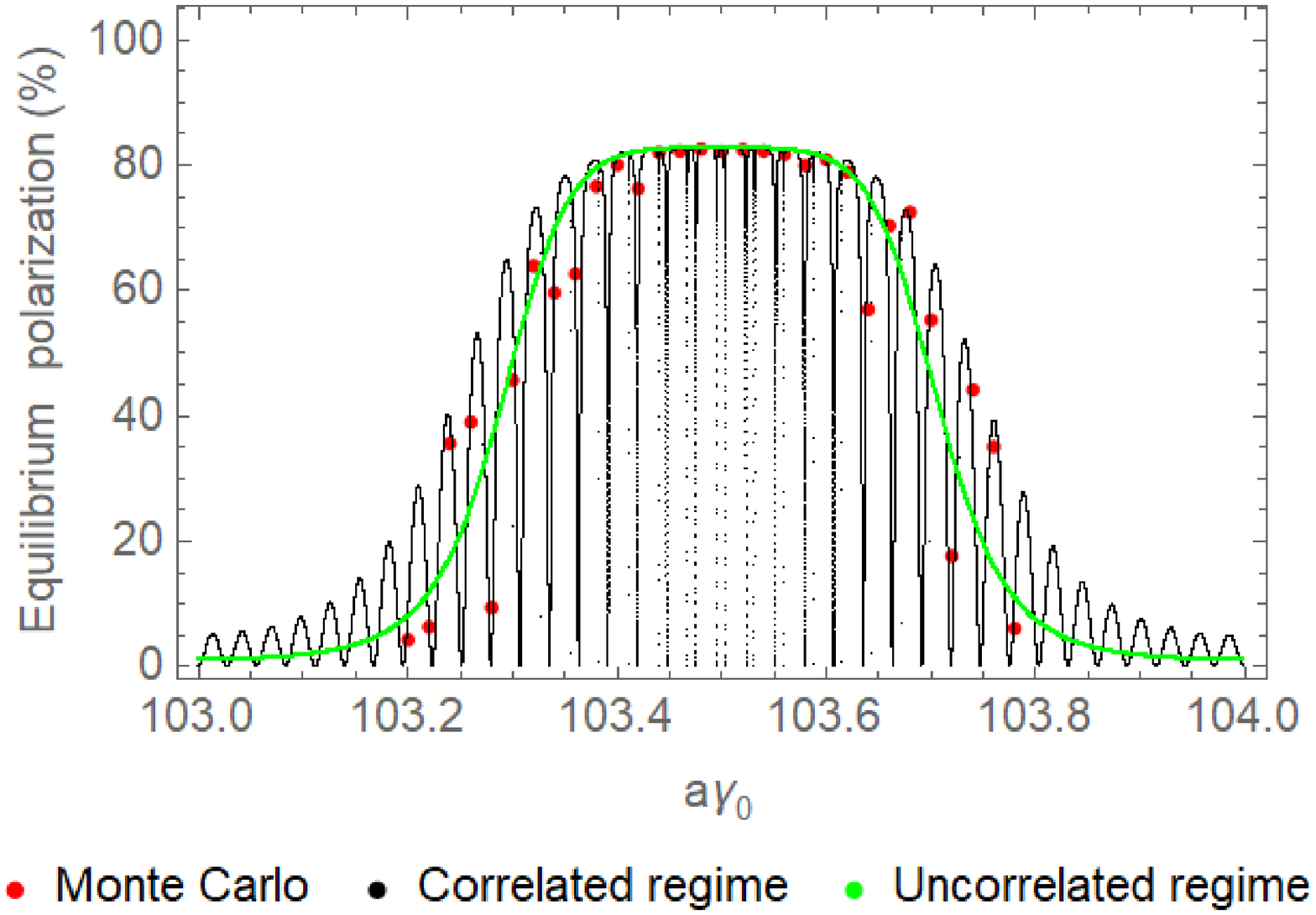} \\
\end{minipage}
}

\subfigure[Case 2: $\theta_+=0.0056$~rad]{
\begin{minipage}{8cm}
\centering
\includegraphics[width=7.5cm]{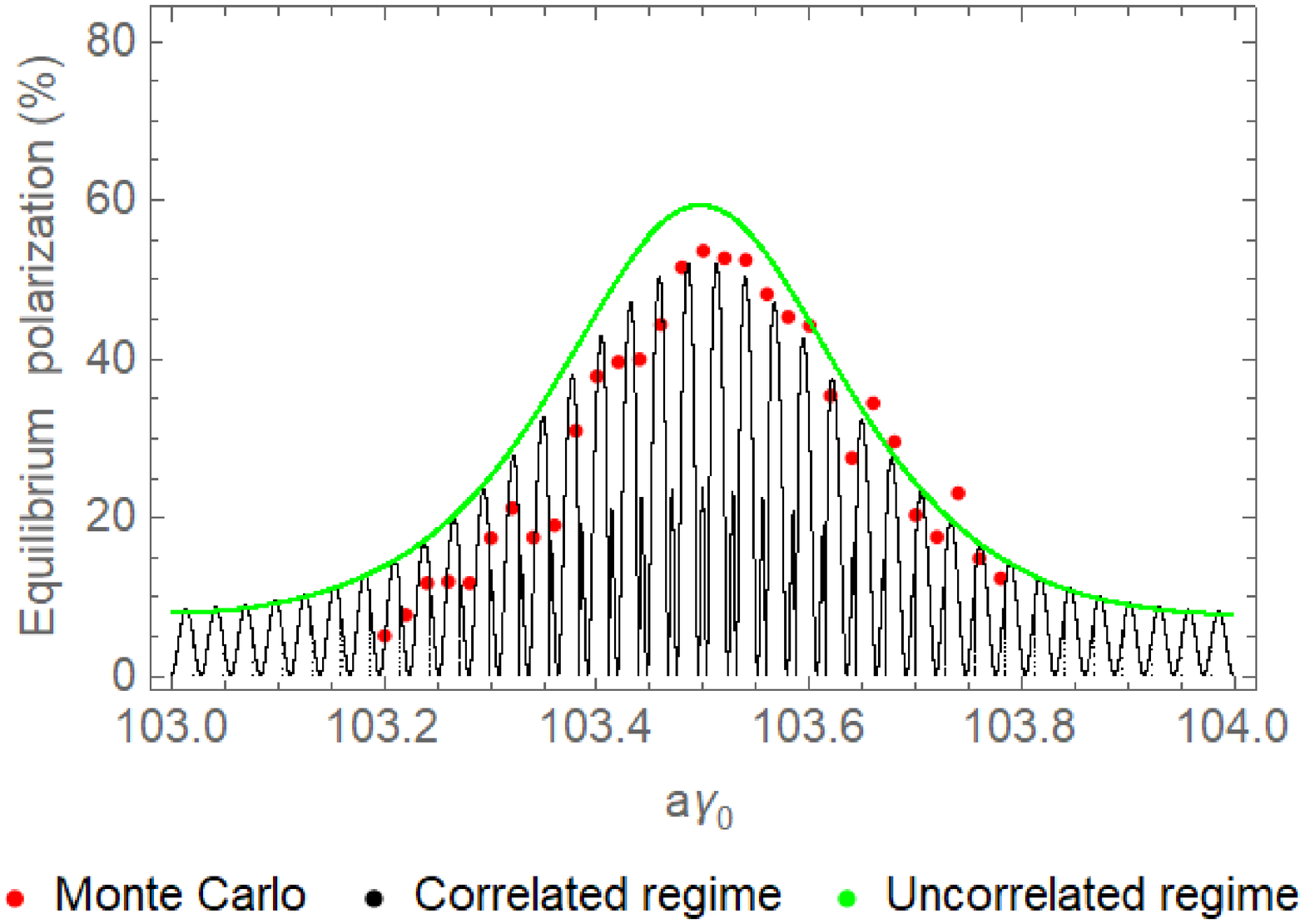} \\
\end{minipage}
}

\subfigure[Case 3: $\theta_+=0.0080$~rad]{
\begin{minipage}{8cm}
\centering
\includegraphics[width=7.5cm]{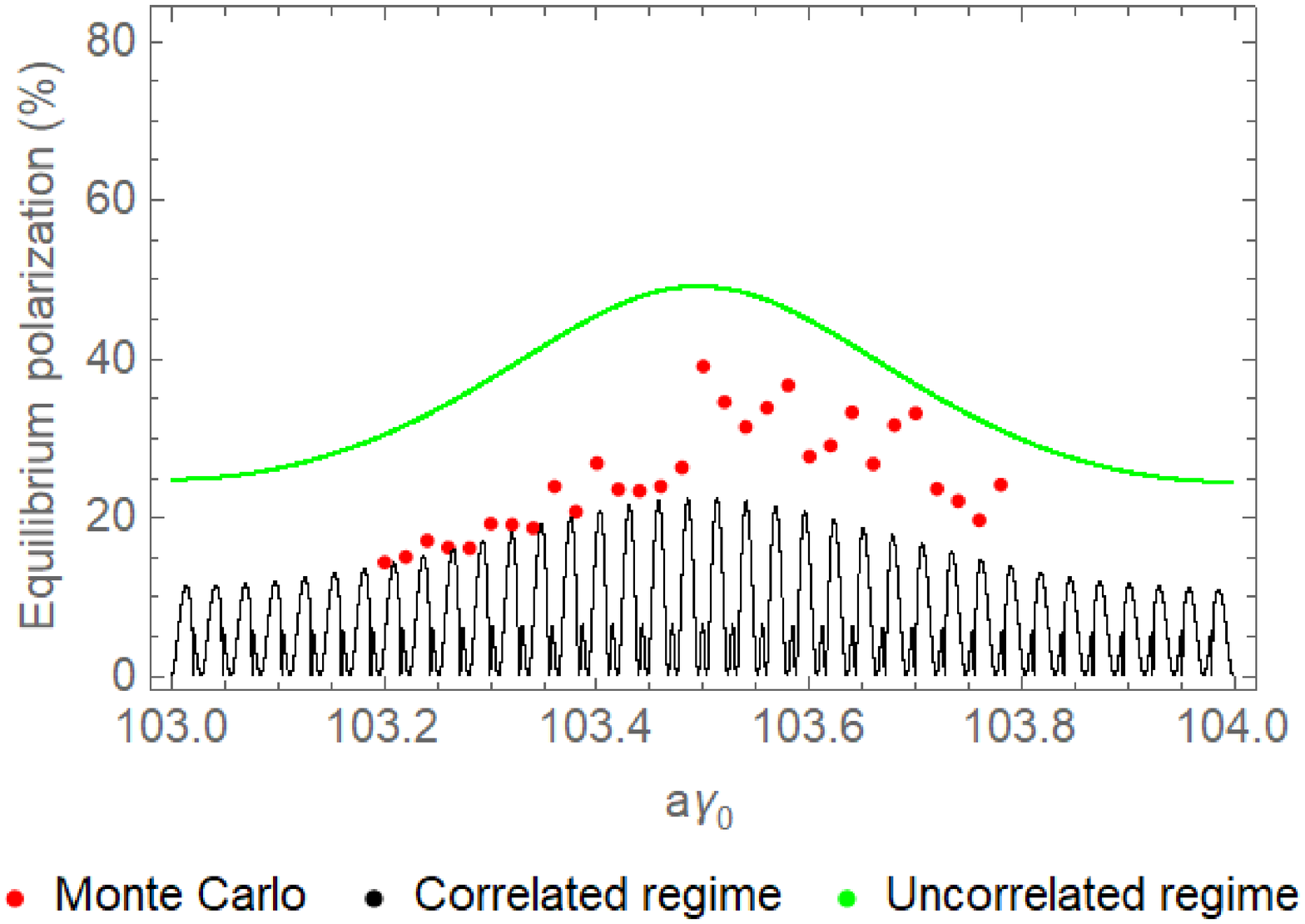} \\
\end{minipage}
}

\caption{ The equilibrium polarization vs. spin tune for the CEPC lattice in the presence of 10 wigglers. The Monte-Carlo simulations
adopt a step size of $a\gamma_0$ of 0.02.}
\label{fig:p_wiggler}
\end{figure}

Fig.~\ref{fig:p_wiggler} shows the equilibrium beam polarization as 
a function of $a\gamma_0$ for the lattices using these three different wiggler settings. The Monte-Carlo simulation results
are compared with the theories of the correlated regime and the uncorrelated regime. 
Fig.~\ref{fig:p_wiggler}(a) shows
the outcomes  of Case 1 where the
Monte-Carlo simulation results fit well with the theory of the correlated regime, with obvious polarization dips in the synchrotron sideband spin resonance regions, which are absent in the theory of the uncorrelated regime.
Fig.~\ref{fig:p_wiggler}(b) shows the outcomes of Case 2 where
the results of the Monte-Carlo simulation  are inconsistent with the valleys
predicted by the theory of the correlated regime.
Fig.~\ref{fig:p_wiggler}(c) shows the outcomes of Case 3 where the results of  Monte-Carlo
simulation  show no obvious resonance structures and are generally between the predicted levels of equilibrium beam polarization of the two theories.

\begin{figure*}[hbt!]
\centering
\subfigure[$U_{0}$ vs. $\theta_+$.]
{
\includegraphics[width=0.35\textwidth]{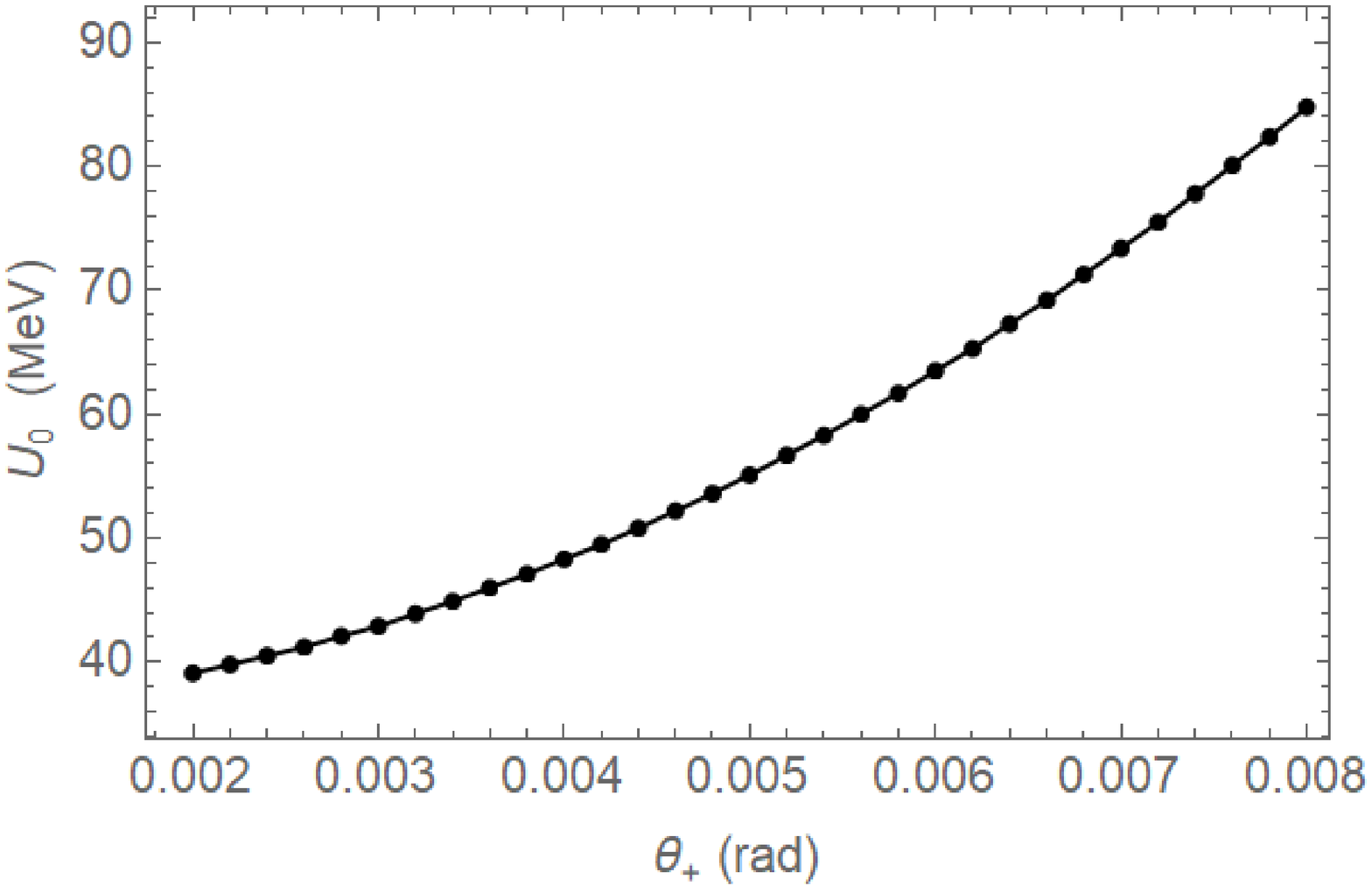}
}
\quad
\subfigure[$\sigma_{\delta}$ vs. $\theta_+$.]
{
\includegraphics[width=0.35\textwidth]{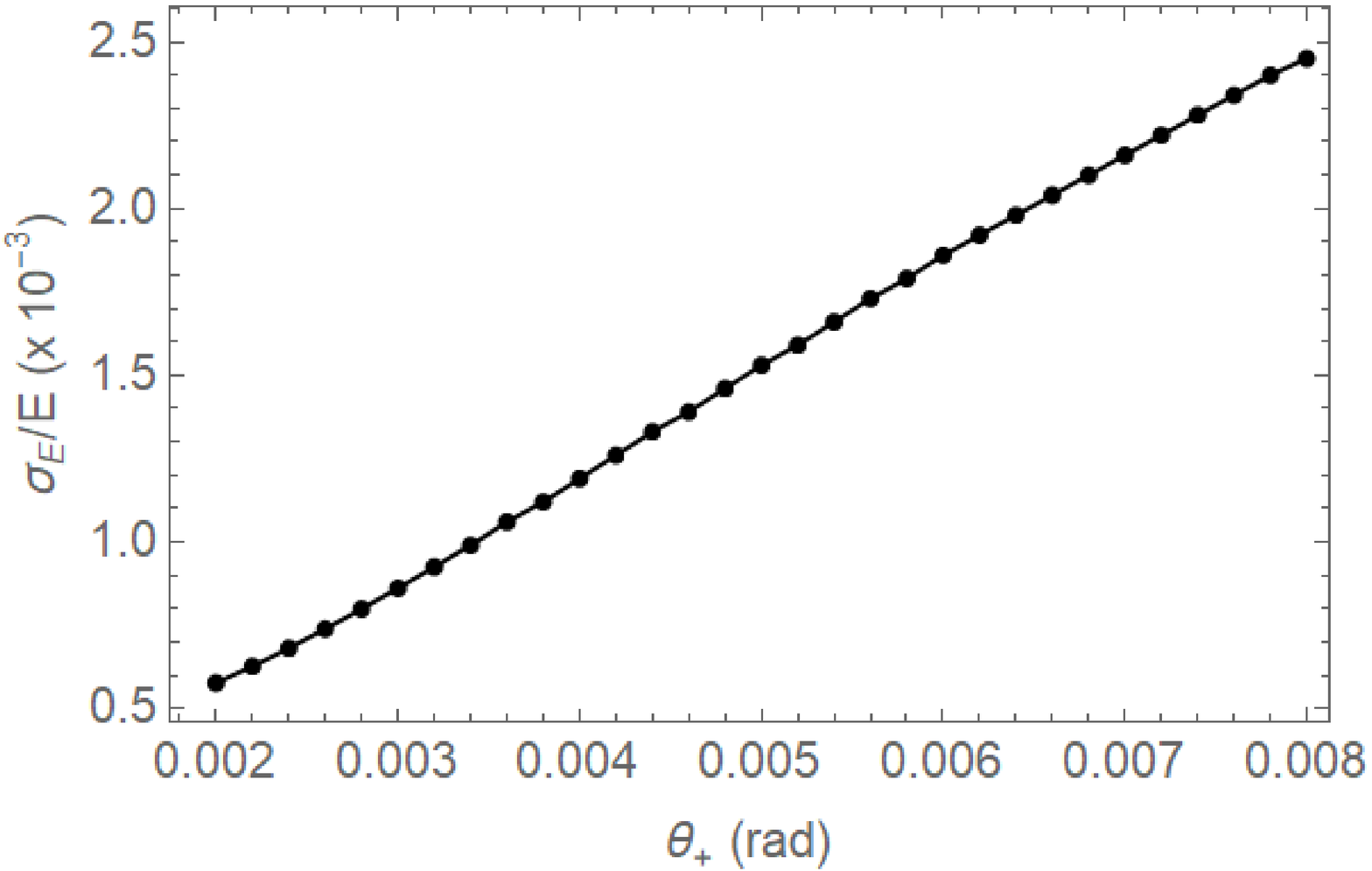}
}
\quad
\subfigure[Polarization time constants vs. $\theta_+$.]
{
\includegraphics[width=0.35\textwidth]{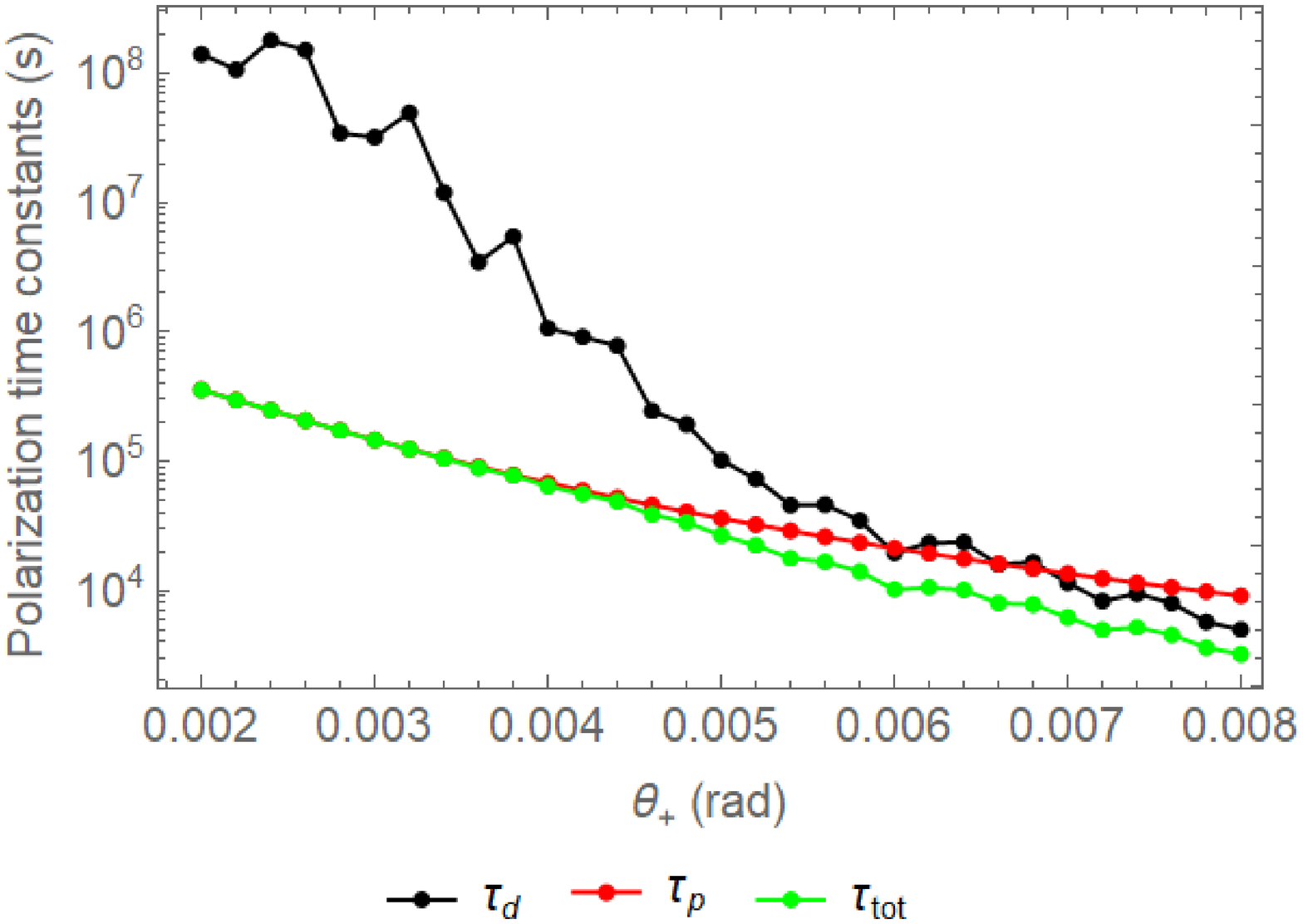}
}
\quad
\subfigure[$P_{eq}$ vs. $\theta_+$.]
{
\includegraphics[width=0.4\textwidth]{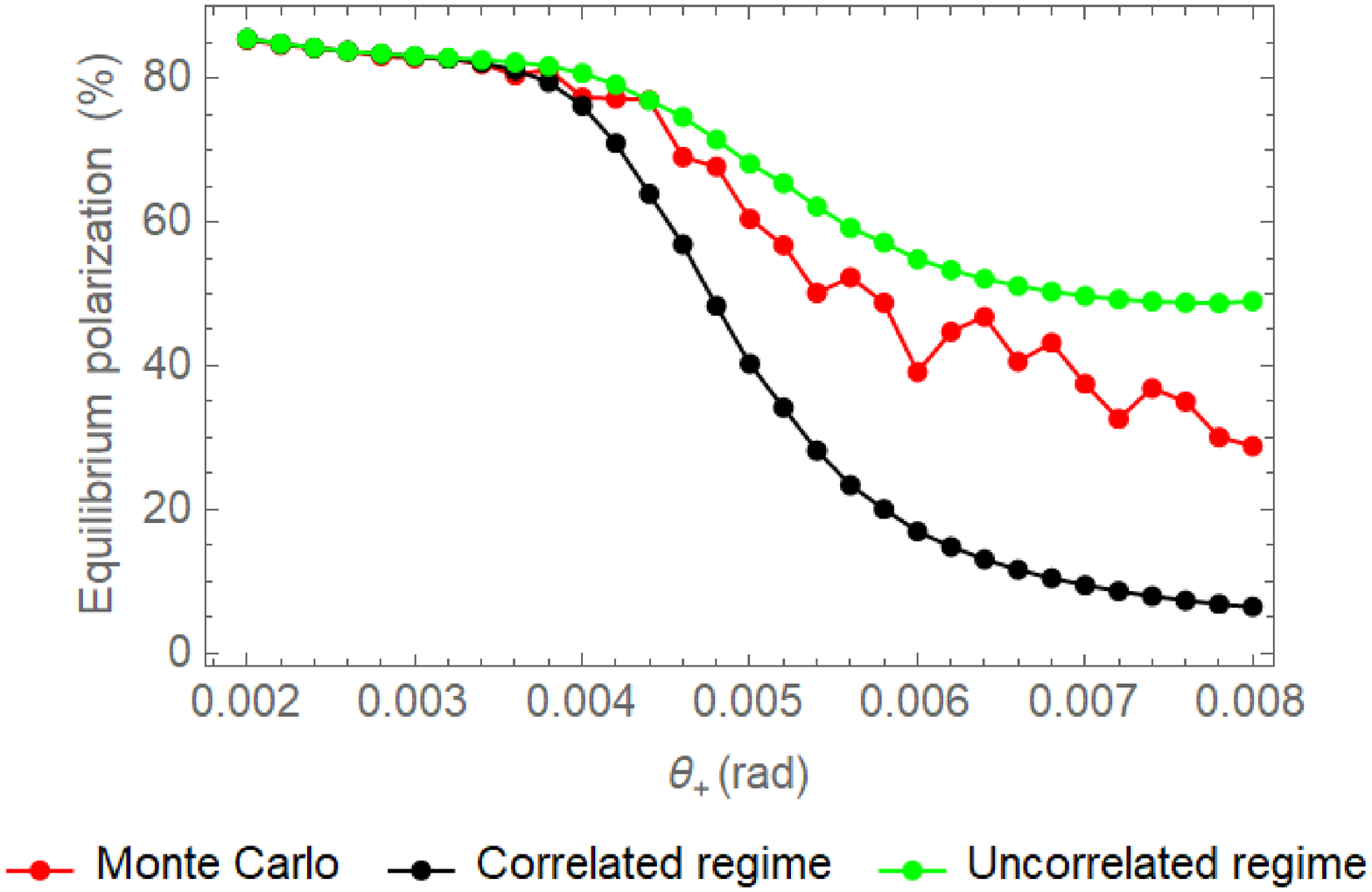}
}
\caption{The energy loss per turn, rms relative energy spread, polarization time and equilibrium polarization VS. $\theta_{+}$ . The step size $\Delta \theta_+=0.0002$ \si{\radian}. }
\label{fig:wigglersVStheta+}
\end{figure*}

These results of the Monte-Carlo simulations indicate deviations from the theory of the correlated regime, when the condition $\kappa\ll1$ on the correlation index no longer holds, but aren't
in agreement with the theory of the uncorrelated regime either. Using even larger $\theta_+$ in the lattice setup leads to an RF bucket 
insufficient for fully capturing the beam particles
in the simulations and thus this parameter space was not explored.

In addition, we fixed the beam energy at 45.6~\si{GeV}, and scanned $\theta_{+}$ to study the dependence of the energy loss per turn, rms relative energy spread, polarization time and equilibrium polarization, as shown in
Fig.~\ref{fig:wigglersVStheta+}. The results of simulations for  the equilibrium polarization match well with the theory of the 
correlated regime for small $\theta_+$ settings,
however, there is an increasing deviation for $\theta_+$ larger than 0.004~rad. In the parameter range covered, the results of Monte-Carlo simulation are between the equilibrium polarization levels
predicted by the two theories, respectively.

\subsection{Influence of RF settings at the Z pole}


In the CDR for the CEPC it is assumed that the ring has
a single frequency RF system of 650~\si{MHz}. Then for a
chosen momentum compaction factor, e.g., according to Eq.~(3.43) and Eq.~(5.66) in Ref.~\cite{sands_physics_1970}, the bunch length 
and synchrotron tune cannot be adjusted separately. However,
earlier studies revealed a novel coherent head-tail
instability (X-Z instability) induced by the beam-beam interaction at a large Piwinski angle~\cite{ohmi_coherent_2017,kuroo_cross-wake_2018}, that could be
enhanced when longitudinal impedance is also taken into account~\cite{zhang_self-consistent_2020,lin_coupling_2022}. 
Thus  it was suggested~\cite{migliorati_interplay_2021} that the instability
could be viably mitigated by including a higher-harmonic
RF system to lengthen the bunches. It has also been suggested that bunch lengthening can lead to a reduction in the center-of-mass energy spread in $e^+/e^-$ collisions. For non-colliding bunches, the 
introduction of harmonic cavity itself does not help to
reduce the rms energy spread. However, for colliding bunches, the rms energy spread of a beam can suffer from a significant 
increase due to beamstrahlung and the beamstrahlung-induced energy spread can to some extent be alleviated by the introduction of harmonic cavities~\footnote{
P. Raimondi, ``Updates on Monochromatization'', talk on FCC-FS EPOL group and FCCIS WP2.5 meeting 15~(24 Nov 2022), and 
D. Shatilov and P. Raimondi, `` FCC-ee parameters and Challenges'', talk on FCCIS Workshop 2022~(06 Dec 2022).
}.
Of course, a reduction of the rms energy spread in the beams also reduces the 
spread in the instantaneous spin-precession rate.

In such a double RF system the synchrotron tune $\nu_z$ at zero amplitude 
is different from that of a single RF system and so is the dependence of $\nu_z$ 
on the synchrotron amplitude~\cite{hof,lee_accelerator_2004}.
Since both the modulation index $\sigma$ and the correlation index $\kappa$ depend
on $\nu_z$, it's expected that
the radiative depolarization effects can be different.

Although there is currently no plan to use a double RF system for the 
CEPC, a study of the effect on depolarization with such a system opens up new directions for study which might bare fruit in the future. 
We therefore
now look in detail at the effects of a double-frequency system 
for the CEPC running near 45.6~GeV.



\begin{figure}[htbp]
\centering
\subfigure[Amplitude dependent synchrotron tune]{
\begin{minipage}{8cm}
\centering
\includegraphics[width=7.5cm]{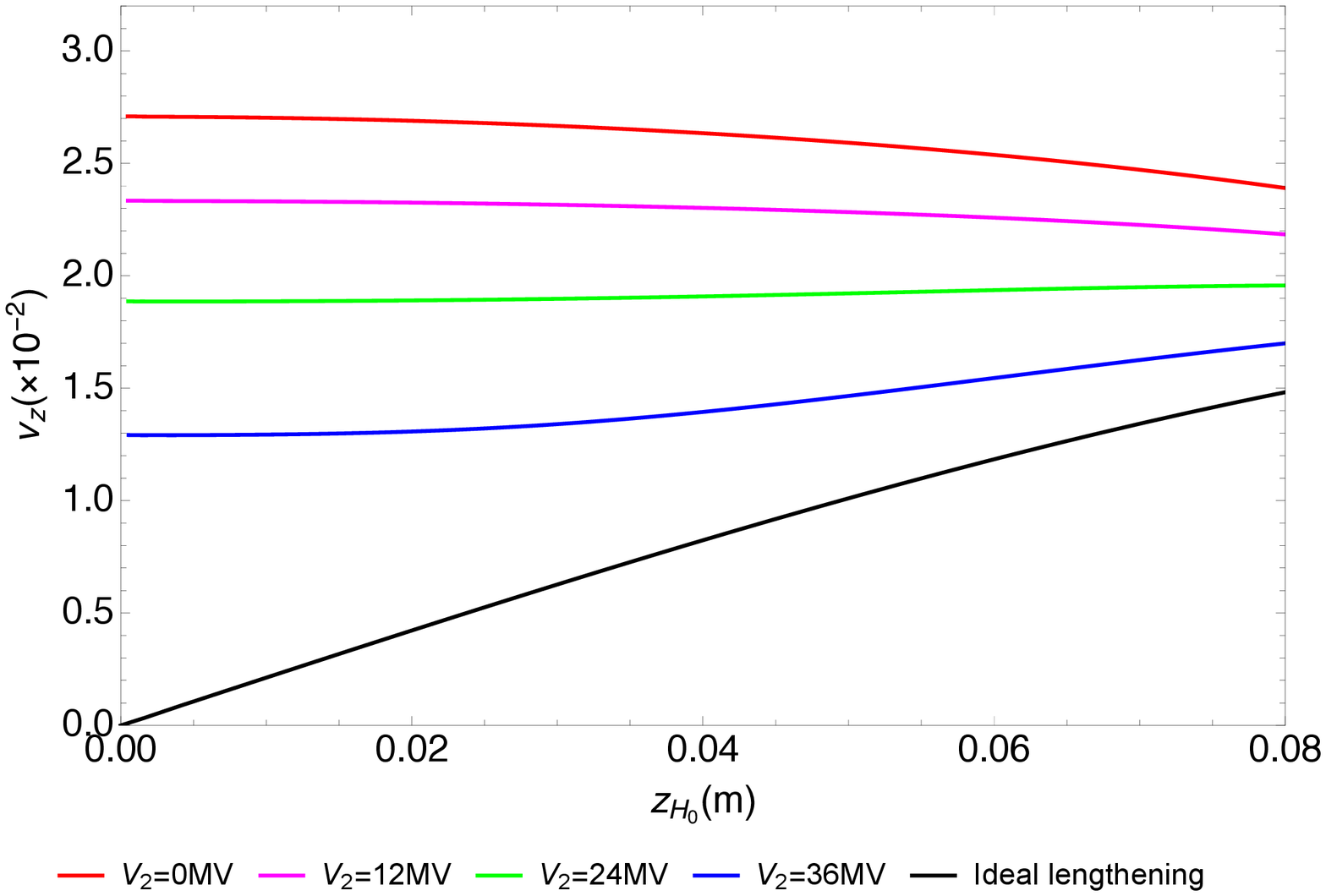} \\
\end{minipage}
}

\subfigure[Beam longitudinal distribution]{
\begin{minipage}{8cm}
\centering
\includegraphics[width=7.5cm]{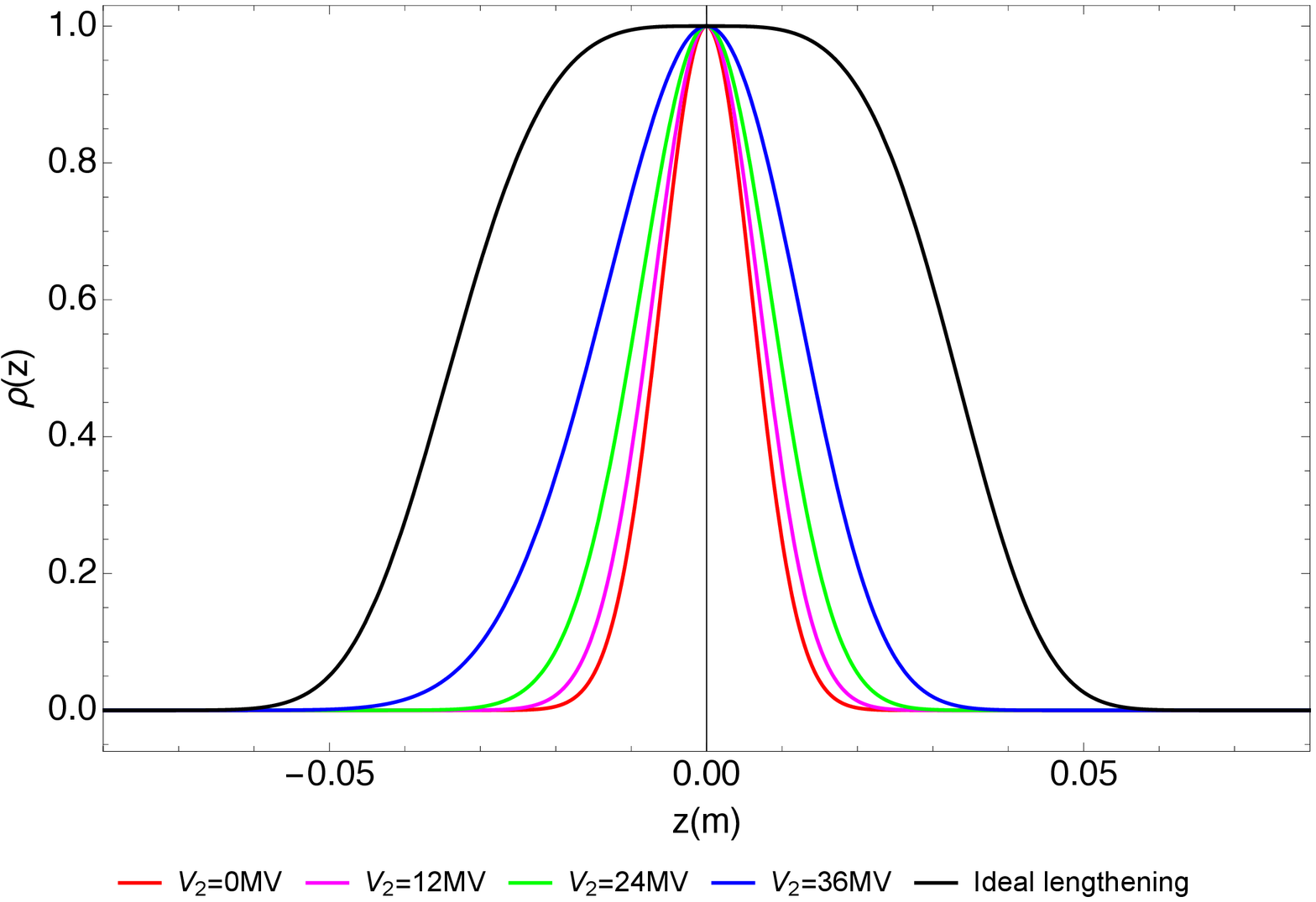} \\
\end{minipage}
}


\caption{The dependence of the synchrotron tune $\nu_z$ on the synchrotron amplitude,
and the longitudinal distribution $\rho(z)$ of beam particles,
for various settings of a double RF system.
The beam energy is set to 45.72 GeV~($a\gamma_0=103.76$) and $V_1=103.372$~MV.
The red curves correspond to Case A with a single RF system alone,
the magenta, blue and blue curves correspond to three instances of Case B with $\phi_{s2}=0$~rad but
different $V_2$, the black curves correspond to Case C at the ideal lengthening condition.
}
\label{fig:double_rf}
\end{figure}


Simulations with the thick-lens SLIM algorithm in the code SLICKTRACK~\cite{barber_polarisation_2005} for a real ring show that synchrotron motion for a single RF system is close to simple harmonic although the RF cavities are localised. Then it is reasonable, as a first step,  to follow the description in Ref.~\cite{lee_accelerator_2004} in which the RF system is distributed uniformly. Moreover, only synchrotron motion is included here  but the applicability of this model has been confirmed
 {\em a posteriori} with  Monte-Carlo simulations with localised cavities as
 described below.

Thus we begin by writing the voltage seen by beam particles in the presence of a double-frequency RF system as
\begin{equation}
V_{\text{tot}}(\phi)=V_1 \sin{(\phi + { \phi}_{s1})} + V_2 \sin{(n\phi + { \phi}_{s2})}
\end{equation}
where $\phi$ is the RF phase of a particle relative to that of the reference particle,  
$V$ and ${\phi}_s$ are the total voltage and RF synchronous phase for each RF frequency, with
subscripts ``1" and ``2" denoting the fundamental and the harmonic RF system, respectively, and the integer $n$ is the ratio of the harmonic RF frequency to the fundamental RF frequency.
The parameters of the double-frequency RF system must be set to compensate for the synchrotron radiation energy loss per turn $U_0$, this requires
\begin{equation}
    eV_1\sin{({ \phi}_{s1})} + eV_2 \sin{({ \phi}_{s2})}=U_0
 \label{eq:U0}    
\end{equation}
where $e$ is the charge of the electron. 

There are still 3
degrees of freedom in the parameters of the double RF system.
To simplify the discussion in this paper, we study three different
cases of RF settings, all these cases feature a single RF bucket and beam particles oscillate around
a single ``stable'' fixed point at $(z,\delta)=(0,0)$.
In Case A, we set $V_2=0$~MV, so that the double RF system effectively
reduces to the case of the single RF system.
Then we specify $V_1$ as the only independent variable,
which is related to the rms bunch length and the RF bucket height.
In Case B, we fix $\phi_{s2}=0$~rad, so that $eV_1\sin{({ \phi}_{s1})}=U_0$, 
and thus the setting of the fundamental RF system is the same as
that of a single RF system alone. For a specified $V_1$,
we choose various positive values of $V_{2}$ that lead to different rms
bunch lengths and different dependences of synchrotron tune on
the synchrotron amplitude. Note that this
reduces to Case A if $V_{2}=0$~MV.
In Case C, we consider a special type of settings called the ``ideal
bunch lengthening'' condition, with a vanishing slope of the
RF wave at the synchronous phase. This requires both the first and second derivatives of the total RF voltage at the synchronous phase be set to zero~\cite{hof}:
\begin{eqnarray}
    \frac{\partial V_{\text{tot}}}{\partial \phi}|_{\phi=0}&=&0 \nonumber \\
        \frac{\partial^2 V_{\text{tot}}}{\partial \phi^2}|_{\phi=0}&=&0
\end{eqnarray}
so that
\begin{eqnarray}
    \label{eq:ideal_lengthening}
    V_2&=&\sqrt{\frac{V_1^2}{n^2}-\frac{U_0^2}{n^2-1}} \nonumber \\
    \phi_{s1}&=&\pi-\arcsin(\frac{n^2}{n^2-1}\frac{U_0}{V_1})  \nonumber \\
    \phi_{s2}&=&-\arcsin(\frac{1}{n^2-1}\frac{U_0}{V_2})
\end{eqnarray}
We specify $V_1$ as the only independent variable, all other RF parameters
are determined using Eq.~\ref{eq:ideal_lengthening}.

In the following studies, we 
use the CEPC lattice with the wiggler settings of Case 1, 
and consider a fundamental RF system running at 650~\si{MHz} and a harmonic RF system running at 1.3~\si{\GHz}.

But before describing our numerical results for the three different cases of RF settings, we direct the reader to Appendix~(Section~\ref{sec:double_rf})
where the mathematical details are presented together with 
illustrative plots of the regions of stable motion in $\delta$ and $\phi$
and indications of the positions of the unstable fixed points and the 
turning points.  Then, with that formalism, we  evaluate 
the synchrotron tune $\nu_z$ at various synchrotron amplitudes as well as
the equilibrium longitudinal distribution $\rho(z)$ of particles 
under the influence of damping and excitation by synchrotron radiation.
For that we choose a beam energy of 45.72~GeV~($a\gamma_0=103.76$) and fix $V_1=103.372$~MV. Then  $\phi_{s1}=2.693$~rad in Case A
and Case B. For Case C, $\phi_{s1}=2.526$~rad, $V_2=44.750$~MV and $\phi_{s2}=-0.340$~rad. 

The dependences of $\nu_z$  on the orbital amplitudes of the three cases
are shown in Fig.~\ref{fig:double_rf}(a) where, as indicated in 
Appendix~(Section~\ref{sec:double_rf}), the amplitudes are represented
by the measure $z_{H_0}$.
As shown, a larger $V_2$ in Case B leads to 
a lower $\nu_z$ at zero synchrotron amplitude relative to that of Case A, and approaches a vanishing 
$\nu_z$ at zero synchrotron amplitude in Case C. 
The latter is expected
given that, as mentioned earlier, the slope of the RF wave is zero at 
the synchronous phase.
In addition, $\nu_z$ is essentially proportional to the amplitude of $z$ at small amplitude in Case C and the spread of $\nu_z$ among beam particles is much larger compared to that
of Case A and Case B. 
As shown in Fig.~\ref{fig:double_rf}(b), the longitudinal distribution $\rho(z)$ in the presence of damping and synchrotron radiation 
is Gaussian in Case A,
and features a ``flat-top'' and a much wider distribution in Case C.
The distribution  $\rho(z)$ becomes wider for increasing
$V_2$ in Case B and falls between the distributions of Case A and Case C.
In addition, if we choose an even larger $V_2$~(for example 50~MV) in Case B, then
$(z,\delta)=(0,0)$ is no longer a single stable fixed point that beam particles oscillate around.
Instead, there are two stable fixed points within the single RF bucket, so that
the longitudinal distribution
follows a ``double-hump'' shape, and the distribution of $\nu_z$ in terms of $z$ would be more complicated.
This ``over-stretched'' regime is discussed in Ref.~\cite{lee_accelerator_2004}, but is beyond the scope of this paper.

The applicability of the model with a uniform distribution of the RF system is illustrated in Appendix~(Section~\ref{sec:double_rf}) where 
one sees good agreement for  Case C between the predictions of the model and the distributions in $\delta$ and $z$ from a Monte-Carlo simulations with the two kinds of localised cavities and full synchro-betatron motion. 

Now we are ready to present the influence of the various settings of the double
RF system on the radiative depolarization effects. For Case A with a single RF system alone,
the results of equilibrium polarization for the CEPC have
already been shown in Fig.~\ref{fig:p_wiggler}(a).
Although different settings of the RF voltage and phase were used
relative to those of Case A in Fig.~\ref{fig:double_rf},
the Monte-Carlo simulation results agree with the 
prediction of the theory of the correlated regime.

For Case C at the ideal bunch lengthening condition, as already shown in Fig.~\ref{fig:double_rf}(a), 
the synchrotron tune $\nu_z$ is zero at zero amplitude. 
A direct application of Eq.~(\ref{eq:highersideband}), by plugging
$\nu_z=0$ at zero amplitude, then leads to a vanishing enhancement factor so that the expected equilibrium beam polarization of
the theory of the correlated regime is the same as the result of the first-order theory.
Meanwhile, the correlation index $\kappa$ is infinite and it is expected that the theory of the uncorrelated regime is applicable. 

\begin{figure}[hbt!]
\centering
\includegraphics[width=8.5cm]{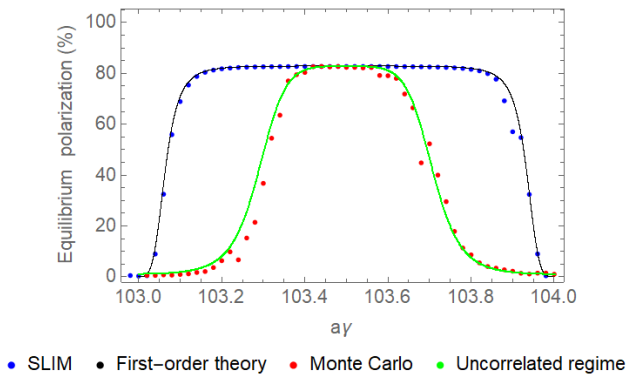}
\caption{The equilibrium polarization for the CEPC with the wiggler setting Case 1 and the double RF system under the ideal lengthening condition.}
\label{fig:har}
\end{figure}

We chose a beam energy of 45.6~GeV, and selected
the following parameters of the double RF system to match the ideal bunch lengthening condition:
$V_1=112.217$~MV, $\phi_{s1}=2.592$~rad, $V_2=50.047$~MV and $\phi_{s2}=-0.297$~rad. To simplify the treatment,
this RF setting was used in the simulations at different beam energies in the range of $a\gamma_0=103$ to $a\gamma_0=104$. Then the ``ideal bunch lengthening'' condition is exactly met for $a\gamma_0=103.5$, with minor deviations for other beam energies.
Fig.~\ref{fig:har} shows the equilibrium beam polarization for the CEPC.
On one hand, the results of SLIM calculations with equal
numbers of both kinds of localized cavities
match well with the prediction of the first-order theory. On the other hand, the results of Monte-Carlo simulations are in good agreement with the predictions for the theory of the uncorrelated regime, and are quite different from the predictions of the first-order theory.
These results are indeed consistent with the theory of the uncorrelated regime 
encapsulated in Eq.~(\ref{eq:uncorrelated}).

We also studied
the influence of the double RF settings in Case B to the radiative depolarization
effects. Here, we fixed the beam energy
at $ E=45.72$ \si{\GeV} ($a\gamma_0 \approx 103.76$), 
where there is a more obvious contrast in the behaviour of the equilibrium beam polarization.
We fixed the settings at $V_1=103.372$~MV, $\phi_1=2.693$ rad and $\phi_2=0$,
and scanned the total voltage of the harmonic RF system $V_2$ from $0$~MV up to $45.07$~MV, thus obtaining a range of synchrotron tunes at zero amplitude of synchrotron motion, from 0.028 down to 0.005.

\begin{figure}[hbt!]
\centering
\includegraphics[width=8cm]{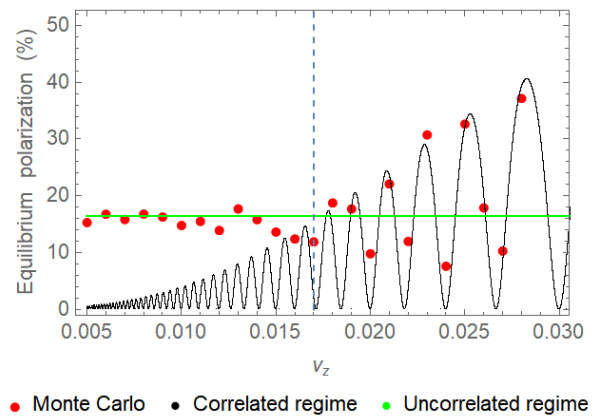}
\caption{The equilibrium polarization vs. synchrotron tune at zero amplitude for the CEPC lattice with the wiggler setting Case 1 at $a\gamma_0 \approx 103.76$. 
The blue dashed line corresponds to the condition $\kappa=1$. 
We chose a step size of $\Delta\nu_z=0.001$ in the Monte-Carlo simulations.}
\label{fig:pvsnuz}
\end{figure}

The outcome is shown in Fig.~\ref{fig:pvsnuz} where the
equilibrium beam polarization is plotted as a function of the synchrotron tune at zero amplitude and the results of Monte-Carlo simulation are
compared with the predictions of the theories.  
The theory of the uncorrelated regime encapsulated in Eq.~(\ref{eq:uncorrelated})
predicts that the equilibrium beam polarization is independent of the synchrotron tune, in strong contrast with the prediction of the theory of the 
correlated regime, whereby a closer examination of Eq.~(\ref{eq:highertorder}) indicates
that the equilibrium beam polarization should oscillate as the
distance to the nearest higher-order synchrotron sideband spin resonance varies
with $\nu_z$. Thus the dips correspond to certain higher-order synchrotron sideband
spin resonances and  as $\nu_z$ increases, so does the distance between adjacent dips
(and adjacent peaks), as well as the peak equilibrium beam polarization.
The results of Monte-Carlo simulation match well with the theory of the correlated regime for $\nu_z>0.025$, and are in line with the theory of the uncorrelated regime for
$\nu_z<0.01$. Between these two extremes, the simulation results suggest a transition from the correlated regime to the uncorrelated regime.
The vertical dashed line corresponds to $\kappa=1$ to guide the eye, but we can't detect  a clear boundary between the correlated regime and the uncorrelated regime.

In summary, when the synchrotron tune at zero amplitude is very small, in the presence of a harmonic RF system, the Monte-Carlo simulation results 
agree well with the theory of the uncorrelated regime. As the synchrotron tune at zero amplitude becomes larger, the Monte-Carlo simulations indicate a gradual
transition from the uncorrelated regime to the correlated regime, and match well with the theory of correlated regime when $\kappa\ll1$. 

As we already mentioned in Section II.B, the theory of the correlated regime was 
derived under the assumption that the synchrotron motion is approximately harmonic,
which is applicable in Case A.
However, in the presence of a double-RF system, the synchrotron
motion could become nonlinear even at small amplitudes. This is particularly true in Case C.
In Case B, as we increase the voltage of the harmonic RF system $V_2$, the synchrotron
motion at small amplitudes becomes more nonlinear and deviates from Case A. 
However, the nonlinear synchrotron motion of $\phi$ and  $\delta$
can still be represented
by action-angle variables, but in a much more complicated form involving elliptical functions~\cite{lee_accelerator_2004}.
It is unclear if a simple analytical form of $\partial \hat{n}/\partial \delta$ can be obtained
following the approach of derivation in Ref.~\cite{yok} in this context. But it is clear that
the application of the theory of correlated regime is limited to the case where
the assumption of approximately harmonic synchrotron motion is still valid.

\section{Conclusion and outlook \label{con}}
This paper presents a detailed investigation of the radiative depolarization effects in the CEPC. Both theoretical evaluations and
Monte-Carlo simulations are employed and compared, for a CEPC CDR lattice seed with detailed error modeling and after
dedicated corrections, and for various lattice settings. 

The contribution from the first-order ``parent" synchrotron spin resonances to radiative depolarization is explicitly derived
and its connection to the underlying integer spin resonance strengths is established, for ultra-high energy electron storage rings. This agrees with the expressions in Ref.~\cite{dks} for a vanishing synchrotron tune. SLIM simulations for the CEPC lattice seed agree well with this first-order theory for various  beam energies, and also indicate that the first-order ``parent" betatron spin resonances are much weaker compared to the first-order ``parent" synchrotron spin resonances. 

There are two distinct theories~\cite{dks} that describe the influence of synchrotron oscillations on  radiative depolarization at 
ultra-high beam energies. The theory of the correlated regime highlights the enhancement of the ``parent" spin resonances by higher-order
synchrotron sideband spin resonances, while the theory of the uncorrelated regime predicts vanishing resonance structures and a
generally higher equilibrium beam polarization level compared to the theory of the correlated regime. Monte-Carlo simulations have been
conducted for various  beam energies, various settings of asymmetric wigglers at the Z energy, as well as for various settings of
harmonic RF cavities. On one hand, 
these simulation results are in line with the theory of the correlated regime for a correlation index $\kappa\ll1$.
On the other hand, the simulation of the double-RF system in the ideal lengthening condition are consistent with the theory
of the uncorrelated regime, while $\kappa$ is infinite by definition. Between these two extremes, the Monte-Carlo 
simulations of the equilibrium beam polarization are generally between the predictions of these two theories, but 
are not in line with either.  
This indicates a gradual evolution from the correlated regime to the uncorrelated regime. A potential theoretical development
is foreseen, to achieve a more complete description of the radiative depolarization, that merges into these two theories at extremes.

In that context we make the following observations.
Electron and positron storage rings need accelerating cavities in order
to replace the particle energy lost by synchrotron radiation.
Then the particle dynamics is unavoidably intrinsically time-dependent
and that is recognised in \cite{DK72, K74, DK75, yok} among many others. Thus, for example, the vector $\hat n$ of the ISF is a function of all six canonical orbital variables as well as $\theta$.
Moreover, as in \cite{beznosov_wave_2020}, it can be argued that for phase-space equilibrium and for the relatively slow variation
of the polarization, the local polarization vector ${\vec P}_{\rm loc}(\vec u; \theta)$ at each point in six-dimensional phase space is parallel to $\hat n(\vec u;\theta)$ defined on six-dimensional phase space. This is the so-called ISF approximation. However, the ISF in \cite{eq,dks} is defined to be explicitly time-independent. Then in \cite{dks} synchrotron motion is added to the dynamics by hand. So what, then, is the direction of ${\vec P}_{\rm loc}(\vec u; \theta)$ at orbital equilibrium in that case?
In any case, we expect the ISF approximation to apply in the regime of non-resonant spin diffusion but it perhaps needs verification for the regime of resonant spin diffusion. 
Moreover, if the ISF is not relevant in the regime of 
resonant spin diffusion, what is the status of the terms 
$[1-\frac{2}{9}(\hat{n}\cdot \hat{s})^2]$ and $\hat{b} \cdot \hat{n}$ in the DK formula, Eq.~(\ref{eq:dk})? Should the averages involving $\hat n$ in those terms simply be replaced by averages involving $\hat n_0$ as is usually done anyway in practical evaluations of the depolarisation-free 
asymptotic polarization?

Fortunately, we can now evaluate the validity of the theory behind resonant spin diffusion and uncorrelated resonance crossing at ultra-high energy by following the history of individual spins using Monte-Carlo simulations. 
So $\nu_0$ could be set close to, or far away, from resonances and the effects on single spins of large synchrotron
amplitude or large photon energy could be studied in detail and perhaps 
then used to predict the behaviour of ensembles.

Monte-Carlo codes like Bmad can also be used to check and support analytical
calculations. For example, 
it has been seen that some perturbative calculations involving
${\partial \hat{ n}}/{\partial \delta} $ of the rate of depolarization for non-resonant spin diffusion need not converge
at high energy as more and more synchrotron sidebands are included \cite{mane_synchrotron_1990,mane_polarization_1992_2}. Then Monte-Carlo simulations should be used instead.
Of course, this matter is unrelated to the smoothing away of synchrotron sidebands in the regime of uncorrelated resonance crossing.

We should also note that the so-called Bloch equations 
\cite{heinemann_bloch_2019,beznosov_wave_2020,Heinemann:2020pzl}
foreseen in \cite{DK75} have the potential to expose deviations from the DK formula, Eq.~(\ref{eq:dk}), starting from first principles.

We are aware of developments in Bmad, in particular
its facility to gain speed by using pre-established spin-orbit maps~\cite{sagan_using_2022} 
for tracking between dipoles where radiation takes place, instead of tracking element-by-element. The increase in computing speed provided by this advanced feature, so-called "sectioning", should make it possible in future 
studies to track with many more particles, to launch finer parameter scans as well as to simulate the resonant depolarization process.

The results presented here refer to a single set of error seeds. Future work,
facilitated by sectioning, will employ several sets of error seeds in order to
obtain a better overview. For example, ensembles of distortions of $\hat{n}_0$ and deviations of $\nu_0$ from $a\gamma_0$ will be available.
The statistical spread of  $\nu_0$  due to closed-orbit distortions will give insights into how well the beam energy,  and then the center-of-mass  energy, can be estimated with RD~\cite{blondel_polarization_2019}. 
In addition, more depolarizing contributors like the detector solenoids, spin rotators to realize
longitudinal polarization,
and more complete machine imperfection sources, shall be included in future studies. These are important
for predicting the achievable beam polarization level, and for establishing methods for  realizing a high beam polarization.
Note that we didn't implement dedicated corrections such as in
harmonic closed-orbit spin matching schemes~\cite{rossmanithCompensationDepolarizingEffects1985,barberHighSpinPolarization1994}, which decrease the distortion of $\hat{n}_0$.
Applications of these correction schemes shall be carefully studied in order to optimize the achievable beam polarization, 
as part of an integrated approach of
performance optimization in these future colliders. 

As far as we know, no other studies for very high energy go as far as those presented here.


\section*{Acknowledgments}
W. H. Xia would like to thank S. Nikitin and F. Meot for early guidance in spin dynamics studies.
The authors are grateful to D. Sagan and E. Forest for kind help with Bmad/PTC. The authors acknowledge the proofreading and suggestions of S. Nikitin. D. P. Barber thanks K. Heinemann and J.A. Ellison for collaboration on many aspects of spin dynamics.
This study was supported by
National Key Program for S\&T Research
and Development (Grant No.~2018YFA0404300 and 2016YFA0400400);
National Natural Science Foundation of China (Grant No.~11975252);  Key Research Program of Frontier Sciences, CAS (Grant No. QYZDJ-SSW-SLH004); Youth Innovation Promotion Association CAS (No. 2021012).

\bibliography{cepc}

\section{Appendix}
\subsection{The calculation of $\vec{e}_y\cdot \hat{k}_0$ \label{eywk}}
$\hat{k}_0$ can be expanded in terms of $\hat{n}_{00}$, $\hat{k}_{00}$ and $\hat{k}_{00}^{\star}$ as
\begin{equation}
\hat{k}_0\approx\hat{k}_{00}+ c_1 \hat{n}_{00}+c_2 \hat{k}_{00} +c_3 \hat{k}_{00}^{\star}.
\label{eq:k0}
\end{equation}
where $|c_1|^2+|c_2|^2+|c_3|^2 \ll 1$.
In fact, it can be shown that $c_3=0$ due to the orthonormality of
$\hat{m}_0$ and $\hat{l}_0$.
Consider that $\hat{k}_{00}$ is in the horizontal plane, while $\hat{n}_{00} \parallel \vec{e}_y$,
\begin{equation}
\vec{e}_y \cdot \hat{k}_0 \approx c_1
\end{equation}

Putting Eq.~(\ref{eq:k0}) into the Thomas-BMT equation
Eq.~(\ref{eq:Yospinmotionrev}) on the closed orbit, we find
\begin{equation}
\frac{\partial c_1}{\partial \theta} \hat{n}_{00}=\Delta \vec{\Omega} \times \hat{k}_0 =i \Delta \vec{\Omega}\cdot \hat{k}_{00}
\end{equation}
then the solution of $c_1$ is
\begin{eqnarray}
c_1 &\approx& 
\substack { {\rm Lim} \\ \epsilon \rightarrow +0}
\left[ i\int^{\theta}_{-\infty}
e^{\epsilon \theta'}
\Delta \vec{\Omega} \cdot \hat{k}_{00}(\theta') d\theta' \right ] \nonumber \\
&\approx&
\substack { {\rm Lim} \\ \epsilon \rightarrow +0}
\left[ i\int^{\theta}_{-\infty}
e^{\epsilon \theta'}
\Delta \Omega_x e^{i\nu_0 \Phi(\theta')} d\theta' \right]
\label{eq:c1eq}
\end{eqnarray}
Here $\Phi(\theta')=R\int _0^{\theta'}\frac{1}{\rho_x}d\theta''$. $\Delta\Omega_x$ is the component of $\Delta \vec{\Omega}$ along the $x$ direction and represents the influence of the radial magnetic field.

Then, the integrand of Eq.~(\ref{eq:c1eq}) can be expanded into Fourier series,
\begin{equation}
\Delta\Omega_x e^{i\nu_0\Phi(\theta')}=\sum_{k=-\infty}^{\infty} {\tilde \omega}_k e^{i(\nu_0 - k)\theta'}
\end{equation}
${\tilde \omega}_k$ is the complex strength of the integer spin resonance,
\begin{equation}
    {\tilde \omega}_k = \frac{1}{2\pi }\int^{2\pi}_{0}\Delta \Omega_x e^{i\nu_0(\Phi(\theta')-\theta')+ik\theta'}d \theta'
\end{equation}
Eq.~(\ref{eq:c1eq}) can be simplified to the following form,
\begin{equation}
\vec{e}_y \cdot \hat{k}_0 (\theta) \approx i\sum_{k=-\infty}^{\infty} \frac{{\tilde \omega}_k e^{i(\nu_0-k)\theta}}{\nu_0-k}
\label{eq:eyk}
\end{equation}

Note that the denominators in this expression are an artefact of the approximations used here and that this expression must be used with care 
when $\nu_0$ is near an integer to avoid invalidating the approximations.

\subsection{$\xi_j$ of an ultra-high-energy electron storage ring\label{xi_0}}
The Fourier harmonic $\xi_j$ of the vertical component of $\vec{\omega}_z$ is
\begin{equation}
    \xi_j=-\frac{1+a\gamma_0}{2\pi}\int_0^{2\pi} R \eta_x G_x e^{ij\theta}d\theta
    \label{eq:xij}
\end{equation}

\begin{figure}[hbt!]
\centering
\includegraphics[width=0.9\columnwidth]{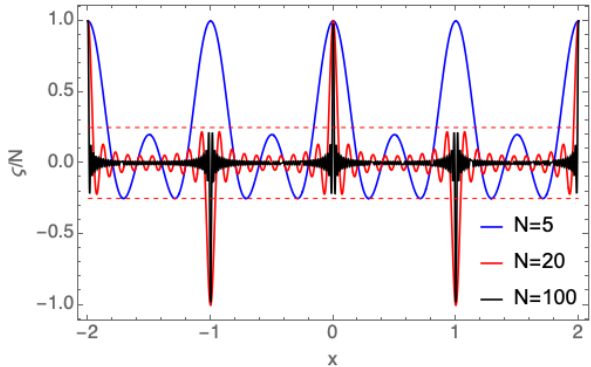}
\caption{The enhancement function $\varsigma(N,x)$ for various $N$. The blue dashed lines indicates $|\varphi/N|=1/4$ as a reference.}
\label{fig:varsigma}
\end{figure}

Let us consider an ultra-high-energy electron storage ring lattice composed of $P$ superperiods, where each superperiod contains a straight section with a length $L_\text{ss}$ which is dispersion free,
and an arc including $M$ FODO cells with half length $L$ with the following structure [~QF/2~~~B~~~QD~~~B~~~QF/2~].

\begin{figure*}[hbt!]
\centering
\subfigure[$|j|<100$]{
\begin{minipage}{\columnwidth}
\centering
\includegraphics[width=0.8\columnwidth]{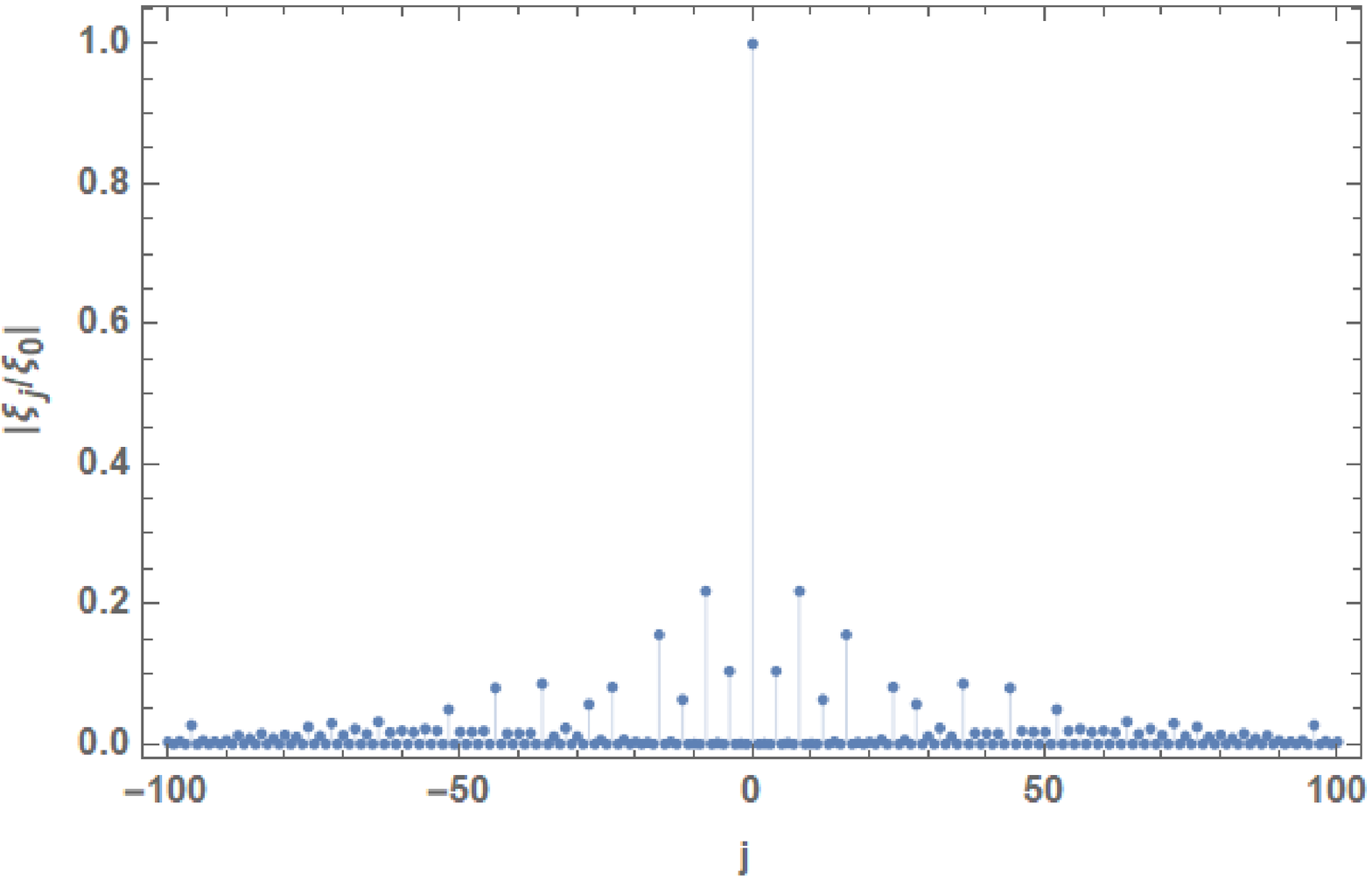} \\
\end{minipage}
}%
\subfigure[$|j|<3000$]{
\begin{minipage}{\columnwidth}
\centering
\includegraphics[width=0.8\columnwidth]{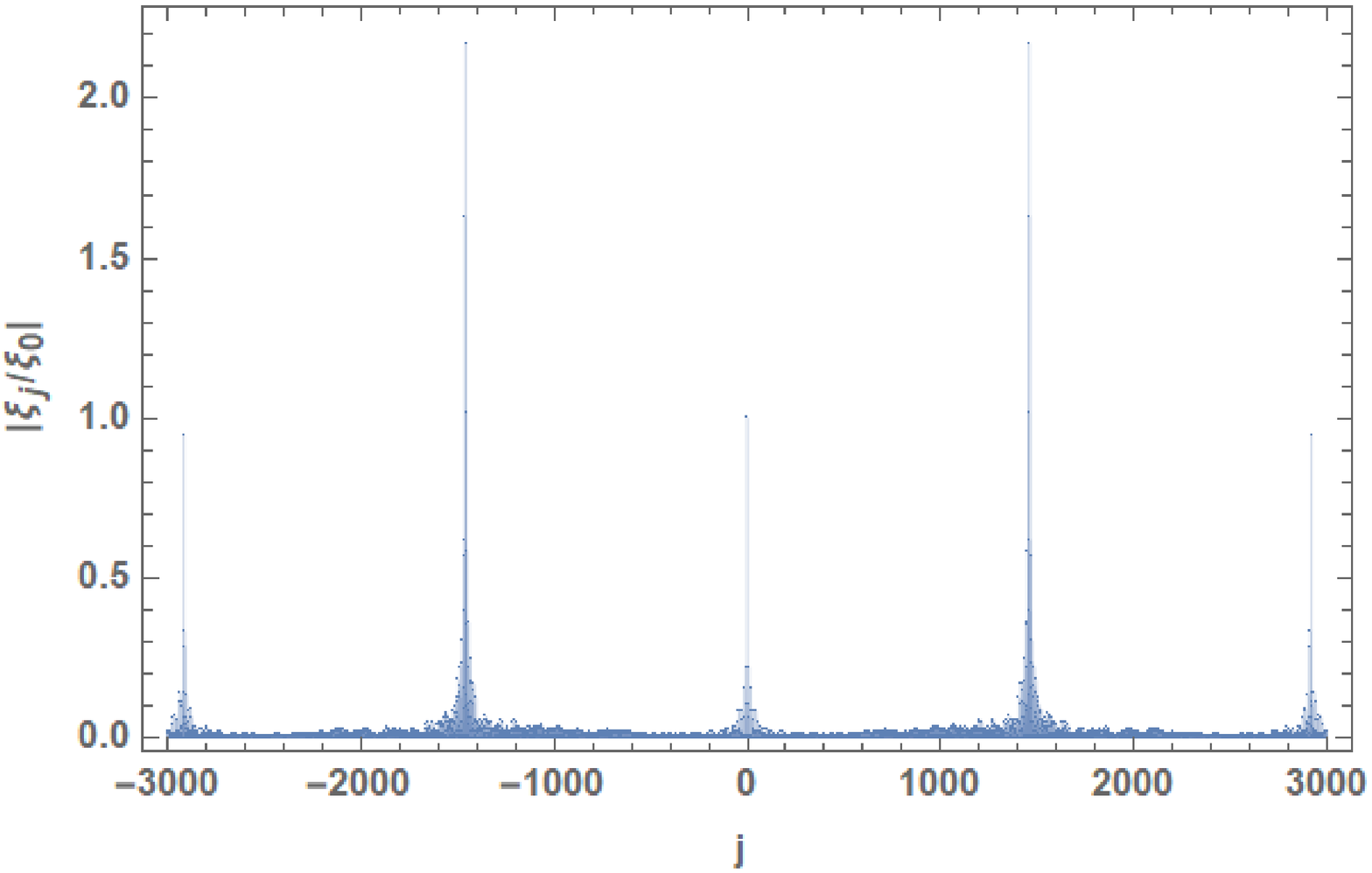} \\
\end{minipage}
}
\caption{$|\xi_j/\xi_0|$ for various $j$ of the CEPC lattice at the Z pole.  Peaks occur when $j$ is near  $kMP/(1-\frac{PL_{SS}}{2\pi R})$, $~k\in \mathbb{Z}$. }
\label{fig:xij}
\end{figure*}

Then, following the treatment of S. Y. Lee~\cite{leeSpinDynamicsSnakes1997}, the Fourier harmonic $\xi_j$ becomes
\begin{eqnarray}
   \xi_j&\approx&-\frac{1+a\gamma_0}{2\pi}[\frac{\eta_{QF,x}}{f_{QF}}\cos(j\frac{L}{R})+\frac{\eta_{QD,x}}{f_{QD}}] \times \nonumber \\
   &&\varsigma(P, \frac{j}{P})\varsigma(M,\frac{j}{MP}(1-\frac{PL_{\text{ss}}}{2\pi{R}})) \times \nonumber \\
   &&e^{ij\frac{L+L_\text{ss}}{R}}e^{ij\frac{P-1}{P}\pi}e^{ij\frac{M-1}{MP}\pi(1-\frac{PL_{\text{ss}}}{2\pi{R}})}
\end{eqnarray}
where the enhancement function $\varsigma(N,x)$ is given by
\begin{equation}
    \varsigma(N,x)=\frac{\sin(N\pi{x})}{\sin(\pi{x})}
\end{equation}
and where $N$ is a positive integer and $x\in\mathbb{R}$. Fig.~\ref{fig:varsigma} shows $\varsigma(N,x)/N$ for several different $N$. The following are some key properties of $\varsigma(N,x)$
\begin{itemize}
    \item  $\varsigma(N,x)=0$ when $N*x\in\mathbb{Z}$ while $x \notin \mathbb{Z}$
    \item  $|\varsigma(N,x)|\rightarrow {N}$, as $x$ approaches an integer $k$ 
    \item  for $N \geq 5$, $|\varsigma(N,x)| \leq \frac{N}{4}$ for $|x-k| \geq \frac{0.8}{N}$, where $k$ is the integer nearest to $x$.
\end{itemize}

Taking the thin lens approximation, the dispersion functions at the center of the 
focusing quadrupole~(QF) and defocusing quadrupole~(QD) of a FODO cell are
\begin{eqnarray}
\eta_{QF,x}&=&\frac{f_{QF}(4f_{QD} -L)\tilde \phi}{2f_{QF}+2f_{QD}-L} \nonumber \\
\eta_{QD,x}&=&-\frac{f_{QD}(-4f_{QF}+L)\tilde \phi}{2f_{QF}+2f_{QD}-L}
\end{eqnarray}
where $f_{QF}$ and $f_{QD}$ are the focal lengths of QF and QD, respectively. 
$\tilde \phi$ is the dipole bending angle in each half FODO cell: $\tilde \phi=\pi/(PM)$.
Note that with $L \ll R$, for small $j$,
\begin{equation}
    \frac{1+a\gamma_0}{2\pi}[\frac{\eta_{QF,x}}{f_{QF}}\cos(j\frac{L}{R})+\frac{\eta_{QD,x}}{f_{QD}}]\approx \frac{1+a\gamma_0}{PM}
\end{equation}

For $j=0$, the enhancement functions amount to $PM$, 
so that $\xi_0 \approx -(1+a\gamma_0)$. 
Then we consider $|\xi_j/\xi_0|$ with $|j|>0$, i.e., the influence of the enhancement functions.


At ultra-high beam energies, the total length of straight sections $PL_{\text{ss}}$ only occupies a small fraction of the ring circumference $2\pi{R}$, since 
 a high filling factor of dipoles is required to reduce the synchrotron radiation energy loss. Taking the CEPC lattice as an example, the fraction is 0.18. 

$\varsigma(P, \frac{j}{P})$ becomes $P$ only for $j=kP, ~~k\in \mathbb{Z}$ and is otherwise 0. In contrast, $|\varsigma(M,\frac{j}{MP}(1-\frac{PL_{\text{ss}}}{2\pi{R}}))|$ approaches $M$ when $\frac{j}{MP}(1-\frac{PL_{\text{ss}}}{2\pi{R}})$ approaches an integer $k$, but is less than $M/4$ when $|j| \geq P$ and $|j|<MP$. 
In addition, only a few $k$ near the integer part of $\nu_0$ have significant influence in Eq.~(\ref{eq:1stvert}). The contribution of terms with very large $|j|$ is further suppressed by the $(\nu_0-k+j)^2$ term in the denominator.
Therefore, we can retain only $\xi_0$ in Eq.~(\ref{eq:1stvert}) as a reasonable approximation.

As an illustration, Fig.~\ref{fig:xij} shows the results of simulation for $|\xi_j/\xi_0|$ for various $j$, using the CEPC lattice. The CEPC lattice includes 8 interleaved arc sections and straight sections. Each arc section
contains 145 regular FODO cells and 4 FODO cells with half bending angles as dispersion suppressors.  In effect $P=8$ and $M=149$. Then peaks of $|\xi_j/\xi_0|$ occur when $j$ is near $kMP/(1-\frac{PL_{SS}}{2\pi R})$, $~k\in \mathbb{Z}$, agreeing with the observation in the lower plot of Fig.~\ref{fig:xij}.
However, the 8 straight sections have different functions and thus different lengths. There are also chromatic-correction sections with dipoles in the final focus systems. These are different from the model that we analyzed, and somehow break the symmetry so that Fourier harmonics other than $j=8k$ have comparable strengths, as shown in the upper plot of Fig.~\ref{fig:xij}.
Nevertheless, the conclusion of our analysis still 
applies for the CEPC lattice, namely that the contribution of $\xi_0$ in Eq.~(\ref{eq:1stvert}) is much larger  compared to other Fourier harmonics.

\subsection{More detailed theory of a double-frequency RF system
\label{sec:double_rf}}

Our treatment of double RF systems follows the treatment in Chapter 3 of Ref.~\cite{lee_accelerator_2004}, but here
and elsewhere we use a different convention for  $\phi$, whereby $\phi$
is the RF phase of a particle
relative to that of the reference particle, rather than the phase coordinate $\phi_{\textrm{Lee}}$ relative to
the fundamental RF system in Ref.~\cite{lee_accelerator_2004}. Then $\phi=\phi_{\textrm{Lee}}-\phi_{s1}$.

Then with time $t$ as an independent variable, 
without considering radiation damping and quantum excitation,
the equations of motion are
\begin{eqnarray}
\label{eq:eq_motion}
\frac{d\phi}{dt}&=&h_1\omega_0\eta\delta \nonumber \\
\frac{d\delta}{dt}&=&
\frac{e\omega_0}{2\pi E_0 \beta^2}
    \Bigl[ V_1\sin(\phi+\phi_{s1}) + V_2\sin(n\phi+\phi_{s2}) \nonumber \\
    &&-\frac{U_0}{e} \Bigr]
\end{eqnarray}
where $\omega_0=c/R$, $\eta=\alpha_c-1/\gamma_0^2$, $\alpha_c$ is the momentum compaction factor and $h_1$ is the
harmonic number of the fundamental RF system.
Equivalently, a particle's longitudinal motion is described by the Hamiltonian
\begin{eqnarray}
\label{eq:Hamiltonian}
    H(\phi, \delta)&=&\frac{h_1\omega_0\eta}{2}\delta^2+
    \frac{e\omega_0}{2\pi E_0 \beta^2}U(\phi) \nonumber \\
    U(\phi)&=& 
    V_1\cos(\phi+\phi_{s1})+\frac{V_2}{n}\cos(n\phi+\phi_{s2}) \nonumber \\
    &&+\phi\frac{U_0}{e} -  V_1\cos(\phi_{s1})-\frac{V_2}{n}\cos(\phi_{s2})
\end{eqnarray}
This is a constant of motion for each particle and it is helpful
to think of the term with $\delta^2$ as being analogous to a kinetic 
energy whereas the term with $U(\phi)$ can be thought of as a potential energy while the Hamiltonion is analogous to a total energy.

To understand the synchrotron motion, one starts by solving for the fixed points,
i.e., the solution of $(\phi, \delta)$
such that 
\begin{eqnarray}\label{eq:fixed_point}
\frac{d\phi}{dt}&=&0 \nonumber \\
\frac{d\delta}{dt}&=&0
\end{eqnarray}
In our convention, we choose $(\phi,\delta)=(0,0)$ as one
such fixed point. Then the parameters of the double-frequency RF system must be set to compensate for the synchrotron-radiation energy loss per turn $U_0$, this leads to Eq.~(\ref{eq:U0}).
In the case of a single RF system, this is
a stable fixed point around which
small amplitude
particles
execute synchrotron oscillations. Besides,
there is an unstable fixed point at $(\phi,\delta)=(\pi-2\phi_{s1},0)$,
which corresponds to the maximum value of the Hamiltonian, i.e., boundary of the RF bucket.

\begin{figure}[hbt!]
\centering
\includegraphics[width=0.9\columnwidth]{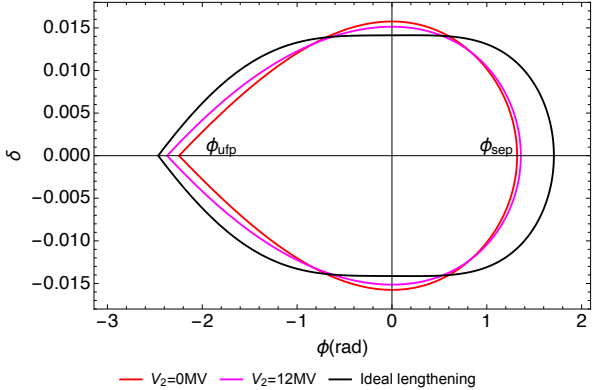}
\caption{The separatrices for various settings of a double-RF system.
The locations of $\phi_{\textrm{ufp}}$ and $\phi_{\textrm{sep}}$
are also labelled.
The red curve corresponds to Case A with a
single RF system alone, the magenta curve
corresponds to one instance of Case B with $\phi_{s2}=0$ rad but
$V_2=12$~MV, the black curve corresponds to Case C at the
ideal lengthening condition.}
\label{fig:rfbucket}
\end{figure}

In the case of
a double RF system, the number and locations of the fixed points can be obtained by numerically solving Eq.~(\ref{eq:fixed_point}). 
In this
paper, we focus on the cases with two fixed points, 
a stable fixed point at $(\phi,\delta)=(0,0)$ and an unstable fixed point 
at $(\phi,\delta)=(\phi_{\textrm{ufp}},0)$. The torus that
passes through the unstable fixed point is called the separatrix. This separates
phase space into regions of bound and unbound motions. We only study  particle motion around
the stable fixed point, and within the boundary of the RF bucket enclosed by the separatrix, i.e., just the bound motion.
Besides the unstable fixed point, there is another special point on the separatrix,
$(\phi,\delta)=(\phi_{\textrm{sep}},0)$, namely the turning point
at which the ``kinetic energy" is zero and the ``potential energy" is
the "total energy". Since the beam energy range is well above the transition energy in this paper, i.e., $\eta>0$, then
$-\pi<\phi_{\textrm{ufp}}<0<\phi_{\textrm{sep}}<\pi$.
Separatrices for Cases A, B and C with our settings are shown in  Fig.~\ref{fig:rfbucket}
together with their unstable fixed points and their turning points.
We see that for Case C the stable region in $\phi$ is indeed larger
than in cases A and B. All three are mirror-symmetric across the 
$\phi$ axis.

Longitudinal phase space trajectories of particles follow
Hamiltonian torii $H(\phi, \delta)=H_0$, where $H_0$ is a constant of motion on a torus.
The trajectories are also mirror-symmetric across the 
$\phi$ axis. Then with Eq.~(\ref{eq:Hamiltonian}), an  $H_0$ is the ``potential energy" for the
phase coordinates 
$\hat{\phi}_{-}$ and $\hat {\phi}_{+}$ with
~($\phi_{\textrm{ufp}}<\hat{\phi}_{-} < 0 <\hat {\phi}_{+} <\phi_{\textrm{sep}}$) 
at which ``kinetic energy" is zero, namely at $\delta=0$. On the other hand, the maximum ``kinetic energy" 
written as 
${h_1\omega_0\eta}{\hat{\delta}}^2/2$ occurs at the minimum of 
the ``potential energy", namely at $\phi = 0$ where $U(\phi) = 0$. 
Then
\begin{equation}
\label{eq:H0}
H_0=\frac{h_1\omega_0\eta}{2}{\hat{\delta}}^2=\frac{e\omega_0}{2\pi E_b \beta^2}U(\hat{\phi}_{-}) = \frac{e\omega_0}{2\pi E_b \beta^2}U(\hat{\phi}_{+}) 
\end{equation}
so that $H_0$ expresses the amplitude of the synchrotron motion
in terms of $\hat \delta$,  ${\hat \phi}_{-}$ or ${\hat \phi}_{+}$.

Eq.~(\ref{eq:Hamiltonian}) can be written as
\begin{eqnarray}
\frac{h_1\omega_0\eta}{2}\delta^2 = H_0 -
\frac{e\omega_0}{2\pi E_0 \beta^2} U(\phi)) \nonumber 
\end{eqnarray}

Then the synchrotron oscillation period can be
obtained by using this  to integrate the first equation of Eq.~(\ref{eq:eq_motion}):
\begin{equation}
    \label{eq:synchrotron_period}
    T_{z} = 
    \oint d\phi \left(  2 h_1 \omega_0 \eta \left[ H_0 - \frac{e\omega_0 }{2 \pi E_0 \beta^2} U(\phi) \right] \right)^{-1/2} 
\end{equation}
This can be numerically evaluated for a specified $H_0$,
so
that we can then obtain the synchrotron tunes, $\nu_z=\frac{2\pi}{T_z \omega_0}$,
plotted in Fig.~9{a}. 
For each $H_0$, a measure of the maximum longitudinal
distance of a particle from the center of the bunch is 
$z_{H_{0}} \equiv \frac{({\hat\phi}_{+} -  {\hat\phi}_{-})R}{2 h_1}$ and this, being more recognisable, is used  instead of $H_0$ as the horizontal axis in Fig.~9(a).

In the presence of radiation damping and quantum excitation,
the longitudinal phase-space distribution settles to an equilibrium given by the static solution of the Fokker-Planck equation~\cite{jowett_introductory_1987}, namely,
$\tilde{\psi}(\delta,\phi)\propto\exp(-\rm{constant}\times H(\phi,\delta))$.
This distribution is approximately Gaussian in $\delta$ with an rms relative energy spread $\sigma_{\delta}$,
while the distribution of $\phi$ is
\begin{equation}
\label{eq:phase_distribution}
    \tilde{\rho}(\phi)=A_{\rho} \exp
    \left(-\frac{eU(\phi)}{2\pi E_0 h_1 \eta \sigma_{\delta}^2}\right)
\end{equation}
where $A_{\rho}$ is a normalization factor chosen so that $\int_{\phi_{\textrm{ufp}}}^{\phi_{\textrm{sep}}} \tilde{\rho}(\phi)d\phi=1$.
The first-order and second-order beam moments can be numerically evaluated by
\begin{eqnarray}
    \label{eq:rms_bunch_length}
    \bar{\phi}&=&\int_{\phi_{\textrm{ufp}}}^{\phi_{\textrm{sep}}}
    \phi \tilde{\rho}(\phi) d\phi \nonumber \\
    \Sigma_{\phi}&=&\int_{\phi_{\textrm{ufp}}}^{\phi_{\textrm{sep}}}
    (\phi-\bar{\phi})^2 \tilde{\rho}(\phi) d\phi
    \label{eq:Sigmas}
\end{eqnarray}
The longitudinal distributions, $\tilde{\rho}(\phi)$, are shown 
in Fig.~9(b), and are obtained by numerically  evaluating the r.h.s. of  Eq.~(\ref{eq:phase_distribution}).
The horizontal axis in Fig.~9(b) is the longitudinal coordinate $z=-\beta c \Delta t=\frac{\phi R}{h_1}$.
The rms bunch length is then
\begin{equation}
    \sigma_z=\frac{\sqrt{\Sigma_{\phi}} R}{h_1}
    \label{eq:sigma_z}
\end{equation}
Note that the second-order beam moments at equilibrium are usually
calculated using the beam-envelope matrix as in ~\cite{ohmi_beam-envelope_1994}. 
That method is still applicable for calculation of 
the rms energy spread, but it fails to deliver the correct rms bunch
length in the presence of a double-RF system, because $U(\phi)$ now contains components higher than second order in $\phi$, which are not properly taken into account by that method.
In this case, one can use Eq.~(\ref{eq:Sigmas}) and Eq.~(\ref{eq:sigma_z})
to numerically evaluate $\sigma_z$, or use Monte-Carlo simulation to obtain
$\sigma_z$.

\begin{figure}[!hbt]
\centering

\subfigure[Distribution of relative energy deviation $\delta$.]{
\begin{minipage}{8cm}
\centering
\includegraphics[width=7.5cm]{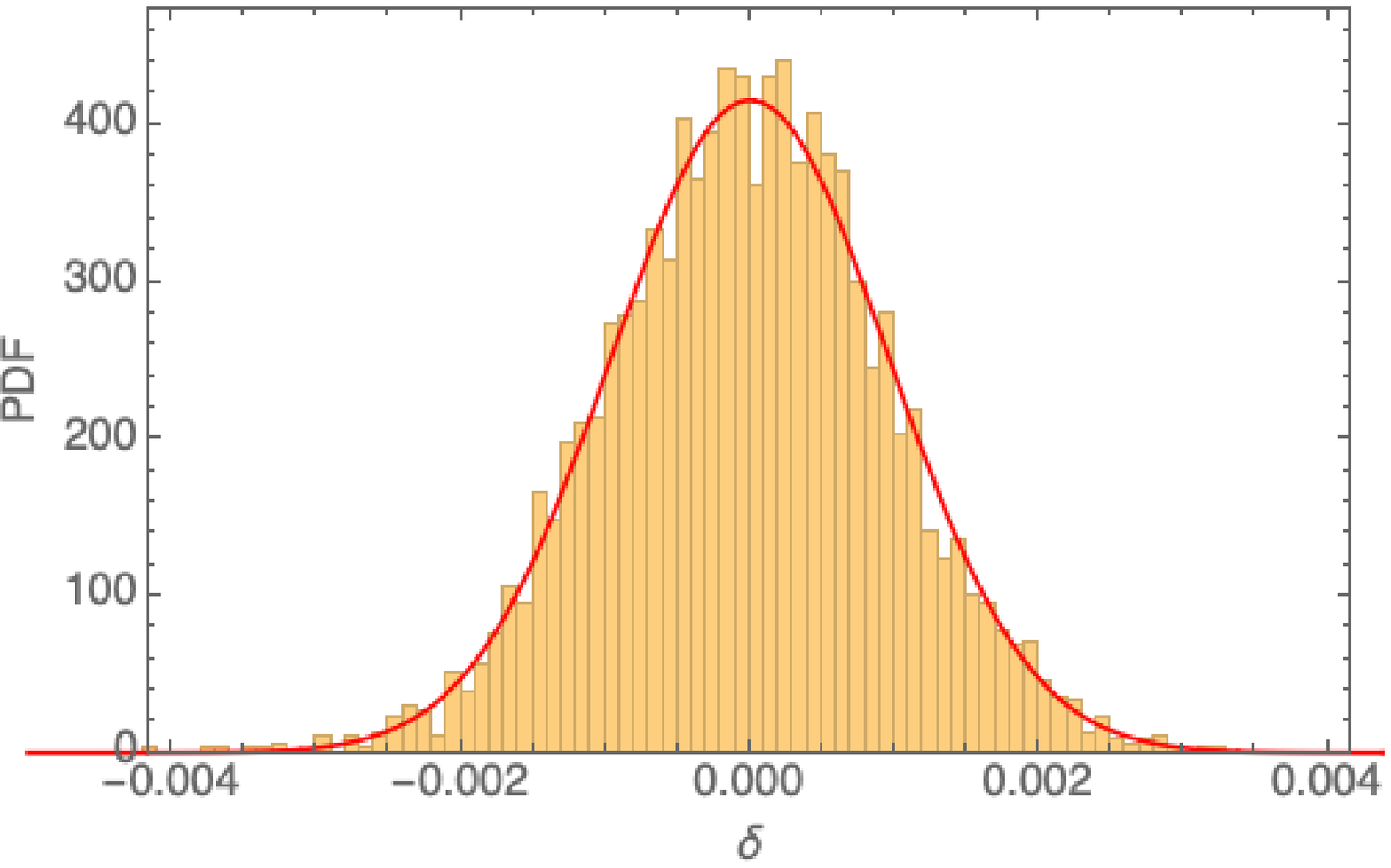} \\
\end{minipage}
}
\subfigure[Distribution of relative longitudinal positron $z$.]{
\begin{minipage}{8cm}
\includegraphics[width=7.5cm]{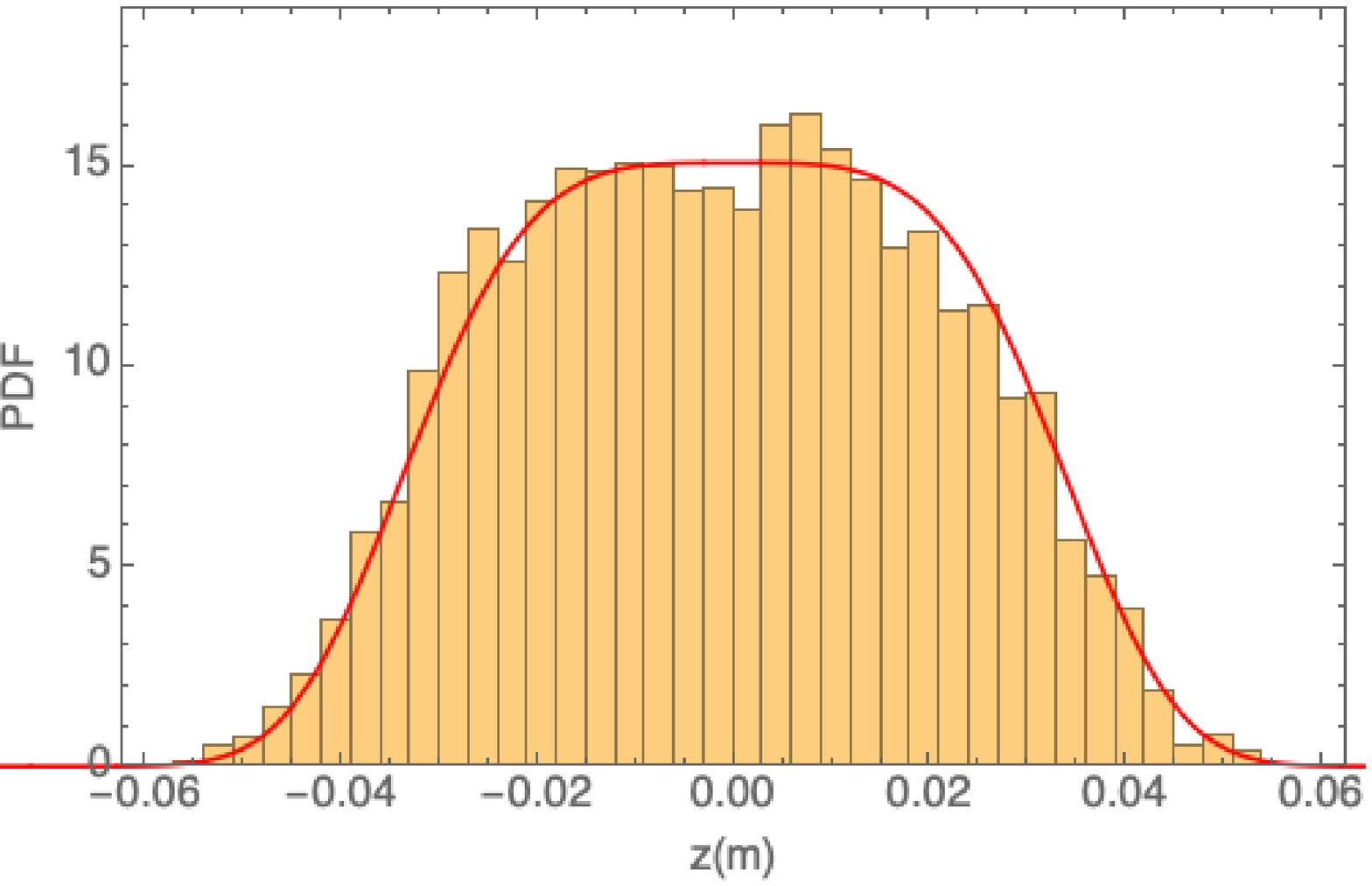} \\
\end{minipage}
}
\caption{ The longitudinal beam distributions of Case C at the ideal
lengthening condition. The red curves are the theoretical probability
density function~(PDF), the histograms are obtained from the final
beam distribution of a Monte-Carlo 
simulation, where 4000 particles are tracked for 10 damping times~(about 25000 turns).
The beam energy is set to 45.72 GeV~($a\gamma_0=103.76$) and $V_1=103.372$~MV.}
\label{fig:double_rf_distribution_comparison}
\end{figure}

In any case,
Fig.~\ref{fig:double_rf_distribution_comparison} shows the
probability density functions~(PDF) of the beam distributions in $\delta$
and $z$, for Case C at the ideal lengthening condition, where
the theoretical distributions match well with the histograms from the 
final beam distribution of a Monte-Carlo simulation with localised cavities
and full synchro-betatron motion tracking with 4000 particles for 10 damping times~(about 25000 turns). The beam energy is set to 45.72 GeV~($a\gamma_0=103.76$) and $V_1=103.372$~MV.

\balance

\end{document}